\documentclass[structabstract]{aa}
\usepackage{xspace}
\usepackage{graphicx}
\usepackage{natbib}
\usepackage{deluxetable}
\usepackage{amssymb}
\usepackage{longtable}
\usepackage{caption}
\usepackage{subfig}
\newcommand{\lsim}{\ensuremath{\,\lesssim\,}\xspace}
\newcommand{\gl}{\ensuremath{\ell}\xspace}
\newcommand{\gb}{\ensuremath{{\it b}}\xspace}

\newcommand{\lb}{\ensuremath{(\gl,\gb)}\xspace}

\newcommand{\kms}{\ensuremath{\,{\rm km\,s^{-1}}}\xspace}

\newcommand{\micron}{\ensuremath{\,\mu{\rm m}}\xspace}

\newcommand{\kpc}{\ensuremath{\,{\rm kpc}}\xspace}

\newcommand{\degree}{\ensuremath{\,^\circ}\xspace}

\newcommand{\msun}{\ensuremath{\,M_\odot}\xspace}     
 
\newcommand{\hii}{{\rm H\,{\scriptsize II}}\xspace}
\newcommand{\hiismall}{{\rm H\,{\scriptsize II}}\xspace}

\newcommand{\co} {\ensuremath{^{\rm 12}{\rm CO}}\xspace}

\newcommand{\phn}{\phantom{0}}

\begin{document}

\title{The Dust Properties of Bubble H{\large\,II} Regions as seen by
  $Herschel$\thanks{{\it Herschel} is an ESA space observatory with
    science instruments provided by European-led Principal
    Investigator consortia and with important participation from
    NASA.}}

\author{L.~D.~Anderson\inst{1, 7} \and
  A.~Zavagno\inst{1} \and
  L.~Deharveng\inst{1} \and
  A.~Abergel\inst{2} \and
  F.~Motte\inst{3} \and
  Ph.~Andr\'e\inst{3} \and
  J.-P.~Bernard\inst{4} \and 
  S.~Bontemps\inst{5} \and
  M.~Hennemann\inst{3} \and
  T.~Hill\inst{3} \and
  J.~A.~Rod\'on\inst{1} \and
  H.~Roussel\inst{6} \and 
  D.~Russeil\inst{1}}

\institute{Laboratoire d'Astrophysique de Marseille UMR\,6110, CNRS, Universit\'e de Provence, 38 rue F. Joliot-Curie, 13388 Marseille, France \and
  Institut d'Astrophysique Spatiale, UMR\,8617, CNRS, Universit\'e Paris-Sud\,11, 91405 Orsay, France  \and
  Laboratoire AIM Paris-Saclay, CEA/DSM–CNRS–Universit\'e Paris Diderot, IRFU/Service d’Astrophysique, CEA Saclay, 91191 Gif-sur-Yvette, France \and
  Centre d'\'etudes spatiales des rayonnements (CESR), Universit\'e de Toulouse (UPS), CNRS, UMR\,5187, 9 avenue du colonel Roche, 31028 Toulouse cedex 4, France \and
  CNRS/INSU, Laboratoire d'Astrophysique de Bordeaux, UMR\,5804, BP 89, 33271 Floirac cedex, France \and
  Institut d'Astrophysique de Paris, UMR\,7095 CNRS, Universit\'e Pierre \& Marie Curie, 98 bis Boulevard Arago, 75014 Paris, France \and
Current Address: Department of Physics, West Virginia University,
Morgantown, WV 26506, USA.\\Loren.Anderson@mail.wvu.edu}

\date{Received / Accepted}

\abstract{Because of their relatively simple morphology, ``bubble''
  \hii\ regions have been instrumental to our understanding of star
  formation triggered by \hii\ regions.  With the far-infrared (FIR)
  spectral coverage of the $Herschel$ satellite, we can access the
  wavelengths where these regions emit the majority of their energy
  through their dust emission.}  {We wish to learn about the dust
  temperature distribution in and surrounding bubble \hii\ regions and
  to calculate the mass and column density of regions of interest, in
  order to better understand ongoing star formation.  Additionally, we
  wish to determine whether and how the spectral index of the dust
  opacity, $\beta$, varies with dust temperature.  Any such
  relationship would imply that dust properties vary with
  environment.}  {Using aperture photometry and fits to the spectral
  energy distribution (SED), we determine the average temperature,
  $\beta$-value, and mass for regions of interest within eight bubble
  \hii\ regions.  Additionally, we compute maps of the dust
  temperature and column density.}  {At $Herschel$ wavelengths
  (70\,\micron\ to 500\,\micron), the emission associated
    with \hii\ regions is dominated by the cool dust in their
  photodissociation regions (PDRs).  We find average dust temperatures
  of $26$\,K along the PDRs, with little variation between the
  \hii\ regions in the sample, while local filaments and infrared dark
  clouds average $19$\,K and 15\,K respectively.  Higher
  temperatures lead to higher values of the Jeans mass, which may
  affect future star formation.  The mass of the material in the PDR,
  collected through the expansion of the \hii\ region, is between
  $\sim300$\,\msun and $\sim10,\,000$\,$\msun$ for the \hii\ regions
  studied here.  These masses are in rough agreement with the
    expected masses swept up during the expansion of the \hii\ regions.
  Approximately 20\% of the total FIR emission is from
    the direction of the bubble central regions.  This suggests that
  we are detecting emission from the ``near-side'' and ``far-side''
  PDRs along the line of sight and that bubbles are three-dimensional
  structures.  We find only weak support for a relationship between
  dust temperature and $\beta$, of a form similar to that caused by
  noise and calibration uncertainties alone.}{}


\titlerunning{Dust in Bubble HII Regions}
\authorrunning{Anderson et al.}
\maketitle

\section{Introduction\label{sec:intro}}
\hii\ regions that appear as a ring at infrared (IR) wavelengths, or a
``bubble'' seen in projection, have been the focus of numerous studies
of triggered star formation because of their relatively simple
morphology.  \hii\ regions expand as they age due to the pressure
difference between the ionized gas and the surrounding neutral medium
\citep[see][]{dyson}.  During this expansion, a layer of collected
neutral material can form on the border of the \hii\ region; within
this collected layer new stars may form.  This process is known as
``collect and collapse'' \citep{elmegreen77}.  It has been shown in
simulations that this layer may contain several thousand solar masses
\citep{hosokawa06}, a fact that has been confirmed for individual
\hii\ regions \citep{deharveng03, zavagno06, zavagno07, pomares09}.

The bubble morphology is common for Galactic \hii\ regions.
\citet{churchwell06, churchwell07} compiled a catalog of $\sim600$
bubbles detected at 8.0\,\micron\ in the {\it Spitzer} Galactic Legacy
Infrared Mid-Plane Survey Extraordinaire
\citep[GLIMPSE;][]{benjamin03}.  Nearly all identified bubbles enclose
\hii\ regions \citep{deharveng10, bania10, anderson11}, and
furthermore, nearly half of all \hii\ regions have a bubble morphology
\citep{anderson11}.  Most of these bubbles have surrounding material
that appears to have been collected through their expansion
\citep{deharveng10}.  Knowing the temperature of this material is
necessary to better estimate the mass and column density in the
collected layers, which in turn is necessary to determine the
efficiency of triggered star formation.

Although it contains just $\sim\,1$\% of the mass, dust plays a
significant role in the energetics of \hii\ regions.  Dust acts as a
coolant for \hii\ regions -- it absorbs high energy photons and
re-emits in the IR.  \citet{wc89a} and \citet{kurtz94} estimate that
for ultra compact (UC) \hii\ regions, dust absorbs between 42\% and
99\% of the ionizing photons.  The absorption by dust leads to a
slower expansion rate and a smaller physical size, stalling the
expansion of an \hii\ region earlier than it would otherwise.  This
phenomenon has been studied by \citet{mathis71}, \citet{petrosian72},
\citet{spitzer78}, and by \citet{arthur04}.

Observations of \hii\ regions from mid-IR to mm-wavelengths trace the
re-radiated energy from dust.  Such observations can be used to derive
the column density and mass distributions.  Using observations of dust
to estimate the total mass of gas and dust has the advantage of being
applicable over a large range of column densities.  Optically thin
far-IR (FIR) to mm-wavelength observations of dust can be used to
trace the mass distribution in high-density environments where
low-density gas tracers such as CO would freeze out onto dust grains,
and also in low-density environments where high-density gas tracers
would not be detectable.

Until recently, the dust temperatures of \hii\ regions were poorly
constrained.  Previous authors have used data at 12, 25, 60, and
100\,\micron\ from the {\it Infrared Astronomical Satellite} (IRAS) to
derive flux ratios (colors), and thus infer a dust temperature.  These
studies found that for nearly all \hii\ regions the flux at
100\,\micron\ is greater than the flux at 60\,\micron, implying dust
temperatures $\la\,30$\,K.  This phenomenon occurs for giant \hii\
regions (those with Lyman continuum photon emission rates $N_{\rm lyc}
> 10^{50}\,{\rm s}^{-1}$) \citep{conti04}, optically visible \hii\
regions \citep{chan95}, and ultra compact (UC) \hii\ regions
\citep{wc89a, crowther03}.
One of the warmer \hii\ regions known is M17, for which
\citet{povich07} found that the 60\,\micron\ flux is greater than the
100\,\micron\ flux.  With one data point on the Rayleigh-Jeans part of
the spectral energy distribution (SED), they were able to better
constrain the dust temperature.  They found a dust temperature in the
photodissociation region (PDR) of M17 of $\sim40$\,K and a dust
temperature of the ``interior'' in the direction of the ionized gas of
$\sim\,100$\,K.



With the advent of the {\it Herschel Space Observatory}
\citep{pilbratt10}, we have access to the crucial FIR regime long-ward
of the IRAS bands, at high angular resolution.  At the longest IRAS
wavelength band of 100\,\micron, the in-scan resolution was
$\sim3\arcmin$ and the out-of-scan resolution was $\sim 4\arcmin$.
The photometric bands of {\it Herschel} span 70\,\micron\ to
500\,\micron, at resolutions from $6\arcsec$ to $37\arcsec$.  Thus,
while only very rarely were \hii\ regions observed to have an IRAS
data point on the Rayleigh-Jeans side of the SED, {\it Herschel} can
provide a well-sampled SED with multiple data points in the
Rayleigh-Jeans side.  Furthermore, the resolution of {\it Herschel}
allows us to determine dust temperature variations within
\hii\ regions, which was largely not possible with IRAS.

The derived dust properties tell us about the grain population, which
in turn tells us about the physical conditions of the environment.
Aside from the temperature, we may also learn the wavelength
dependence of the dust opacity, $\kappa_\nu$.  This relationship can
be modeled as a power law \citep[cf.][]{hildebrand83}:
\begin{equation}
\kappa_\nu = \kappa_0 (\nu / \nu_0)^\beta\,,
\label{eq:kappa}
\end{equation}
where $\kappa_0$ is the opacity at frequency $\nu_0$ and $\beta$ is
the spectral index of the dust opacity.  This relationship does
not hold a shorter wavelengths in the mid-infrared (MIR) where the 
dust emission profile is more complicated \citep[see model in][]{compiegne11}.

Many authors have empirically found support for an inverse
relationship between the dust temperature and $\beta$.  For example,
such a relationship was found in PRONAOS observations of Galactic
cirrus and star-forming regions \citep{dupac03}, ARCHEOPS sub-mm point
sources (including numerous star formation regions) \citep{desert08},
BOOMERANG observations of Galactic cirrus \citep{veneziani10}, {\it
  Herschel} Hi-Gal \citep{molinari10} observations of the Galactic
plane \citep{paradis10},  {\it PLANCK} observations of nearby
  molecular clouds \citep{planck11a}, {\it PLANCK} observations of
  cold cores \citep{planck11b}, {\it BLAST} observations near the Vela
  complex \citep{martin11}, and individual \hii\ region environments
observed with {\it Herschel} \citep{anderson10a, rodon10}.  It
  has also been found in laboratory work of dust similar to that of
  the ISM \citep{mennella98, coupeaud11}.  Such a relationship,
however, can be falsely produced by measurement noise and line of
sight temperature variations \citep{shetty09a, shetty09b, juvela12}.
\citet{malinen10} find that a false relationship is caused in the
  vicinity of a heating source.  Classically, we would expect that
$\beta$ is between 1.0 and 2.0 \citep[see][and references
  therein]{tielens87} -- for a blackbody, $\beta$ is equal to zero.
The $\beta-T_d$ relationship has been proposed to arise naturally from
the disordered structure of amorphous dust grains \citep{meny07}.

Using {\it Herschel} science demonstration phase (SDP) data,
\citet{anderson10a}, derived dust temperatures and dust $\beta$-values
for the Galactic \hii\ region RCW\,120.  They found temperatures of
the dust associated with the \hii\ region from 20 to 30\,K, with
colder nearby patches of 10\,K associated with infrared dark clouds
(IRDCs).  A similar result was found by \citet{rodon10} for Sh\,104,
again using {\it Herschel} SDP data.  The results of both papers
were consistent with a relationship between $\beta$ and $T_d$, with the coldest regions having
$\beta$-values near 3.0.  In the present work, we extend these
analyses to a sample of eight ``bubble'' \hii\ regions observed by
{\it Herschel}.

\section{HII Region Sample\label{sec:hii_sample}}
Our sample includes eight bubble \hii\ regions: Sh\,104, W5-E,
Sh\,241, RCW\,71, RCW\,79, RCW\,82, G332.5$-$0.1, and RCW\,120.  With
the exception of W5-E and G332.5$-$0.1, all \hii\ regions in our
sample were listed as collect and collapse candidates in
\citet{deharveng05}.  All but one region, G332.5$-$0.1, have been
identified from their optical emission and thus extinction and the
amount of intervening dust along the line of sight is low.  The basic
parameters for the regions are given in Table~1, which lists the right
ascension, declination, Galactic longitude, and Galactic latitude of
the approximate center position, the approximate angular diameter, the
assumed distance, and the physical diameter.  Our sample targets span
a range of angular diameters from $3\arcmin$ to $34\arcmin$, distances
from 1.3\,kpc to 4.7\,kpc, and physical diameters from $2$\,pc to
20\,pc.  Three-color {\it Herschel} images composed of data from
500\,\micron, 250\,\micron, and 100\,\micron\ respectively in the red,
green, and blue channels are shown in Figure~\ref{fig:herschel}; these
data are discussed in \S\ref{sec:data}.  The contours in
  Figure~\ref{fig:herschel} show the radio continuum emission from the
  NRAO VLA Sky Survey \citep[NVSS;][]{condon98}, Sydney University
  Molonglo Sky Survey \citep[SUMSS;][]{bock99}, or the Green Bank
  6\,cm survey \citep[GB6;][]{gregory96}.  The radio continuum emission traces the ionized gas
  of the \hii\ regions.  These contours are only meant to show the
  strongest radio continuum components.  Due to the sensitivity of
  these surveys, much of the diffuse emission is not detected, including in some
  cases emission from the interior of the bubbles.


Throughout we refer to the direction of the central region of the bubbles as the
``interior,'' and the collected material delineating the bubble
structure as the ``PDR.''  For all regions in our sample, the interior
is spatially coincident with the ionized gas.  We use the term ``IRDC''
for any filament that is detected in absorption at 8.0\,\micron.

\begin{table*}\scriptsize
{\small \caption{Basic properties of the observed \hii\ regions}}
\begin{center}
\begin{tabular}{lccccccc}
\hline\hline
Name & RA (J2000) & Dec. (J2000) & $l$ & $b$ & d & Ang. Diam. & Phys. Diam. \\
& hh:mm:ss & hh:mm:ss & deg. & deg. & kpc & arcmin. & pc \\
\hline
Sh\,104	&	20:17:42	& $+$36:45:26	& \phantom{0}74.761	& $+$0.618	& 4.0	& \phantom{0}7	& \phantom{0}8\\
W5-E	&	02:59:41	& $+$60:32:01	&           138.073	& $+$1.488	& 2.0	&           34	&           20\\
Sh\,241	&	06:03:58	& $+$30:15:25	& 	    180.872	& $+$4.110	& 4.7	& \phantom{0}3 	& \phantom{0}6\\
RCW\,71	&	12:50:21	& $-$61:34:57	&           302.803	& $+$1.289	& 2.1	& \phantom{0}3	& \phantom{0}2\\
RCW\,79	&	13:40:17	& $-$61:44:00	&           308.681	& $+$0.592	& 4.0	&           10	&           12\\
RCW\,82	&	13:59:29	& $-$61:23:40	&           310.984	& $+$0.409	& 3.4	& \phantom{0}6	& \phantom{0}6\\
G332.5$-$0.1 &16:16:59        & $-$50:48:14   &           332.523	& $-$0.136	& 4.2	& \phantom{0}4	& \phantom{0}5\\
RCW\,120&	17:12:24	& $-$38:27:44	&           348.252	& $+$0.474	& 1.3	& \phantom{0}8	& \phantom{0}3\\
\hline
\end{tabular}
\label{tab:sample}
\end{center}
\end{table*}

\begin{figure*}[!ht]
  \centering
  \subfloat{\includegraphics[width=3.4 in]{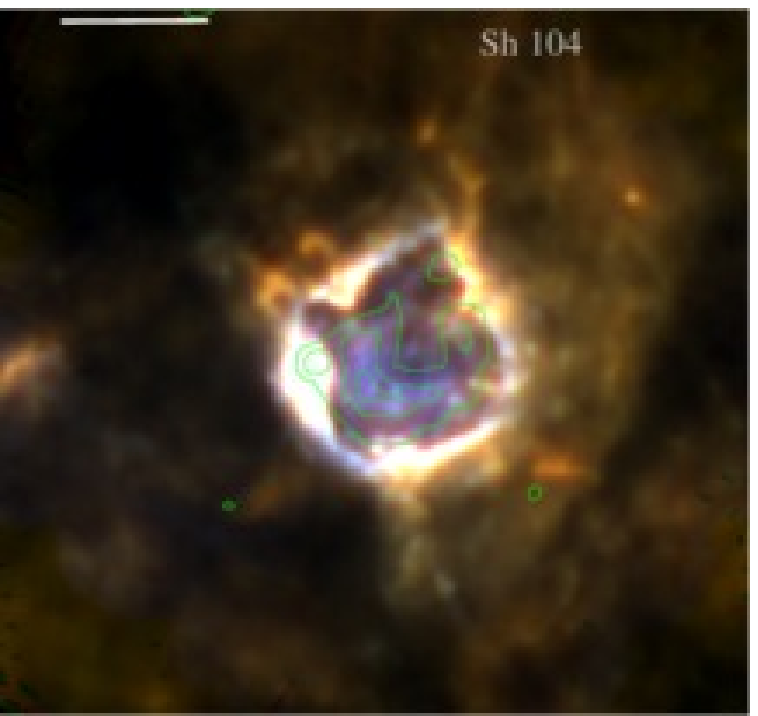}
    \hskip 0.00000001cm
    \includegraphics[width=3.4 in]{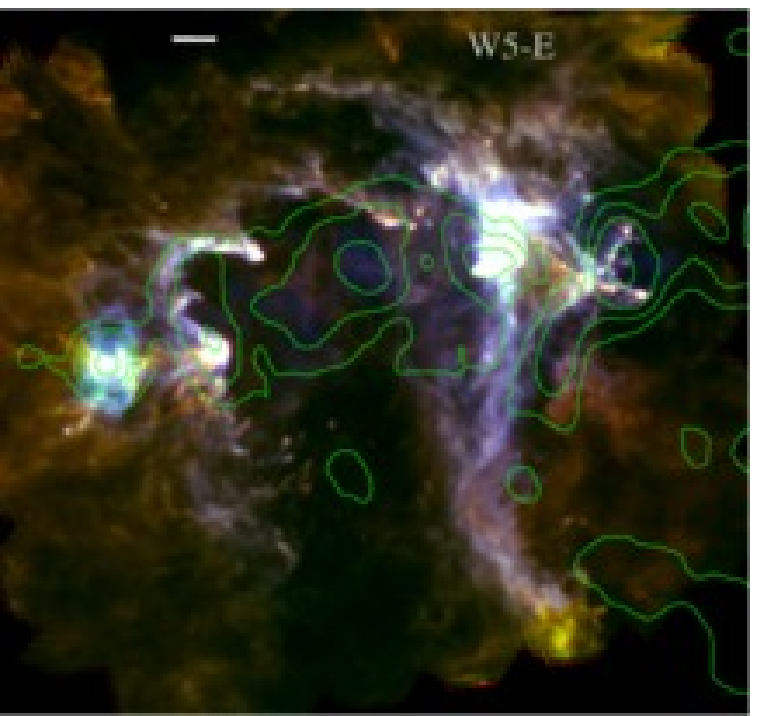}
  }
  \qquad
  \subfloat{\includegraphics[width=3.4 in]{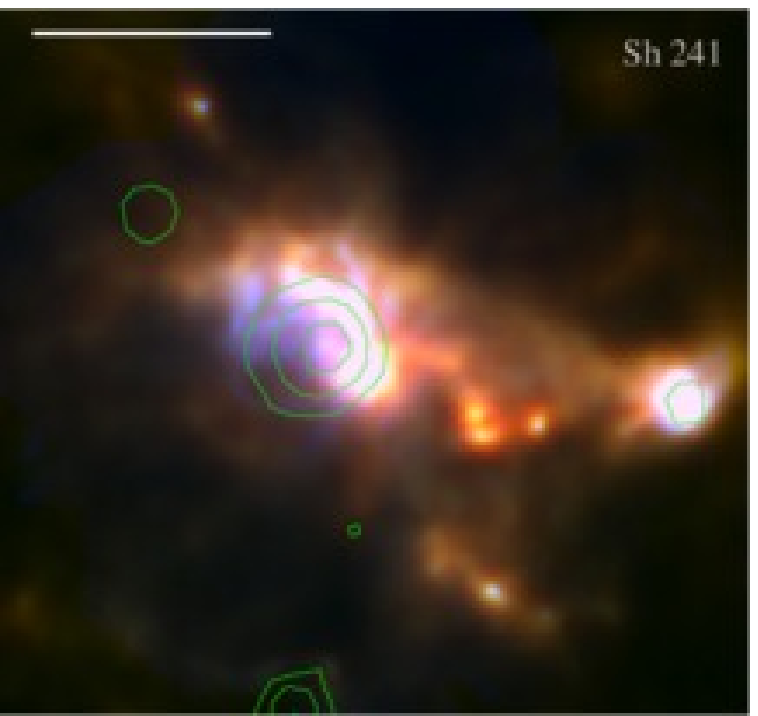}
    \hskip 0.00000001cm
    \includegraphics[width=3.4 in]{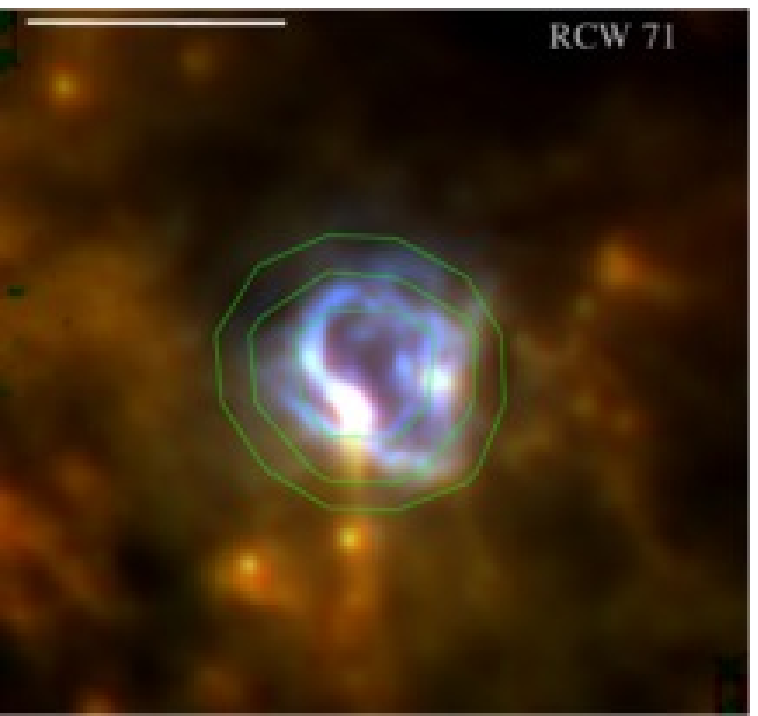}
  }
  \caption{Three-color {\it Herschel} images for the eight observed
    regions.  For each source the 500\,\micron\ data are shown in red,
    the 250\,\micron\ data in green, and the 100\,\micron\ data in
    blue.  The images are oriented in RA and Dec. such that north is
    up and east is left.  The white scale bar in the upper left corner
    is $5\arcmin$ in length.  The coldest regions in each field appear
    red while the warmest appear blue.  Contours show radio
      continuum emission from the NVSS (for Sh\,104, Sh\,241, and
      RCW\,120), SUMSS (for RCW\,79, RCW\,82, and G332) or GB6 (for
    W5-E and RCW\,71).}
 \label{fig:herschel}
\end{figure*}

\begin{figure*}[!ht]
  \ContinuedFloat
  \centering
  \subfloat{\includegraphics[width=3.4 in]{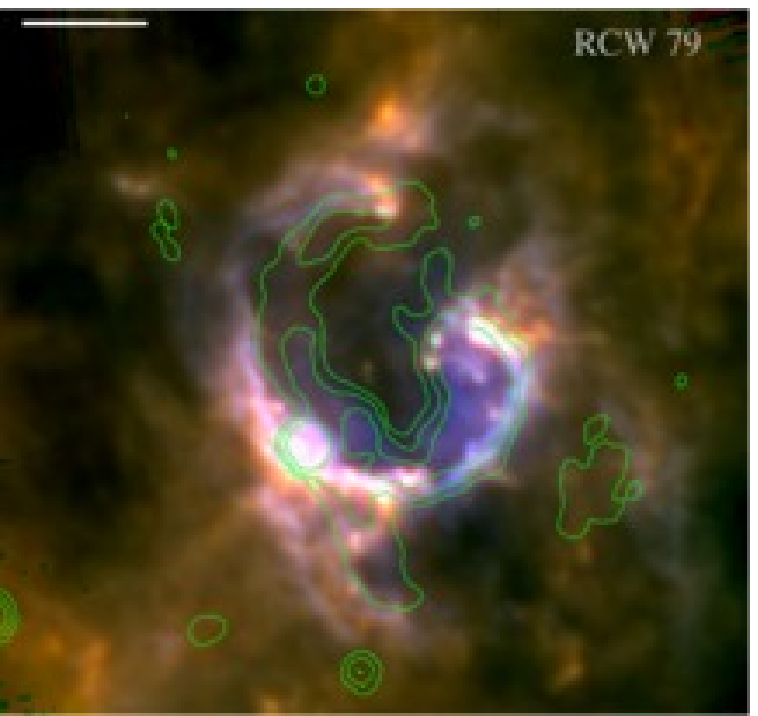}
    \hskip 0.00000001cm
    \includegraphics[width=3.4 in]{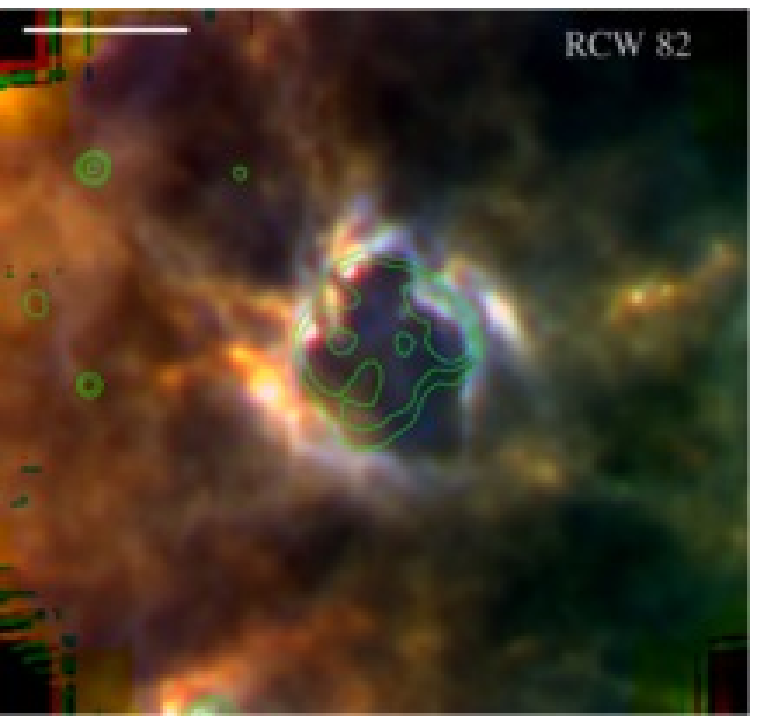}
  }
  \qquad
  \subfloat{\includegraphics[width=3.4 in]{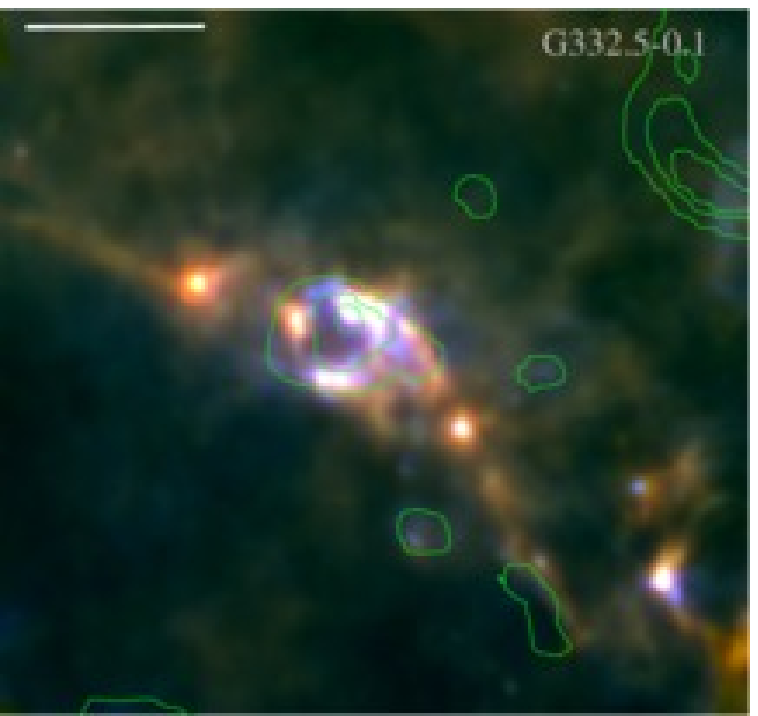}
    \hskip 0.00000001cm
    \includegraphics[width=3.4 in]{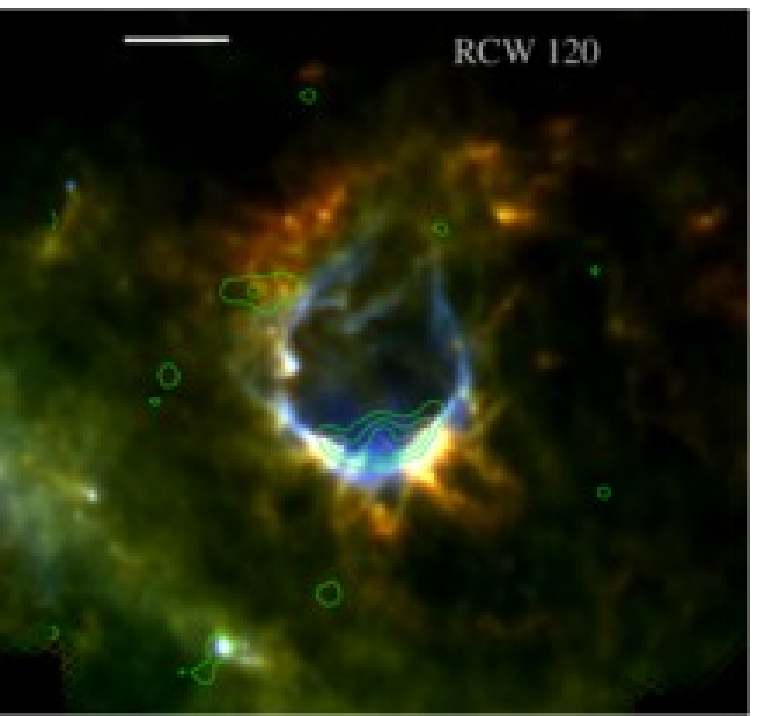}
  }
  \begin{flushleft}
    \caption{ -- continued.}
  \end{flushleft}
\end{figure*}

\subsection{Sh\,104}
Sh\,104 \citep{sharpless59} is a $\sim7\arcmin$-diameter \hii\ region.
It has a shell of material located along the PDR that was detected in
CO\,($2\rightarrow 1$) observations by \citet{deharveng03}.  They
estimate that the total mass in the shell is $\sim6,\,000$\,\msun.  CS
observations of the same field show four regularly-spaced dense
molecular clumps on the border of Sh\,104.  One clump, of mass
$\sim670\msun$, is spatially coincident with an \hii\ region
detected in the NVSS.  This
region is seen on the eastern PDR of Sh\,104 in
Figure~\ref{fig:herschel}.

\citet{russeil03} lists a kinematic distance for Sh\,104 of
$4.5^{+0.7}_{-0.8}$\,kpc and a spectroscopic, or spectro-photometric,
distance, of $3.1\pm0.9$\,kpc.  \citet{deharveng03} used 4\,kpc in
their analysis, which we adopt for the present work.  Sh\,104 is
excited by an O6V star \citep{crampton78, lahulla85}.  Given the
angular diameter, the physical diameter is $\sim\,8$\,pc.

Sh\,104 has strong 100\,\micron\ emission in its interior region (see
\ref{sec:warm}).  Also, it is the most ``complete'' of the
\hii\ regions in the sample in that the PDR is continuous and
surrounds the entire region.  These two phenomina may be related as
the completeness of the PDR may indicate that radiation pressure and
stellar winds are not particularly strong.  In this scenario, the
increased emission at 100\,\micron\ is a sign that the radiation is
trapped in the bubble interior.  There are numerous cold filaments
that begin on the border of Sh\,104 and lead radially away.


\subsection{W5-E}
W5-E \citep{westerhout58} is a $\sim34\arcmin$-diameter outer Galaxy
\hii\ region that has been the focus of numerous studies.  It is part
of the W3-W4-W5 complex; W4 and W3 are to the west of W5.  Our
observations are of W5-East, one part of the W5 region; the other
part is W5-West, which is a second bubble that appears to be
interacting with W5-East.  Here we refer to
``W5-East'' as ``W5-E.''

Here we adopt a distance of 2.0\,kpc for W5-E.  This distance is the
same as the maser parallax distance assigned to W3(OH),
$2.04\pm0.07$\,kpc \citep{hachisuka06}, and agrees well with estimates
from spectroscopic parallax measurements \citep{becker71, moffat72,
massey95}.  The main exciting star of W5-E is HD\,18326, which is of
spectral type O7V \citep{walborn73}.

The {\it Herschel} image in Figure~\ref{fig:herschel} shows a nearly
complete bubble, open to the south.  This southern opening is likely
due to a density gradient in the ambient medium, as can be inferred
from the CO observations in \citet{heyer98}.  There are also numerous
``bright-rimmed clouds'' (BRCs) and ``elephant trunks'' pointing toward the bubble
interior.  These features were seen in {\it Spitzer} observations of
the region \citep{koenig08}, and are prime sites for investigating
triggered star formation.
Our observations contain three BRCs cataloged by \citet{sugitani91}: BRC12, BRC13, and BRC14.

Point sources detected by {\it Spitzer} were analyzed by
\citet{koenig08}, who found two distinct populations of protostars; a
population of Class~II young stellar objects (YSOs) toward the
interior of W5-E and a population of Class~I YOSs in the surrounding
molecular material.  They hypothesize that these populations are
separated by age and that triggering may have caused the creation of
the Class~I population.

A detailed study of star formation in W5-E using these same {\it Herschel} data is given
in a companion paper, \citet{deharveng12}.

\subsection{Sh\,241}
Sh\,241 \citep{sharpless59} is a $\sim3\arcmin$-diameter
\hii\ region.  The region itself shows only faint H$\alpha$ emission
(Pomar\`es et al., 2012, in prep.), although there is a dense adjacent
molecular core that has been the focus of numerous studies in CS and
HCN \citep{plume92, pirogov99, shirley03, wu10}.  It has also been
mapped at 350\,\micron\ \citep{mueller02}, a molecular outflow has
been detected \citep{wu99}, and an H$_2$O maser has been detected
\citep{cesaroni88, henning92}.  \citet{moffat79} find that the
spectroscopic distance to Sh\,241 is $4.7$\,kpc.

Sh\,241 is the least complete of the bubble \hii\ regions in our
sample.  Figure ~\ref{fig:herschel} shows that Sh\,241 is open to the
south and defined in the north by material that emits strongly in the
FIR.  There are numerous condensations in the field, especially to the
west and south, and there is a separate \hii\ region to the west
detected in the NVSS.  We note that there is faint emission seen
  to the south in Figure ~\ref{fig:herschel} that may be part of a
  secondary PDR of Sh\,241.  It is unclear if this is truly a
  secondary PDR however and the reported $3\arcmin$ size refers only
  to the more compact emission seen in at the center of the {\it
    Herschel} data.

\subsection{RCW\,71}
RCW\,71 \citep{rcw60} is a $\sim3\arcmin$-diameter \hii\
region.  Compared to the other \hii\ regions in the sample, little is
known about RCW\,71.  The {\it Herschel} images in
Figure~\ref{fig:herschel} show numerous small condensations to the south and
west of RCW\,71 that are bright at SPIRE wavelengths.
The CO velocity of these condensations is similar to that of RCW\,71
and are therefore likely associated (M. Pomar{\`e}s, private communication).
The PDR of RCW\,71 seen in
Figure~\ref{fig:herschel} is filamentary.  
The spectroscopic distance of the exciting star of RCW71, HD311999, 
is 2.11 kpc while its near/far
kinematic distances are 3.05/6.15 kpc \citep{russeil03}.
The spectroscopic distance is in better agreement with distances to
nearby young stellar clusters \citep{russeil98} and we adopt
the spectroscopic distance of 2.1\,\kpc for the present work.  This gives a
physical diameter for RCW\,71 of $\sim2$\,pc.

\subsection{RCW\,79}
RCW\,79 \citep{rcw60} is a $\sim12\arcmin$-diameter \hii\ region.
Using 1.2\,mm continuum observations, \citet{zavagno06} found a
fragmented layer of neutral material along the PDR of RCW\,79 with a
total mass of $\sim2,\,000$\,\msun.  The mass of the most
massive condensation is $\sim1,\,000$\,\msun.  \citet{zavagno06}
detected several Class~I YSOs in the most massive condensations.
The bright compact source seen to the south-east in Figure~\ref{fig:herschel}
is a separate \hii\ region detected in radio continuum emission with
SUMSS.
Figure~\ref{fig:herschel} shows that RCW\,79 is open to the northwest.
To the south, there is what appears to be a second ionization front.

\citet{russeil03} find a kinematic distance of $4.8^{+0.9}_{-1.0}$\,kpc for
RCW\,79 and a spectroscopic distance of $4.0\pm$0.6\,kpc.  We adopt 4\,kpc
for the present work, which leads to a physical diameter of 12\,pc.
\citet{martins10} found that RCW\,79 is ionized by a cluster of a
dozen O~stars.

\subsection{RCW\,82}
RCW\,82 \citep{rcw60} is a $6\arcmin$-diameter \hii\ region.
\citet{pomares09} studied the molecular emission traced with CO isolopologues surrounding RCW\,82
and found a fragmented shell of emission with a total mass of
$\sim10,\,000$\,\msun. The most massive condensations have masses of
$\sim2,500$\,\msun.  The YSO population identified by \citet{pomares09} is
not evenly distributed, but is concentrated on the border of RCW\,82,
indicating that triggered star formation may have lead to their
formation.

RCW\,82 appears along an IRDC filament detected on both sides of the
\hii\ region.  The CO data in \citet{pomares09} show that this
filament is at the same velocity as RCW\,82, and that the filament and
\hii\ region are therefore associated.  There are numerous other
filaments seen in Figure~\ref{fig:herschel} that lead radially away
from RCW\,82; most are not seen in absorption at 8.0\,\micron.

There is some uncertainty as to the distance to RCW\,82.
\citet{russeil03} lists a kinematic distance of $4.3\pm0.7$\,kpc and a
spectroscoptic distance of $2.9\pm0.9$\,kpc.  \citet{pomares09} use a
distance of $3.4\pm0.9$\,kpc, which we adopt for the present work.
Given the angular diameter and this assumed distance, the physical
diameter of RCW\,82 is 6\,pc.  \citet{martins10} find that RCW\,82 is
ionized by two O9--B2V/III stars.

\subsection{G332.5$-$0.1}
G332.5$-$0.1 is a $3\arcmin$-diameter \hii\ region.  
It is located along a prominent IRDC running east-west, detected
in emission at longer wavelengths with {\it Herschel}, that is associated
with G332.5$-$0.1 in velocity (Deharveng et al., 2012b, in prep.).   
This filament, seen in Figure~\ref{fig:herschel}, has numerous
sources within it detected by {\it Herschel}.  G332.5$-$0.1
is unique in our sample in that hot dust at 100\,\micron\ is
faint in its interior.

On the northern border of G332.5$-$0.1 there is an associated UC \hii\
region that has been the focus of numerous studies as part of the Red MSX Source
survey \citep[RMS;][]{urquhart08}.  Just off the western border of
G332.5$-$0.1 there is another region of extended radio continuum
emission detected at 843\,MHz with SUMSS and also at 24\,\micron\ with
the MIPSGAL survey \citep{carey09} -- it is likely a distinct compact \hii\
region associated with G332.5$-$0.1.

The recombination line velocity of G332.5$-$0.1, $-50$\,\kms\ from
\citet{caswell87} places G332.5$-$0.1 at a kinematic distance of
3.7\,\kpc.  This assumes the near kinematic distance, which seems
likely given its association with IRDCs (IRDCs at the far distance
would be difficult to detect due to a lack of background behind the
cloud).  \citet{russeil05} have a slightly revised distance of
4.2\,kpc for G332.5$-$0.1 based on nearby \hii\ regions with a similar
velocity; we use 4.2\,kpc here which gives a physical diameter of
$\sim5$\,pc.

\subsection{RCW\,120}
RCW\,120 \citep{rcw60} is an $8\arcmin$-diameter \hii\ region that has
been the focus of numerous recent studies.  It has a massive,
fragmented layer of neutral material seen along its PDR traced at mm
\citep{zavagno07} and sub-mm \citep{deharveng09} wavelengths.  Of the
eight mm-condensations located by \citet{zavagno07}, five are found on
the PDR, indicating that this material has been collected during the
expansion of the \hii\ region.  \citet{deharveng09} estimated that the
mass of the collected layer is $\sim2,\,000$\msun.

RCW\,120 is among the nearest \hii\ regions to the Sun.
\citet{russeil03} lists a kinematic distance of
$1.8^{+0.6}_{-0.7}$\,kpc and a stellar distance of $1.3\pm0.4$\,kpc
for RCW\,120; we adopt 1.3\,kpc for the present work.  Given the
angular size, the physical size of RCW\,120 is $\sim3$\,pc.  A
single star is ionizing RCW\,120: CD$-38\degree11636$.
\citet{georgelin70} found that its spectral type is O8 from
spectroscopic measurements while \citep{crampton71} find that it is an
O9.  More recently, using near-IR integral field spectroscopy,
\citet{martins10} found that RCW\,120 is ionized by a O6-8V/III star.

There is significant active star formation in the surroundings of
RCW\,120.  \citet{zavagno07}, \citet{deharveng09}, and
\citet{zavagno10a} all located YSOs in the field.  The most massive
condensation is known as ``Condensation\,1'' and is located to the
south-west in Figure~\ref{fig:herschel}.  Within Condensation\,1,
\citet{deharveng09} found an evenly spaced chain of eleven Class~I or
flat spectrum sources parallel to the IF.  The authors suggest that it
is an example of the Jeans gravitational instability.  Using the same
{\it Herschel} data shown here, \citet{zavagno10a} detected a
massive YSO towards Condensation~1 with a stellar mass of 8--10\msun.  This condensation
was detected previously at 70\,\micron\ by {\it Spitzer}
\citep{deharveng09}.  \citet{zavagno10a} suggest that this is the first
detection of a massive Class~0 object formed by the collect and
collapse process on the border of an \hii\ region.

There are also numerous IRDCs seen in the periphery of RCW\,120, many
with point sources observed in the mid-infrared suggestive of active
star formation.  The largest IRDC, G348.40+00.47, is observed to the
north-east of RCW\,120 in Figure~\ref{fig:herschel}.
\citet{jackson08} measure a CS velocity of $-6.8$\,\kms\, for 
G348.40+00.47, which is close to the recombination line velocity of
the ionized gas in RCW\,120, $-12$\,\kms\ \citep{caswell87}.  The IRDC and the \hii\
region are therefore likely associated.  We assume the
smaller IRDCs to the north and west are also at the same distance as RCW\,120.
There are numerous cold filaments to the south of
RCW\,120 observed by {\it Herschel}, some of which are also seen in
absorption at 8.0\,\micron.

\section{Observations and Data\label{sec:data}}
In addition to the {\it Herschel} data that is the main focus of this
work, we utilize auxilliary data from {\it Herschel} Hi-Gal, {\it
  Spitzer} GLIMPSE, {\it Spitzer} MIPSGAL, and APEX-LABOCA ATLASGAL.
These auxillary data are available for four of our objects: RCW\,79,
RCW\,82, G332.5$-$0.1, and RCW\,120.

There are numerous point sources detected at all wavelengths for the
\hii\ regions in our sample.  We make no attempt to remove them
from the data.  This is primarily because we cannot with confidence separate
the sources from the local background at all wavelengths, while
ensuring that we are locating the same source at all wavelengths.
For example, an embedded source in a small IRDC can easily be
identified at MIR wavelengths, but at the longest {\it Herschel}
wavelengths with lower angular resolution, its emission 
extends over the entirety of the IRDC.  If we were to remove this source,
we would likely be removing a large fraction of the flux from the IRDC
as well, preferentially for data with lower angular resolution.

\subsection{{\it Herschel} HOBYS and ``Evolution of Interstellar Dust''}
Our sources were observed by the {\it Herschel Space Observatory} with
the PACS \citep{poglitsch10} and SPIRE \citep{griffin10}
instruments\footnote{The instrument parameters and calibration for
  PACS and SPIRE listed here are given in the PACS Observers' Manual,
  HERSCHEL-HSC-DOC-0832 Version 2.1 and the SPIRE Observers’ Manual,
  HERSCHEL-HSC-DOC-0789 Version 2.1.  Up-to-date versions of both
  manuals can be found here:
  http://herschel.esac.esa.int/Documentation.shtml .}  as part of the
HOBYS \citep{motte10} and ``Evolution of Interstellar Dust"
\citep{abergel10} guaranteed time key programs.  Data were taken in
five wavelength bands from 100\,\micron\ to 500\,\micron:
100\,\micron\ and 160\,\micron\ for PACS at FWHM resolutions of
$6.7\arcsec$ and $11\arcsec$, and 250\,\micron, 350\,\micron, and
500\,\micron\ for SPIRE at FWHM resolutions of $18\arcsec$, $25\arcsec$,
and $37\arcsec$.  For both instruments, the observations were composed
of two orthogonal scans.  For all observations with the PACS
instrument, the scan-speed was $20\arcsec$ per second while for SPIRE
it was $30\arcsec$ per second.  The observational parameters are
summarized in Table~2, which lists for each source
the field size, the total integration time, the observation
identification numbers, and the observation date for observations with
the PACS and SPIRE instruments.  These data are available from
the {\it Herschel Science Archive}\footnote{http://herschel.esac.esa.int/Science\_Archive.shtml}.

Calibration for the 100\,\micron\ and 160\,\micron\ PACS bands using
five standard stars has been found to be good to within $\sim 3\%$ at
100\,\micron\ and $5\%$ at 160\,\micron\ (see PACS Observers' Manual).
The extended source calibration is good to within $\sim10\%$ at {70\,\micron\ and}
  100\,\micron\ and $\sim20\%$ at 160\,\micron; we adopt the latter
  numbers here.  For SPIRE, the calibration is good to within 10\%
for all bands (see SPIRE Observers' Manual).  The SPIRE data are in Jy
per beam and therefore we must know the beam size to convert to a flux
density in Jy.  Because of diffraction effects, the SPIRE beams are
not entirely Gaussian -- for flux calculations we use the modified
beam areas of $423\arcsec^2$, $751\arcsec^2$, and $1587\arcsec^2$ for
the 250\,\micron, 350\,\micron, and 500\,\micron bands
\citep[see][]{sibthorpe11}.  The beam area at 250 is $\sim\,10\%$
larger than would be expected for a purely Gaussian beam while those
at 350\,\micron\ and 500\,\micron\ are approximately that expected.

\begin{table*}\scriptsize
{\small \caption{{\it Herschel} observational parameters}}
\begin{tabular}{l|cccc|cccccc}
\hline\hline
& \multicolumn{4}{c|}{\it PACS} & \multicolumn{4}{c}{\it SPIRE} \\
Name & Size & Time & ObsIDs & Date & Size & Time & ObsID & Date \\
& arcmin. & sec. & & yyyy-mm-dd & arcmin. & sec. & & yyyy-mm-dd \\
\hline
Sh\,104 &                     $20\times20$ &           \phantom{0}1730 & 1342185575, 1342185576 & 2010-10-10 &                     $12\times12$ & \phantom{0}629 & 1342185535 & 2009-10-06 \\
W5-E    &                     $70\times70$ &                     13194 & 1342191006, 1342191007 & 2010-02-23 &                     $65\times60$ &           4309 & 1342192088 & 2010-03-11 \\
Sh\,241	&                     $20\times20$ &           \phantom{0}1730 & 1342218725, 1342218726 & 2011-04-17 &                     $15\times15$ & \phantom{0}770 & 1342192093 & 2010-03-11 \\
RCW\,71 & $\phantom{0}9\times9\phantom{0}$ & \phantom{0}\phantom{0}402 & 1342189389, 1342189388 & 2010-01-16 & $\phantom{0}6\times6\phantom{0}$ & \phantom{0}332 & 1342203561 & 2010-08-23 \\
RCW\,79 &                     $30\times30$ &           \phantom{0}2768 & 1342188880, 1342188881 & 2010-01-03 &                     $19\times19$ & \phantom{0}837 & 1342192054 & 2010-03-10 \\
RCW\,82 &                     $20\times20$ &           \phantom{0}1354 & 1342188882, 1342188883 & 2010-01-03 &                     $12\times12$ & \phantom{0}571 & 1342192053 & 2010-03-10 \\
G332.5$-$0.1&	              $22\times22$ & \phantom{0}\phantom{0}817 & 1342204085, 1342204086 & 2010-09-05 &	                   $10\times10$ & \phantom{0}546 & 1342192055 & 2010-03-10 \\
RCW\,120&                     $30\times30$ &           \phantom{0}3302 & 1342185553, 1342185554 & 2009-10-09 &                     $22\times22$ &           1219 & 1342183678 & 2009-09-12 \\
\hline
\end{tabular}
\label{tab:observations2}
\end{table*}

We reduce all {\it Herschel} data using slightly modified
  versions of the default PACS and SPIRE pipelines within the
Herschel Interactive Processing Environment (HIPE) software, version
7.1.  The level~2 pipeline-reduced data products from both
instruments suffer from striping artifacts in the in-scan directions.
Additionally, the level~2 PACS data have artifacts (flux decrements)
around bright zones of emission caused by the median filtering
baseline removal.  To limit both the striping and the artifacts, we
use the {\it Scanamorphos}
software\footnote{http://www2.iap.fr/users/roussel/herschel/index.html}
\citep{roussel12}, version 9.  {\it Scanamorphos}
estimates the true measured value at each sky position by exploiting
the redundancy in the data.   We use {\it Scanamorphos} without
  the ``Galactic'' option; in this configuration, the large-scale
 gradients of the sky emission are removed.  During a typical {\it Herschel} observation, each sky
position is observed by multiple bolometers in multiple scans.  These
multiple observations of the same sky positions can be used to
estimate the true sky level.  We have found {\it Scanamorphos} to be
very effective at limiting striping without the addition of artifacts
around bright emission zones.  {\it Scanamorphos} also estimates the
photometric uncertainty at each sky position using the weighted
variance of all the samples contributing to each pixel.  These
uncertainty maps are useful because at present there is no propagation
of the errors associated with each processing step in HIPE.  We use
these uncertainty maps in the aperture photometry described in
\S\ref{sec:dustprop}.

The astrometry of the PACS and SPIRE {\it Herschel} data show offsets
relative to each other, but also relative to higher-resolution {\it
  Spitzer} data.  We found that the two PACS bands are well-aligned
with respect to each other, as are the three SPIRE bands.  To mitigate
the astrometry effects, we align the PACS~100\,\micron\ data to the
MIPSGAL data using point sources detected at both 100\,\micron and
24\,\micron\ in the {\it Spitzer} MIPSGAL survey (see below).  We use
the same astrometry solution for the PACS~160\,\micron\ data.  Using
point sources detected at 250\,\micron\ and 160\,\micron, we then
align the SPIRE~250\,\micron\ data to the PACS~160\,\micron\ data and
accept the same solution for the SPIRE~350\,\micron\ and
SPIRE~500\,\micron\ data.  Offsets were $<10\arcsec$ for all regions,
in all PACS and SPIRE bands.

The {\it Herschel} data for RCW\,120 and Sh\,104 were first shown
  in \citet{anderson10a} and \citet{rodon10}, respectively.  The data
  presented in these articles were processed with slightly modified
  versions of the default PACS and SPIRE pipelines within HIPE version
  2.0.  Some of our results differ from those of
  \citet{anderson10a} and \citet{rodon10}, in part as a result
  of this new processing (see Section~\ref{sec:beta_t}).

The new data processing differs from the old processing in that:
  1) baselines (i.e. drifts on timescales larger than the scan leg
  crossing time) are now estimated by {\it Scanamorphos} using the
  redundancy in the data instead of a high-pass filter; 2)
  small-timescale drifts are also derived from the redundancy in {\it
    Scanamorphos} (they were either neglected [for SPIRE], or
  attenuated by Fourier filtering [for PACS] within HIPE); 3)~data
  projection is done using a projection matrix in {\it Scanamorphos}
  for SPIRE data, which prevents biases between beam center and the
  projection center (HIPE used a nearest-neighbor projection); and 4)
  {\it Scanamorphos} includes relative gain corrections for SPIRE,
  taking into account beam variations from bolometer to bolometer, and
  reducing errors in maps of extended emission; 5) the newer version
  of HIPE used here has updated calibration values.

\subsection{{\it Herschel} Hi-Gal}
The Hi-Gal survey \citep{molinari10}, when complete, will map the
entire Galactic plane using
the PACS and SPIRE detectors of {\it Herschel}.  While the Hi-Gal data
are less sensitive than the observations shown here, Hi-Gal uses the
PACS detector to observe at 70\,\micron, a band not utilized with our
observation mode.  Here we use the Hi-Gal 70\,\micron\ data for
RCW\,79, RCW\,82, G332.5$-$0.1, and RCW\,120, the four regions that
overlap with the current Hi-Gal coverage.  These data have a spatial FWHM
resolution of $6\arcsec$.

\subsection{{\it Spitzer} GLIMPSE}
We also use 8.0\,\micron\ data from the {\it Spitzer} Galactic Legacy
Infrared Mid-Plane Survey Extraordinaire
\citep[GLIMPSE;][]{benjamin03}.  GLIMPSE extends from $+60\degr \ge l
\ge -60\degr$, $|b| \le 1\degr$.  The 8.0\,\micron\ IRAC
\citep{fazio04} filter contains, in addition to the continuum emission
of hot dust, polycyclic aromatic hydrocarbon (PAH) bands at
7.7\,\micron\ and 8.6\,\micron.  These molecules emit strongly when
excited by far-UV photons and thus can be used to trace ionization
fronts.  The same 8.0\,\micron\ IRAC band also shows many absorption
features, IRDCs, which have been shown to be dense molecular clouds
\citep[e.g.][]{simon06, pillai06}.  The resolution of the GLIMPSE
  8.0\,\micron\ data is $\sim2\arcsec$.


\subsection{{\it Spitzer} MIPSGAL}
The {\it Spitzer} MIPSGAL survey \citep{carey09} at 24\,\micron\ and
70\,\micron mapped the Galactic plane from $+60\degr \le l \le
-60\degr$, $|b| \le 1.0\degr$ with the MIPS instrument
\citep{rieke04}.  Here we use only data at 24\,\micron.  At
24\,\micron, the emission from \hii\ regions has two components.
First, there is emission from very small grains (VSGs) out of thermal
equilibrium.  This emission is detected in the interior area of the
\hii\ region \citep[see][]{watson08, deharveng10}.  Secondly, there is
thermal emission from the PDR from grains that appear to be in thermal
equilibrium.  These two components have roughly equal fluxes for
Galactic bubbles \citep[][\S\ref{sec:warm}]{deharveng10}.  The
  resolution of the 24\,\micron\ MIPSGAL data is $\sim6\arcsec$.

\subsection{APEX-LABOCA ATLASGAL}
Finally, we use data from the APEX Telescope Large Area Survey of the
Galaxy at 870~$\mu$m \citep[ATLASGAL;][]{schuller09}.  ATLASGAL
extends from $+60\degr \ge l \ge -60\degr$, $|b| \le 1.5\degr$ and has
a spatial FWHM resolution of $19\arcsec$.  The ATLASGAL survey was
performed with the Large Apex BOlometer CAmera (LABOCA), a 295-pixel
bolometer array \citep{siringo09}.  For RCW\,120, instead of ATLASGAL,
we use more sensitive pointed APEX-LABOCA observations first shown in
\citet{deharveng09}. In the processing of APEX-LABOCA data,
large-scale low amplitude structures $>2.5\arcmin$ are filtered out.
The calibration uncertainty in ATLASGAL is $\lsim 15\%$
\citep{schuller09}.




\section{The Dust Properties of Galactic Bubbles\label{sec:dustprop}}
Assuming optically thin emission, the flux from a population of dust grains
can be modeled as a modified blackbody:
\begin{equation} F_\nu \propto \kappa_\nu \, B_\nu(T_d) \, N_{\rm dust}\,,
\label{eq:gray}
\end{equation}
where $F_\nu$ is the flux density per beam,
$\kappa_\nu$ is the dust opacity in cm$^2$\,g$^{-1}$,
$B_\nu(T_d)$ is the Planck function for dust temperature $T_d$ at
frequency $\nu$ (here in Jy\,sr$^{-1}$), and
$N_{\rm dust}$ is the dust column density in cm$^{-2}$.
The dust opacity law we assume for the 
present work is given in Equation~\ref{eq:kappa}.
To derive the total hydrogen column density and mass (gas+dust), we must assume a
gas to dust ratio, $R$.  
The column density of gas and dust is:
\begin{equation} N_{\rm H} = R\,\frac{F_{\nu}}{2.8\,m_{\rm H}\,\kappa_{\nu}\,B_{\nu}(T_d)\,\Omega}\,, \label{eq:column}
\end{equation}
where $m_{\rm H}$ is the mass of a hydrogen atom in g, the factor of 2.8 accounts for
elements heavier than hydrogen, and $\Omega$ is the beam size in steradians.
The total mass of dust and gas in g may then be estimated from the integrated flux
in Jy, $S_\nu$ \citep[cf.][]{hildebrand83}:
\begin{equation}
M = R\,\frac{S_\nu\,D^2}{{\kappa_\nu}\,B_{\nu}(T_{d})}\,,
\label{eq:mass}
\end{equation}
where $D$ is the distance in cm.  The preceeding equations assume that the
measured flux at frequency $\nu$ is entirely due to thermal dust
emission.  For the present work, 
at 350\,\micron, we
assume the functional form for the dust opacity given in
\citet{beckwith90}, $\kappa_\nu~=~10(\nu)^\beta$, where $\nu$ is in
THz and we assume $\beta=2$.  This leads to an opacity at
350\,\micron\ of $\kappa_{350}~=~7.3$\,cm$^{2}$\,g$^{-1}$, which we use throughout.  For all mass
calculations, we assume that the dust-to-gas mass ratio has a value of
100.  These values for $\kappa_\nu$ and $R$ are uncertain, which leads
to large uncertainties in the mass and column density estimates.  For
example, \citet{preibisch93} estimate that using the above formulation
to compute masses results in uncertainties of up to a factor of five.
Errors in distances for some sources may be up to a factor of
  two in extreme cases, further increasing the uncertainty of the mass
  estimates.


%

\begin{figure*}[!ht]
 \centering
 \subfloat{\includegraphics[width=3.4 in]{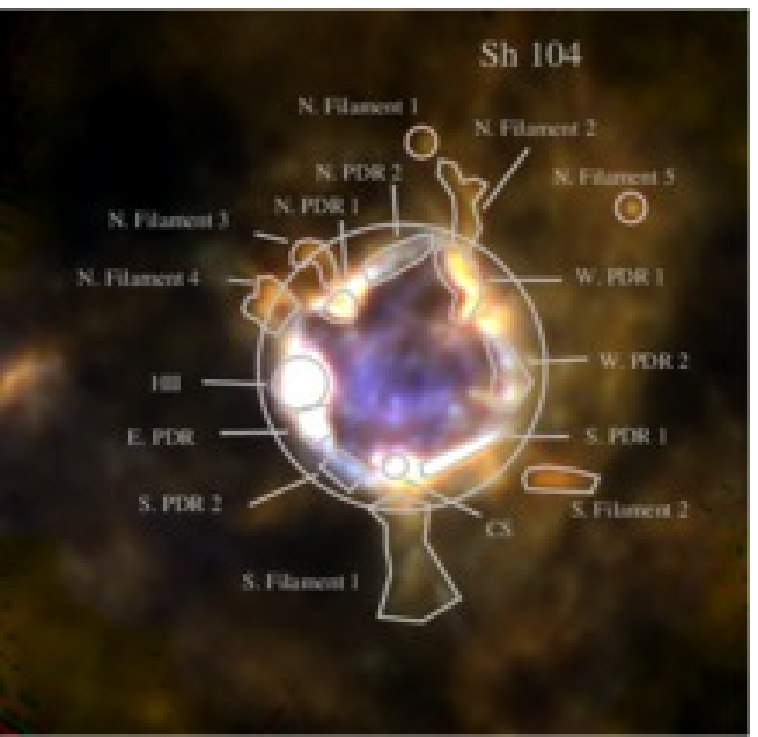}
   \hskip 0.00000001cm
   \includegraphics[width=3.4 in]{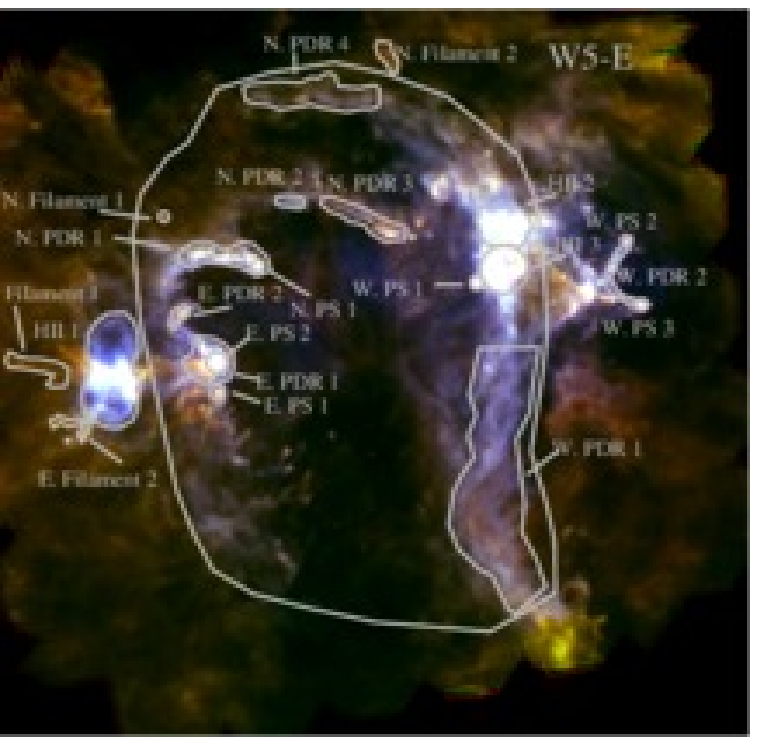}
 }
 \qquad
 \subfloat{\includegraphics[width=3.4 in]{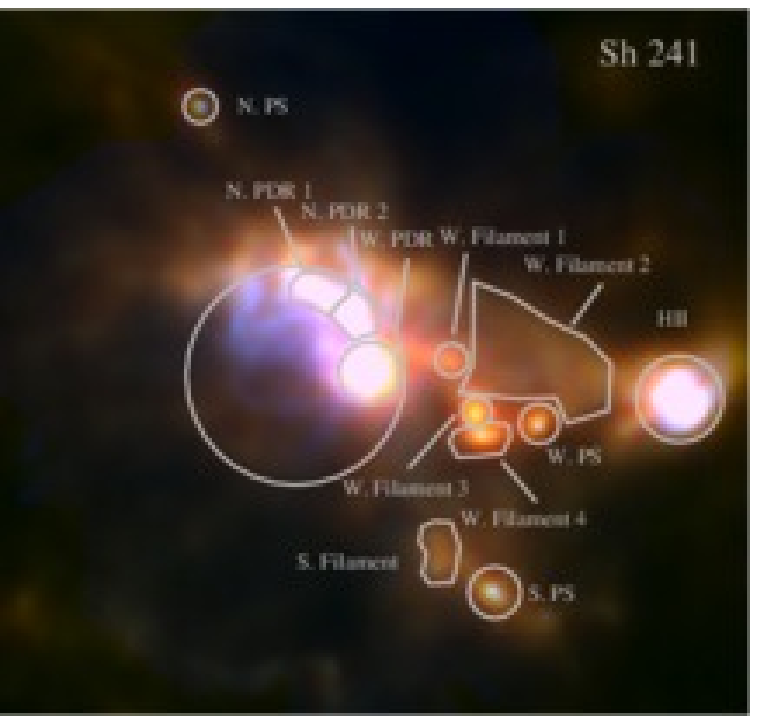}
   \hskip 0.00000001cm
   \includegraphics[width=3.4 in]{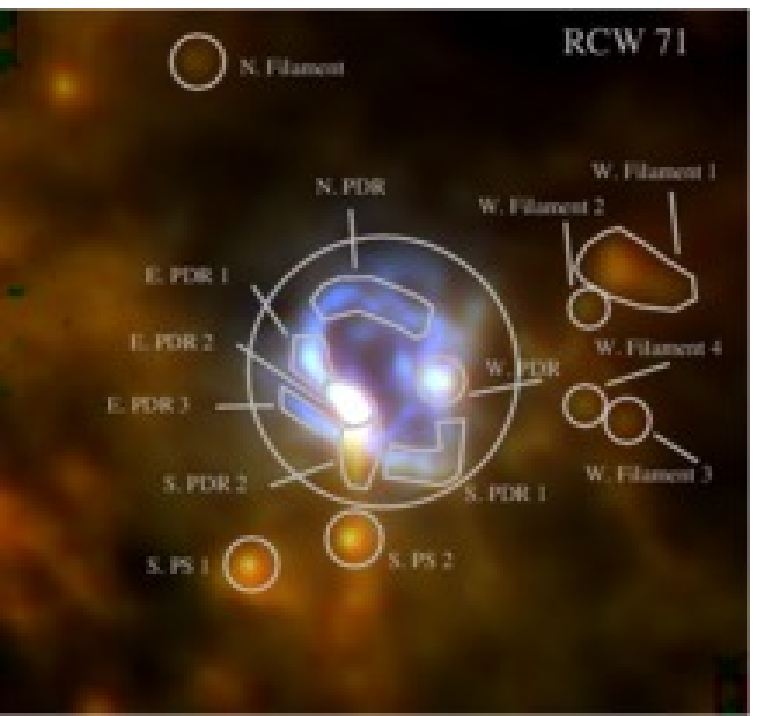}
 }
 \caption{Same as Figure~\ref{fig:herschel}, but showing the regions
   used in the aperture photometry.  As in Figure~\ref{fig:herschel},
   the three-color images are composed of {\it Herschel} data for each
   source with 500\,\micron\ in red, 250\,\micron\ in green and
   100\,\micron\ in blue.  The apertures used for the aperture
   photometry measurements are shown in gray outline.}
 \label{fig:herschel_regions}
\end{figure*}

\begin{figure*}[!ht]
 \ContinuedFloat
 \centering
 \subfloat{\includegraphics[width=3.4 in]{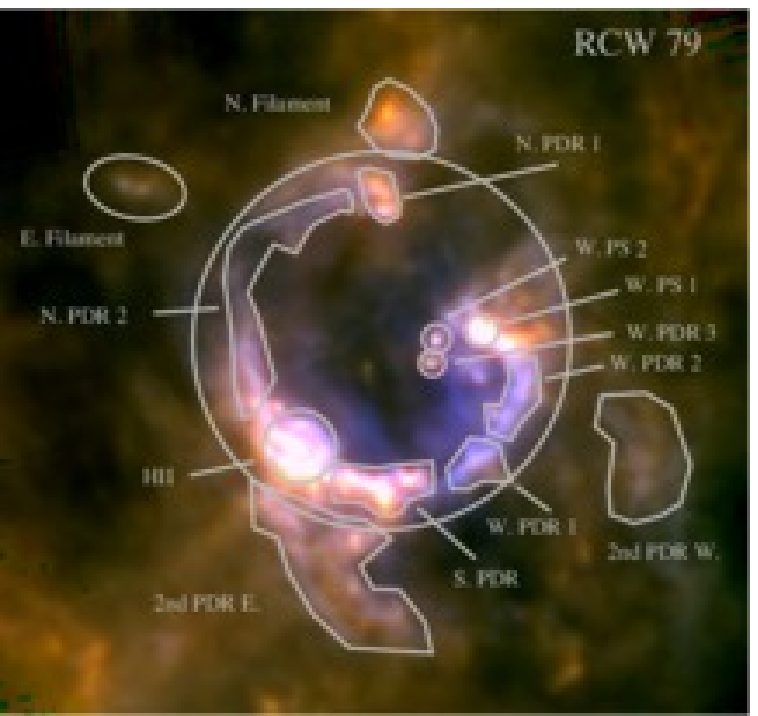}
   \hskip 0.00000001cm
   \includegraphics[width=3.4 in]{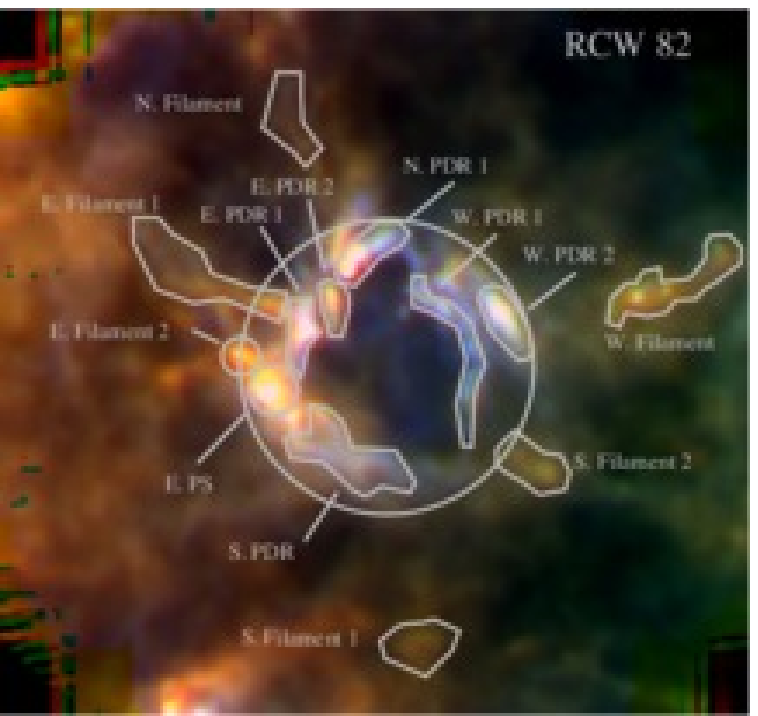}
 }
 \qquad
 \subfloat{\includegraphics[width=3.4 in]{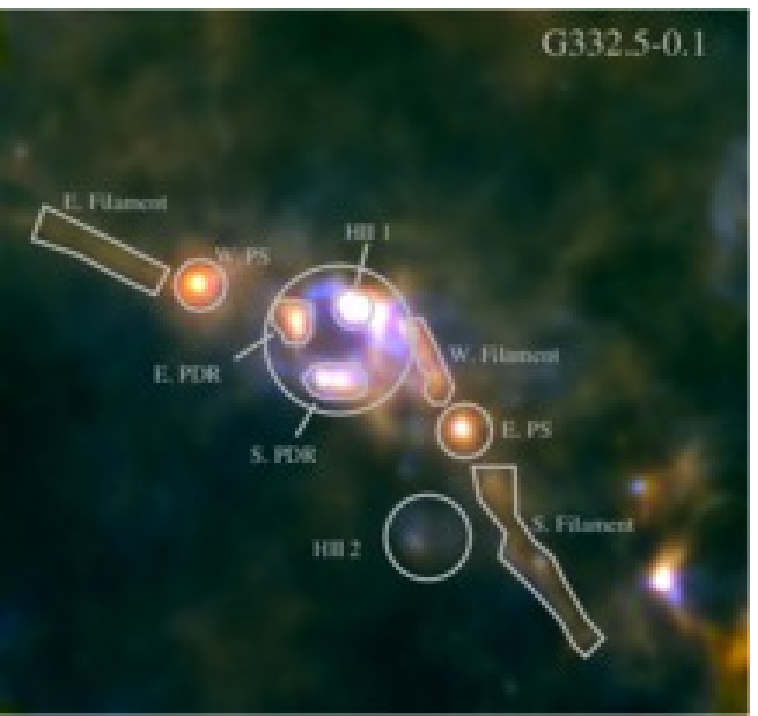}
   \hskip 0.00000001cm
   \includegraphics[width=3.4 in]{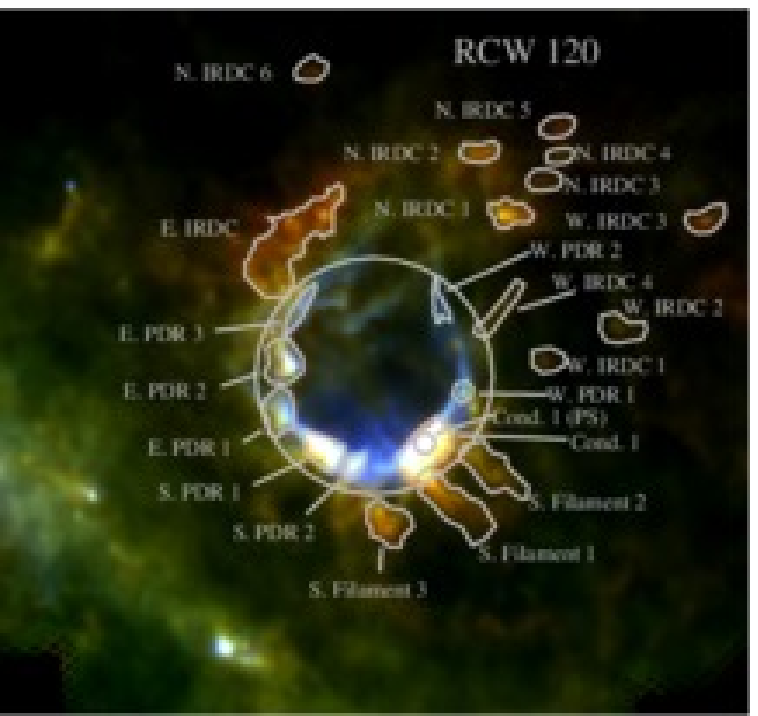}
 }
 \begin{flushleft}
   \caption{ -- continued.}
 \end{flushleft}
\end{figure*}

\subsection{Aperture Photometry\label{sec:apphot}}
Aperture photometry allows us to derive the average dust properties
within an aperture of arbitrary size and shape.  The measured flux
within an aperture contains the combined emission along the line of sight.
Most of the regions studied here are optically visible and thus the
extinction along the line of sight is relatively low.  The contribution to the
fluxes from dust along the line of sight not associated with the \hii\ region is therefore
minimized, but cannot be completely absent.

We define the apertures by eye to include contiguous regions of dust
emission from a volume that can be characterized by a single
temperature value.  This process is subjective; we use the three-color
images shown in Figure~\ref{fig:herschel} to help define regions where
the temperature variation within the region is not severe.  In
addition to specific regions of interest, we define for each
\hii\ region a large aperture that includes all emission associated
with the \hii\ region itself, but not the emission associated with
local clouds or filaments.  The apertures are shown in
Figure~\ref{fig:herschel_regions} on top of the same {\it Herschel}
data shown in Figure~\ref{fig:herschel}.  In total, we define 129
apertures.

We visually categorize the apertures into five classifications based
on their location and appearance in the {\it Herschel} data.  The five
aperture classes are: ``Entire'' for the largest apertures that
include all the emission from the \hii\ region (including that of hte PDR), ``Other \hii''
for the apertures that include emission from other \hii\ regions
  in the fields, ``PDR'' for locations along the PDRs, ``Point
source'' for apertures containing only the emission from sources that
appear point-like at 100\,\micron\ (including possibly their outer
envelope detected at longer wavelengths), and ``Filament'' for
apertures that enclose extended structures detected at
250\,\micron\ that are not along PDRs and that are not
  the longer-wavelength emission of point-like objects detected at
  100\,\micron.  Because filaments seen in absorption at
8.0\,\micron\ must be dense and therefore may be especially cold, we
subdivide the ``Filament'' class into those seen in absorption at
8.0\,\micron (IRDCs), and those that are not seen in absorption at
8.0\,\micron.  Excluding the ``Entire'' and {``Other \hii\''} classes,
which both contain eight apertures, there are 113 apertures.  Of
these, 49 are classified as ``PDR'', 46 as ``Filament''
(with 17 as infrared dark filaments), and 18 as ``Point
source''.

 Many of the apertures for RCW\,120 and Sh-104 were defined
  previously in \citet{anderson10a} and \citet{rodon10}, respectively.
  We have made some modifications to the apertures shown in these
  works: we have changed the shape of some apertures, added new
  apertures to highlight additional
  regions of interest, removed apertures that either cannot be well-characterized
  by a single dust temperature, and given some apertures a
  different name.  The new aperture shapes should better contain
  dust that can be characterized by a single temperature, and are more
  consistant between all regions in the current sample.  We removed apertures in the direction of the
  ``interior'' of the bubbles because of the difficulty of fitting a
  single dust temperature.  The aperture names used here better reflect the location of
  the aperture and again are more consistant with the naming
  convention employed for the other regions in our sample.

 In addition to the ``bubble'' \hii\ regions that we
targeted in our observations, there are eight additional sources in the {\it Herschel} fields that have characteristics of \hii\ regions.
All of these eight sources have spatially
  coincident IR (from {\it Herschel} and {\it Spitzer}) and radio
  emission from the NVSS or SUMSS.
  Together spatially coincident IR and radio emission point to a
  thermal source \citep{haslam87, broadbent89, anderson11}, e.g., a
  planetary nebula (PN) or an \hii\ region.

\citet{anderson12} derive
  IR color criteria for distinguishing between PNe and \hii\ regions.
  To determine the classification of the eight other \hii\ regions, in
  addition to the aperture photometry described below we compute
  aperture photometry using the {\it Wide-field Infrared Survey
    Explorer (WISE)} data at 12\,\micron\ and 22\,\micron.  All the
  \hii\ regions tested satisfy the criterion found by \citet{anderson12}: ${\rm log(}F_{160}/F_{12}) > 1.3$, where $F_{160}$ is the flux at
160\,\micron\ from {\it Herschel} and $F_{12}$ is the flux at
12\,\micron\ from WISE and in fact all have ${\rm log(}F_{160}/F_{12}) > 2$.
    \citet{anderson12} found that 10\% of their sample of PNe satisfy
    ${\rm log(}F_{160}/F_{12}) > 1.3$ and no PNe satisfy ${\rm log(}F_{160}/F_{12}) > 2$.  Together with the spatially coincident
IR and radio emission, these criteria suggest that these sources are {\it bona fide} \hii\ regions.  There are no WISE
        data available for the additional \hii\ region located on the
        PDR of Sh2-104.  Since this source satisfies the criteria for
        UC \hii\ regions in \citet{wc89b} and has detected CS emission
        \citep{bronfman96}, it is almost surely an \hii\ region and not a PN.  We assume throughout following that the eight other
        \hii\ regions are at the same distance as the bubble
        \hii\ regions that were targeted.  This assumption may not hold for
\hii\ regions well-separated in angle from the bubble \hii\ regions, i.e., for
one of the \hii\ regions in the field of G332 and the second \hii\ region
in the field of Sh\,241.

Because our data have a range of resolutions, we smooth all data to
the lowest resolution SPIRE~500\,\micron\ data, which has a resolution
of 37\,\arcsec.  To mediate pixel edge effects, we rebin all data sets
so they have the same pixel size and location using the
Montage software\footnote{http://montage.ipac.caltech.edu/}.

For each \hii\ region, we compute the flux at all available
wavelengths for each aperture.  In this process we utilize the Kang
software\footnote{http://www.bu.edu/iar/kang/}, version 1.3, which contains
routines to perform aperture photometry with apertures of arbitrary
size and shape, and accounts for fractional pixels.  We estimate the errors at each wavelength by summing
in quadrature the calibration uncertainties given in \S\ref{sec:data}
and the photometric error as calculated by {\it Scanamorphos}.  For
{\it Herschel} data, we compute the photometric error by adding in
quadrature the individual errors of all pixels within the aperture
from the error maps produced by {\it Scanamorphos}.  For the other
data sets, we estimate the photometric error using the standard
deviation of the background region multiplied by the square root of
the number of pixels in the aperture.

We subtract from each aperture flux, at each wavelength, a background
value.  For apertures at the center of the frame (mainly the
``Entire'', ``Other \hii,'' and ``PDR'' classes) we use a single background
value at each wavelength for all apertures.  Because apertures in
these classifications generally have high fluxes, we have found that
the exact choice of background value does not severely impact our
results.  A single background value, however, does not approximate the
true background level accross the entire field.  Therefore, for
apertures located away from the center of the frame we define a local
background value.  The removal of a background is necessary because
the true ``zero-point'' for our observations is unknown.  Perhaps more
importantly, however, there are potentially contributions from
multiple emission regions along the line of sight.  In the {\it
  Scanamorphos} data reduction, we remove the general
 gradients of the sky emission within the field of view (by not using the ``Galactic''
  option) and therefore we believe the flat background approximation
is valid; attempts to remove a more complicated background were met
with difficulty due to the relatively small size of our fields.

\begin{figure} \centering
\includegraphics[width=3.5 in]{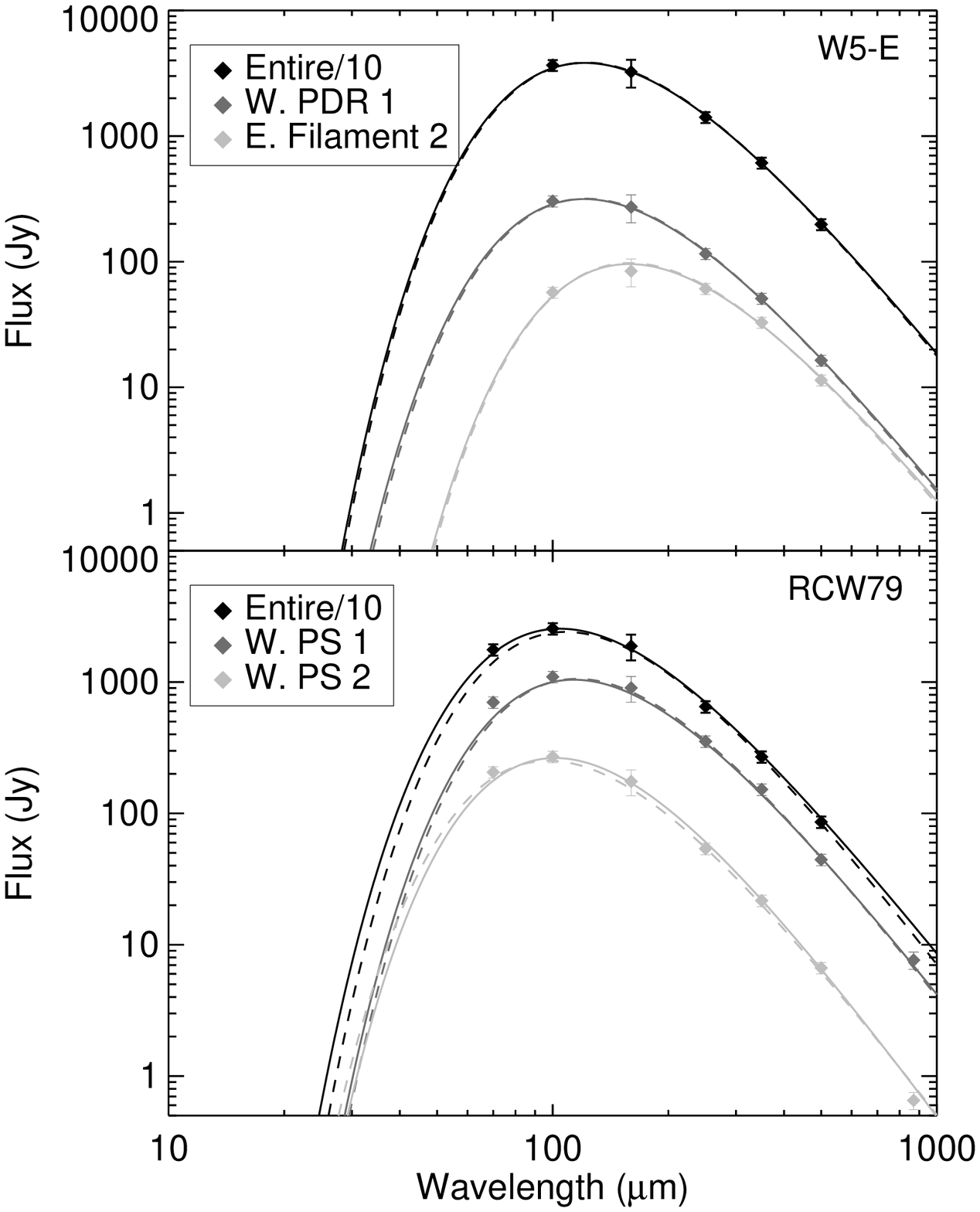}

\caption{Example SEDs fits for regions within W5-E (top) and RCW\,79
  (bottom).  The solid curves show the $\beta=2.0$ trial and the
  dashed curves show the $\beta$-free trial.  We do not have
    70\,\micron\ or 870\,\micron\ data for W5-E, and we exclude the
    870\,\micron\ point from the fit to the ``Entire'' aperture for
    RCW\,79.}

\label{fig:SEDs}
\end{figure}

We construct a spectral energy distribution for each aperture using
the background-corrected fluxes at wavelengths from 70\,\micron\ to
870\,\micron.  Using the ``MPFIT''\footnote{see
  http://purl.com/net/mpfit} least-squares minimization routines in
IDL \citep{markwardt09}, we then fit Equation~\ref{eq:gray} in two
trials: once with the normalization and dust temperature as free
parameters and $\beta$ fixed to a value of 2.0 and once with the
normalization, dust temperature, and $\beta$ as free parameters.  In
the SED fits, we generally include data at both 70\,\micron\ and
100\,\micron, although there is likely some contribution from a warmer
dust component at these wavelengths.  We do, however, exclude data at
70\,\micron\ from IRDC apertures because at this wavelength they may
be optically thick and we exclude 870\,\micron\ data from the
  ``Entire'' apertures because these data are not sensitive to large-scale
  diffuse emission.  Emission from warmer dust is detected mainly
toward the interior of the bubbles in our sample -- it has minimal
impact on the parameters derived here (see \S\ref{sec:warm}).  Using
Equation~\ref{eq:mass}, we also calculate the mass within each
aperture.

The measured flux within a photometric bandpass is a function of the
shape of the SED across the bandpass and of the shape of the bandpass
itself.  Each measured flux is therefore a function of temperature and
$\beta$.  To correct for this issue, we perform a ``color correction''
on the measured fluxes.  During the spectral fitting, we iteratively
apply color correction factors by repeatedly fitting a temperature and
$\beta$ then applying the corresponding correction factors until
successive fit results are unchanged.  For SPIRE, these color
correction factors, given in the SPIRE Observers' Manual, are $<10\%$
for all values of $\beta$ considered here.  For PACS, the color
correction factors are given in \citet{poglitsch10}.  The PACS color
correction factors can be quite large at low temperatures, especially
for the 70\,\micron\ band.  For example, at 10\,K, the correction
factor is 3.65 for the 70\,\micron\ band and 1.71 for the
100\,\micron\ band.  At 15\,K, the respective factors are 1.61 and
1.16.  While these correction factors are large, we find that the net
impact on the derived temperature is small.  For the apertures used
here, the mean absolute temperature difference between apertures fit
without a color correction and the same apertures fit with a color
correction is 0.6\,K, or $\sim3\%$.

The results of the aperture photometry analysis are given in Table~3.
This table lists for each aperture the name shown in
Figure~\ref{fig:herschel_regions}, mean Galactic longitude and
latitude, the angular size of the aperture in square arcminutes, the
derived dust temperature and mass for the two fit trials, the
$\beta$-values for the $\beta$-free trial, and the classification.
The errors in dust temperature and $\beta$ account for calibration
uncertainty and photometric errors.  We show in \S\ref{sec:beta_t}
that they are good estimates of the errors in the derived quantities.
Example fits are shown in Figure~\ref{fig:SEDs} for three apertures
within W5-E and RCW\,79.

\addtocounter{table}{1}

\subsection{Maps of Dust Properties}
The high-angular resolution {\it Herschel} data also allow us to
produce two-dimensional maps showing the distributions of temperature
and column-density.  With such maps, we may examine both small- and
large-scale variations in dust properties.  These maps have
the resolution of the 500\,\micron\ data, $37\arcsec$
and are available from the authors upon request.

\subsubsection{Temperature Maps}
We construct temperature maps for each of the eight regions in our
sample by fitting the SED extracted at a grid of locations using
Equation~\ref{eq:gray}.  When doing so, we use the same chi-squared
minimization routine used for the aperture photometry and again leave
the column density as a free parameter.  In the creation of the
temperature maps, we again include data at 70\,\micron; including the 70\,\micron\ emission differs from
the method of other authors \citep[e.g.,][]{hill11}.  The effect of
including data at this wavelength is explored in \S\ref{sec:warm}.  

While the wavelength coverage of {\it Herschel} and the number of data
points is theoretically sufficient to simultaneously fit for $T_d$ and
$\beta$, we find that the uncertainties in such fits are large and the
correlation between the derived temperatures and the values derived in
the aperture photometry is weak.  We therefore hold $\beta$ fixed to a
value of 2.0 when creating the maps, which is the average value found in the
aperture photometry.


We remove a flat average background level from the smoothed, rebinned
data at each wavelength.  The background value was defined such that
99.99\% of the pixels in the resultant images have an
intensity greater than zero.  This ensures that nearly all locations
have positive flux values.
This method of background subtraction is different from the
method of the aperture photometry but is necessary to ensure that
background-subtracted pixels have a positive intensity value, and thus
that we will be able to determine the dust temperature and column
density over the entire map.



\begin{figure*}[!t]
  \centering
  \subfloat{\includegraphics[width=2.7 in]{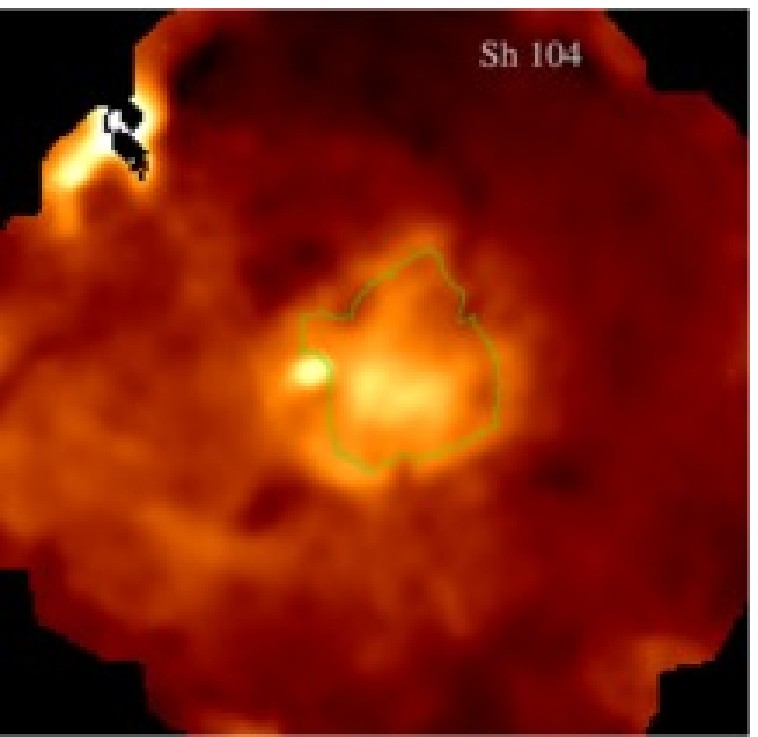}
    \hskip 0.00000001cm
    \includegraphics[width=2.7 in]{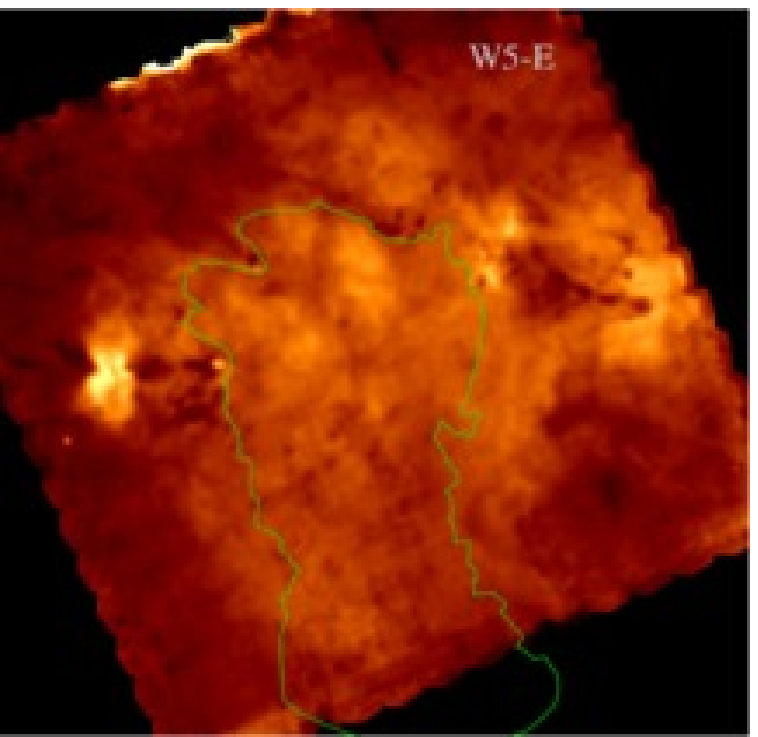}
  }
  \qquad
  \subfloat{\includegraphics[width=2.7 in]{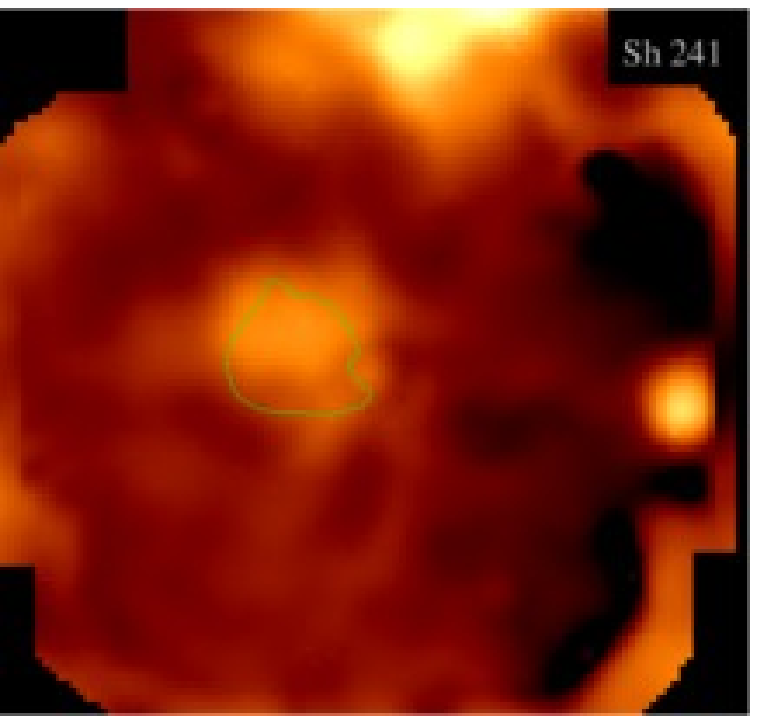}
    \hskip 0.00000001cm
    \includegraphics[width=2.7 in]{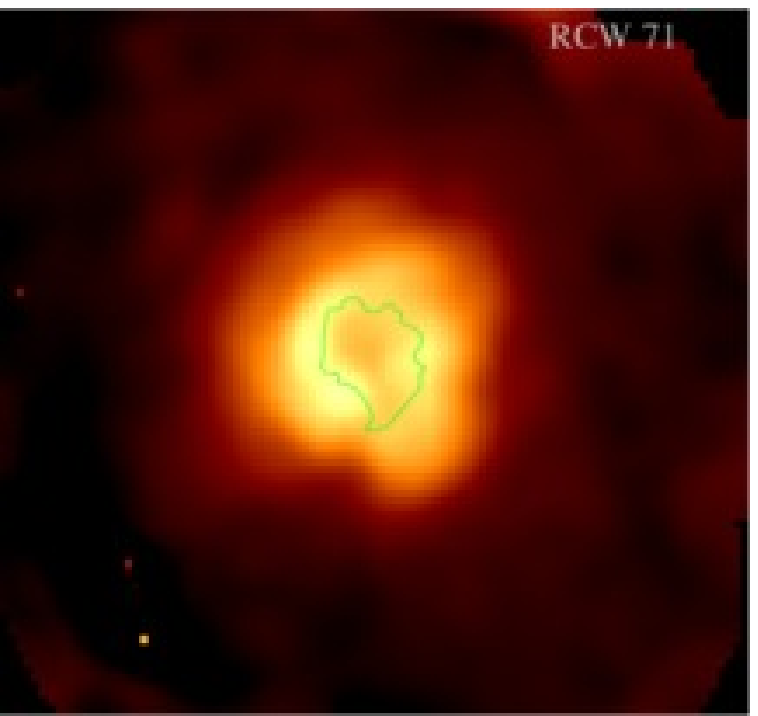}
  }
  \hskip 0.00000001cm
  \includegraphics[width=5.6 in]{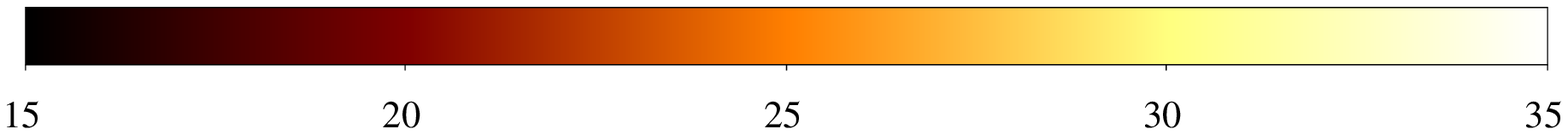}
  \caption{Temperature maps created by fitting the SEDs extracted at
    a grid of locations.  All panels have the same color scale and range from
    15\,K to 35\,K.  The green curves show the approximate extent of the bubble interiors.}
  \label{fig:tempmap}
\end{figure*}

\begin{figure*}[!ht]
  \ContinuedFloat
  \centering
  \subfloat{\includegraphics[width=2.7 in]{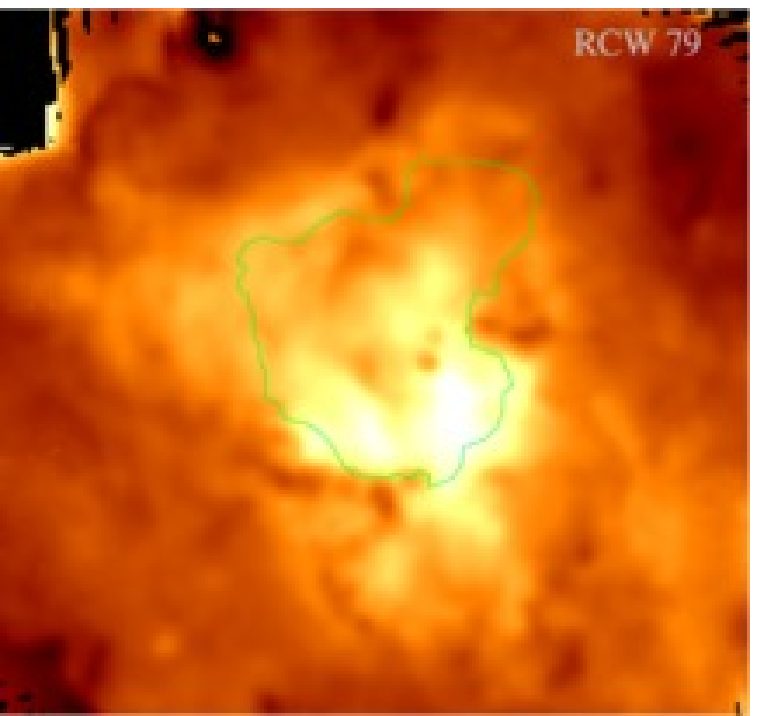}
    \hskip 0.00000001cm
    \includegraphics[width=2.7 in]{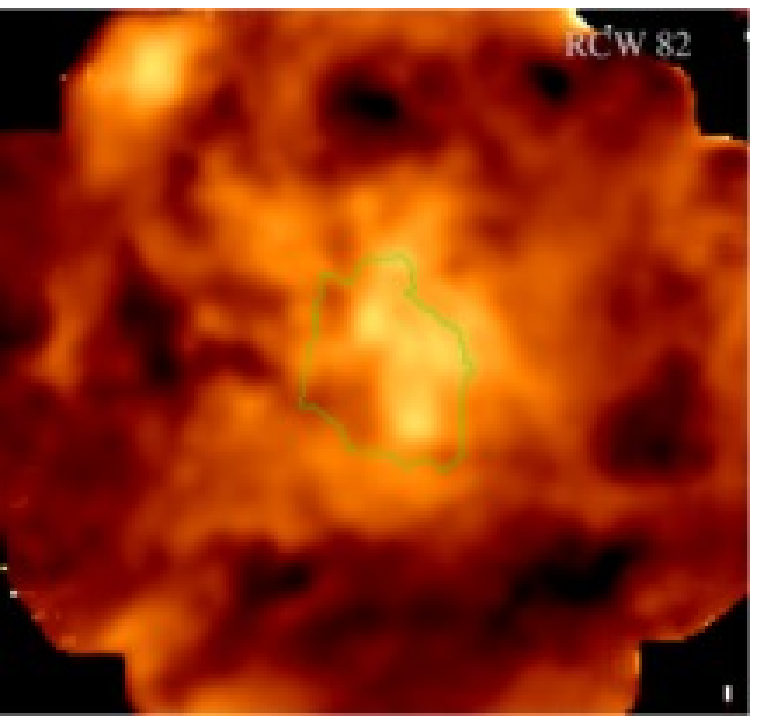}
  }
  \qquad
  \subfloat{\includegraphics[width=2.7 in]{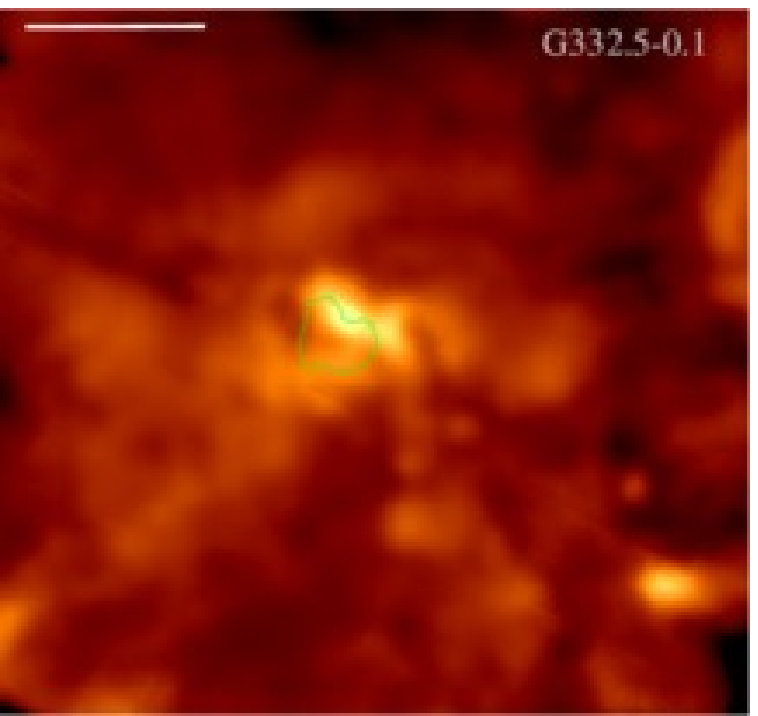}
    \hskip 0.00000001cm
    \includegraphics[width=2.7 in]{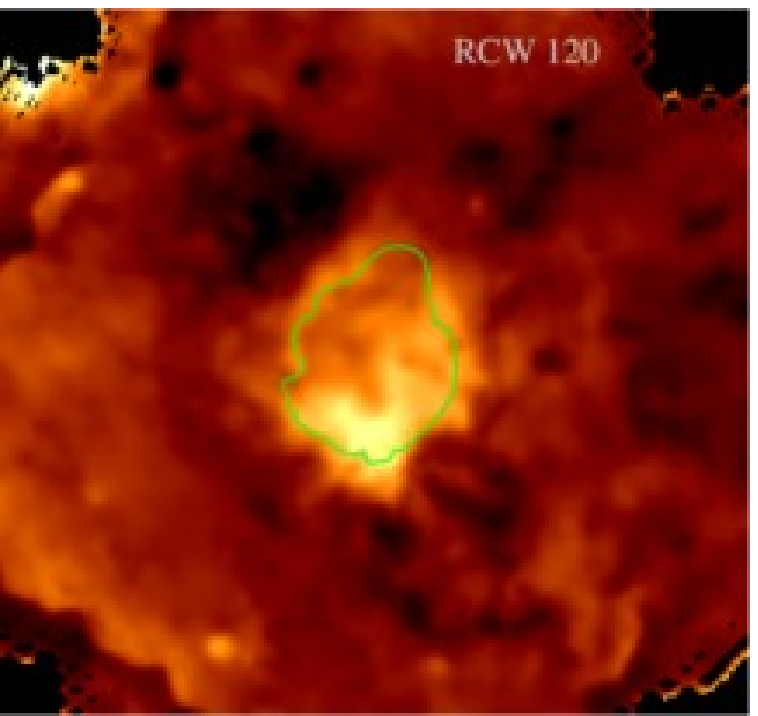}
  }
  \hskip 0.00000001cm
  \includegraphics[width=5.6 in]{temp_colorbar_BB.eps}
  \begin{flushleft}
    \caption{ -- continued.}
  \end{flushleft}
\end{figure*}

We show temperature maps for the eight regions in
Figure~\ref{fig:tempmap} for the same map areas shown in
Figures~\ref{fig:herschel} and \ref{fig:herschel_regions}.  The
highest temperatures, $\sim40$\,K, are found in the direction of the interior regions
of the bubbles and toward other local \hii\ regions.  The coldest
temperatures, $\sim 12$\,K, are found for the IRDCs and other regions
designated as ``filaments'' in the aperture photometry
(\S\ref{sec:apphot}).  Errors in the derived temperatures for
individual pixels are generally less than 10\%.  Large patches of
very high temperatures on the periphery are in general not real, but
are caused by low emission (e.g., to the north-east of Sh\,104, to the north of Sh\,241,
and to the north-east of RCW\,82).

Figure~\ref{fig:tempmap} shows a wide range of temperature structures.
For Sh\,104, Sh\,241, RCW\,79, RCW\,120, and RCW\,82, the bubble
interiors show increased temperatures.  These temperatures are quite
uncertain though and individual SEDs often show contributions from
multiple components (see \S\ref{sec:warm}).  The hottest temperatures
for RCW\,71, however, are in the PDR.  For G332.5$-$0.1 the hottest
dust temperature is also along the PDR, to the north, coincident with
the UC and compact \hii\ regions.  W5-E shows little variation in
temperature although there are warmer locations associated with
the other \hii\ regions in the field.  Variations within the fields of
individual regions are dicussed in Appendix~\ref{sec:individual}.

\begin{figure}[!b]
\centering
\includegraphics[width=2.5 in]{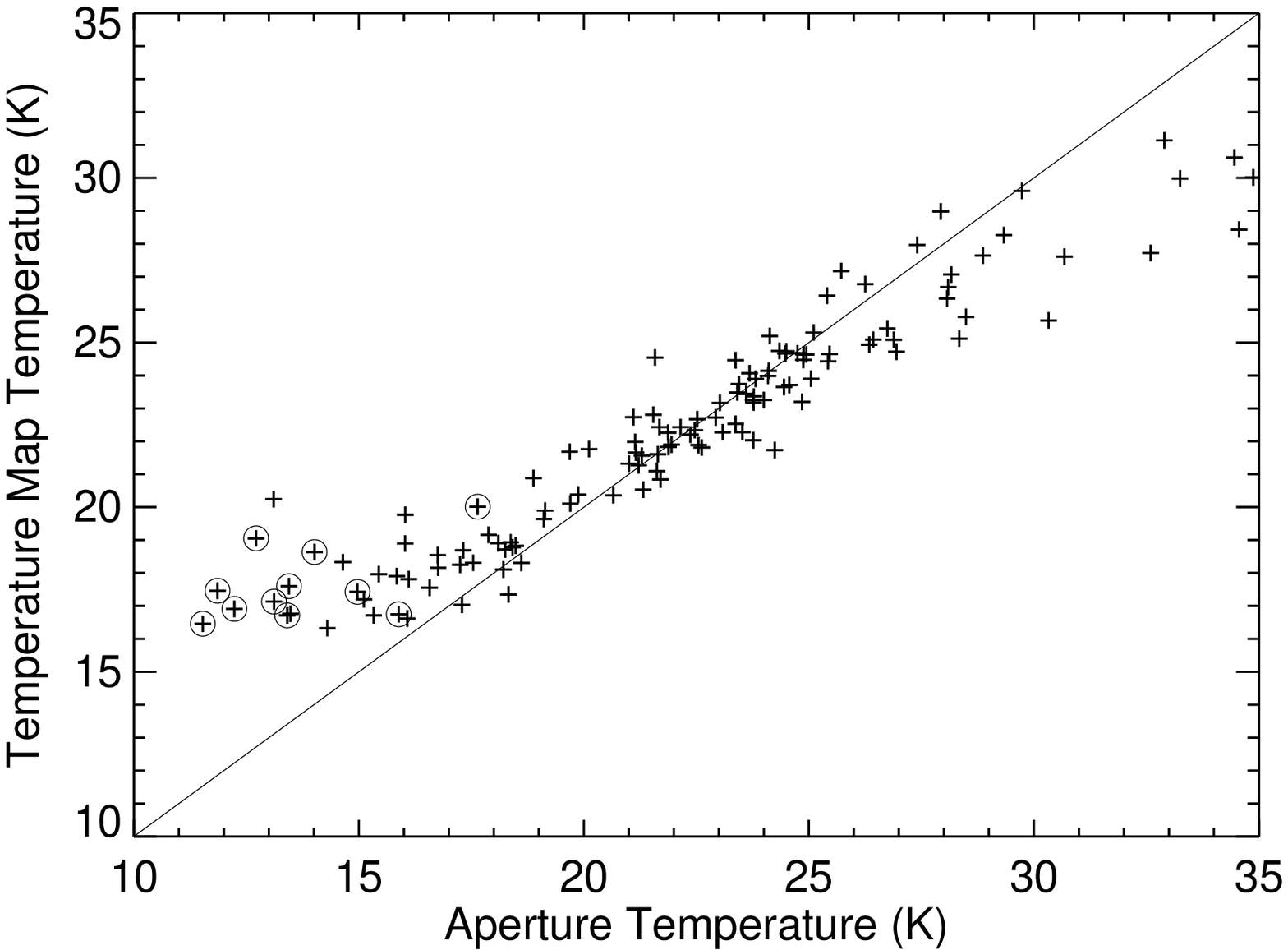}

\caption{The temperature derived using aperture photometry 
versus the average temperature in the temperature map
derived using the same apertures.  The line is not a fit but rather shows a 1:1
relationship.  The circled sources are IRDCs from the field of RCW\,120; they
account for much of the difference at low temperature.}

\label{fig:apphot_test}
\end{figure}

The temperature map values agree with the results of the aperture
photometry, but there are large differences for colder regions.
To compare the two methods of deriving tempatures, we compute the
average dust temperature found in the temperature map within the apertures used previously.
In Figure~\ref{fig:apphot_test}, we plot the temperature derived in
the aperture photometry versus the average temperature map temperature
for each aperture.
We find that the two temperatures are on average 12\% different,
or roughly 2.3\,K.  This difference is pronounced for apertures with temperatures less than 15\,K and warmer than 30\,K,
for which the temperature map values are sytematically higher or
lower, respectively.  These discrepancies are due to the
  differences in background treatment between the two methods.  Because
we evaluate the background individually for each aperture, we
believe the aperture photometry dust temperature values are more accurate. The
circled points in Figure~\ref{fig:apphot_test} are IRDCs in the field
of RCW\,120; they account for much of the difference at low
temperature.  The difference in temperature for these cold regions is
due to our assumption of a flat background in the temperature maps.
For cold regions this assumption causes us to overestimate the
temperature, while the inverse is true for warm regions.  We do
not expect a perfect correlation because the aperture photometry
temperatures are weighted toward the highest emitting regions within
the aperture.  The mean standard deviation of temperature map values
within the apertures is 2.0\,K.

\subsubsection{Column Density Maps}
From the temperature maps, we also create column density maps using
Equation~\ref{eq:column}.  As mentioned above, we
evaluate $\kappa_\nu$ at 350\,\micron, and assume a value of
$\kappa_{350}$ = 7.3\,cm$^2$\,g$^{-1}$ and a dust-to-gas mass ratio of
100.  At each grid location in the column density map, we calculate
the Planck function value $B_\nu(T_d)$ using the derived temperature
from the temperature map.  For the flux at 350\,\micron, we use the
{\it Herschel} SPIRE image data, after removing a background value
defined as before such that 99.99\% of the pixels at
350\,\micron\ have an intensity greater than zero.
The removal of a background value produces a more realistic picture of the 
column density associated with each region, rather than
the total integrated column density along the line of sight.  The resultant maps
are shown in Figure~\ref{fig:column}.



The column density distributions in Figure~\ref{fig:column} shows
enhancements along the bubble PDRs.  These enhancements likely are
from the material swept up during the expansion of the \hii\ region.
For RCW\,71, however, the PDR is barely enhanced relative to the
background.  Other prominent column density enhancements in the fields
are generally caused by filaments, especially IRDCs, and unresolved
sources.  For comparison, the IRDCs in \citet{peretto10b} have typical
column densities of $\gtrsim10^{22}\,{\rm cm^{-2}}$.  The highest
column densities are generally associated with condensations with
detected protostars inside (i.e. Condensation~1 in RCW\,120).  Column
density values toward the bubble interiors are uncertain due to
uncertainties in temperature and because of the low flux values found there.  The spurious
temperature values near the image edges mentioned previously result in spurious
column density values as well (e.g., south-west of Sh\,241 and south-east of RCW\,71). A more complete discussion for
individual regions is given in the Appendix.

PDRs and cold filaments share nearly the same mean column density values.
We calculate the mean column density value shown in Figure~\ref{fig:column}
within the previously defined apertures for regions in the ``PDR'' and
``Filament'' classes.  The average value for the ``PDR'' class
is $1.8\times10^{21}\,{\rm cm^{-2}}$ while it is $1.9\times10^{21}\,{\rm cm^{-2}}$
for the ``Filament'' class.  The above average values were calculated after
first taking the log of the column density values, so large values of
column density would not bias the result.

\section{Discussion}
\subsection{Comparison of Aperture Classes}
\begin{figure*}
\centering
\includegraphics[width=5 in]{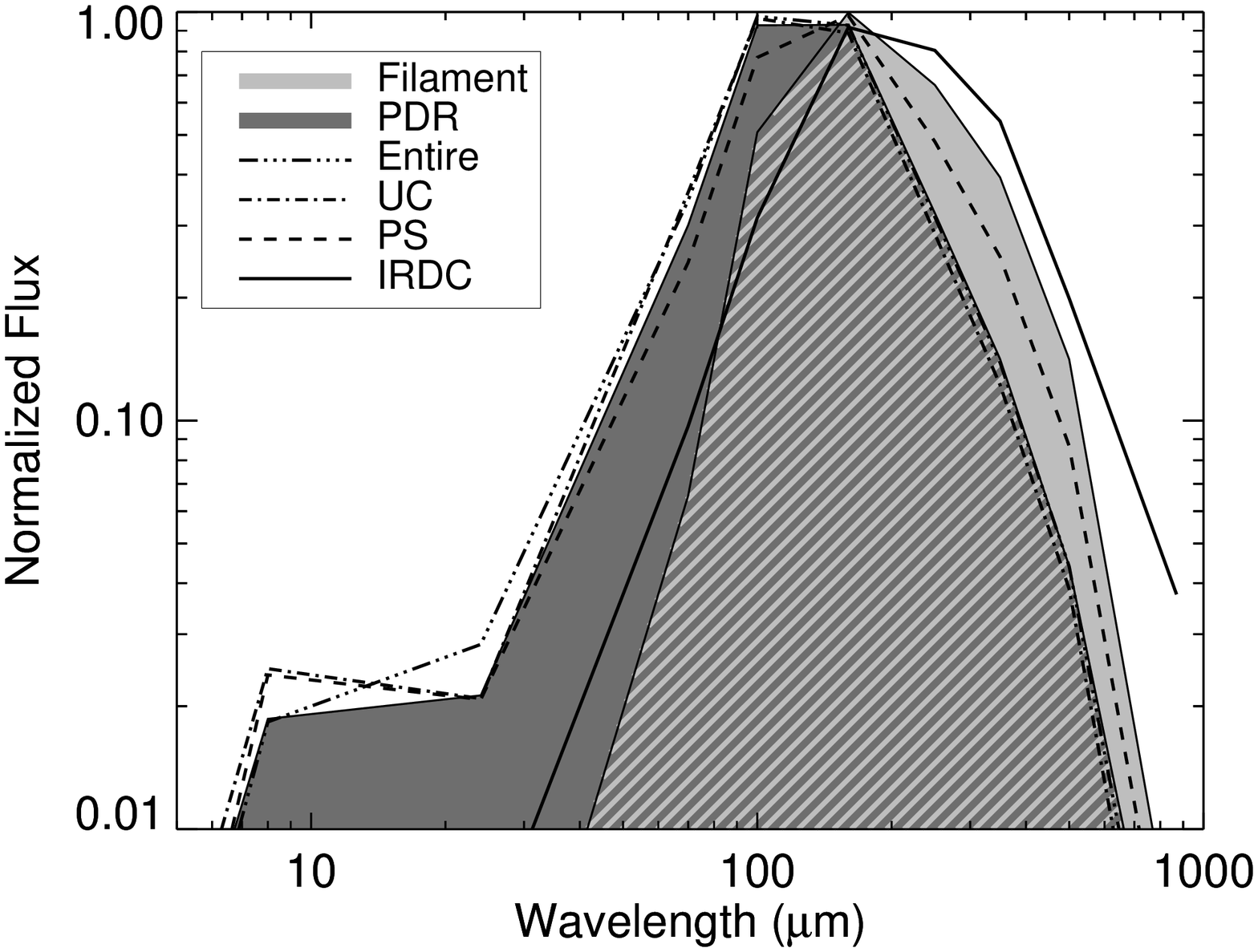}

\caption{The average normalized SEDs for the aperture classes.
  Results from the ``PDR'' class are shown in dark gray while those in
  the ``Filament'' class are shown in light gray.  Where the two
  ``PDR'' and ``Filament'' distributions overlap, we show light gray
  with dark gray hatching.  Lines show the average normalized SEDs for
  the ``Entire'', Other ``H{\scriptsize II},'' ``PS,'' and ``IRDC'' classes; the ``IRDC''
  class SED is similar to that of the ``Filament'' class while the
  SEDs of the ``Entire'', and ``Other \hiismall'' classes are more similar to that
  of the ``PDR'' class.  The ``PS'' class has a mean SED shape in
  between that of the ``PDR'' and ``Filament'' classes.  The
    points at 8.0\,\micron\ and 24\,\micron\ are shown here although
    these data were not used in the fits.}

\label{fig:avg_seds}
\end{figure*}

With the calculated values of $\beta$ and $T_d$ in Table~3, we may
investigate how the dust properties change with location in the
\hii\ region environment using the aperture classifications.  The
easiest way to see differences in dust properties is in the shapes of
the SEDs themselves.  We show average SEDs for the classifications in
Figure~\ref{fig:avg_seds}.  To create Figure~\ref{fig:avg_seds}, we
normalized the SED for each aperture by the highest flux and averaged
all apertures for a given classification.  The curves are therefore
unweighted averages of the normalized fluxes.  The ``Entire,'' ``Other \hii,''
and ``PDR,'' classes have similar shapes, as do the ``Filament'' and
``IRDC'' classes.  The ``PS'' class has a mean SED shape in between
the two groupings.  The similar shapes of the SEDs hint at similar
dust properties for these classifications.  Compared to the other
classifications, the ``Filament'' and ``IRDC'' classes peak at longer
wavelengths and show more emission in the FIR part of the SED.  This
shows that these apertures have colder mean temperatures.

\begin{figure*}[!ht]
 \centering
 \subfloat{\includegraphics[width=3.4 in]{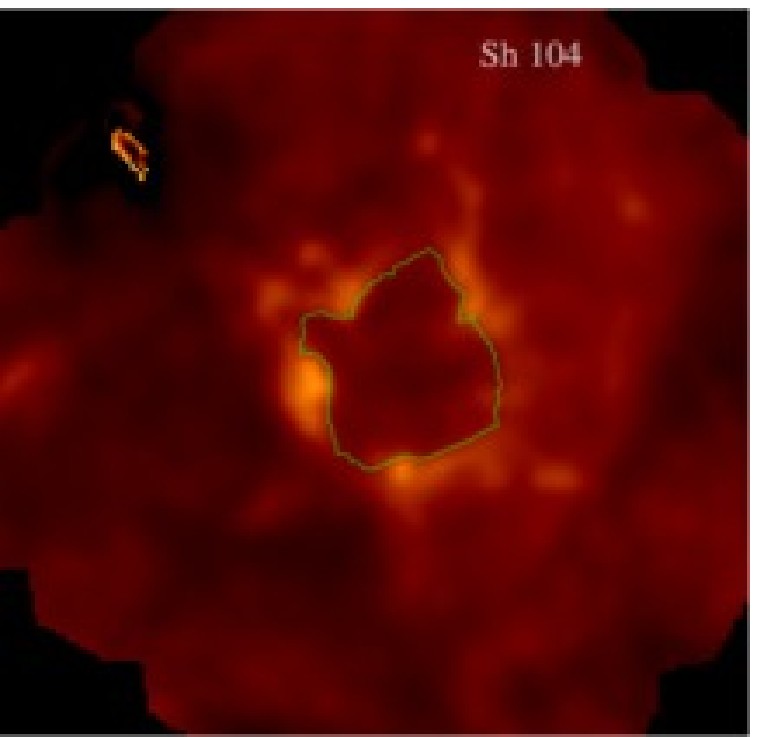}
	\hskip 0.00000001cm
	\includegraphics[width=3.4 in]{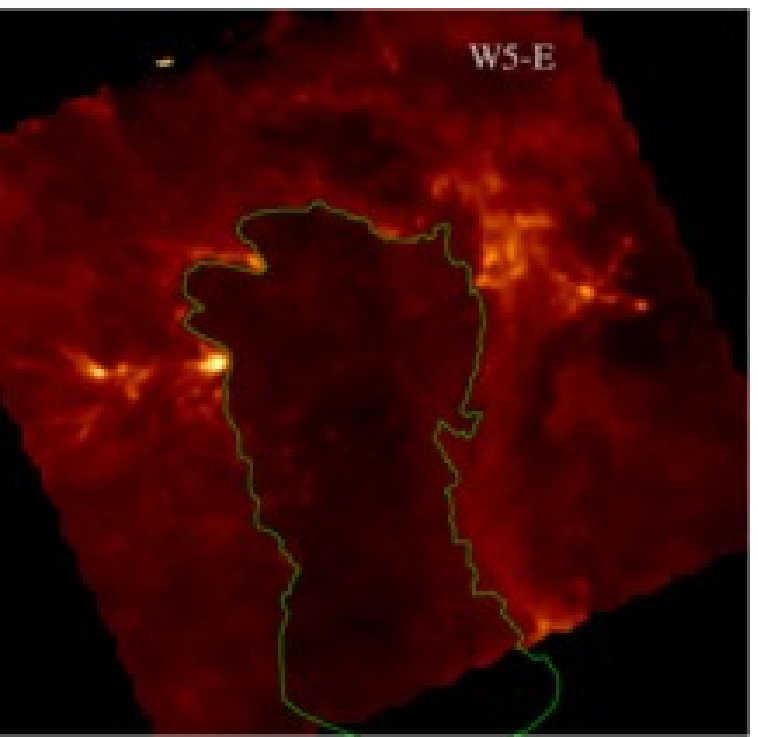}
	}
 \qquad
 \subfloat{\includegraphics[width=3.4 in]{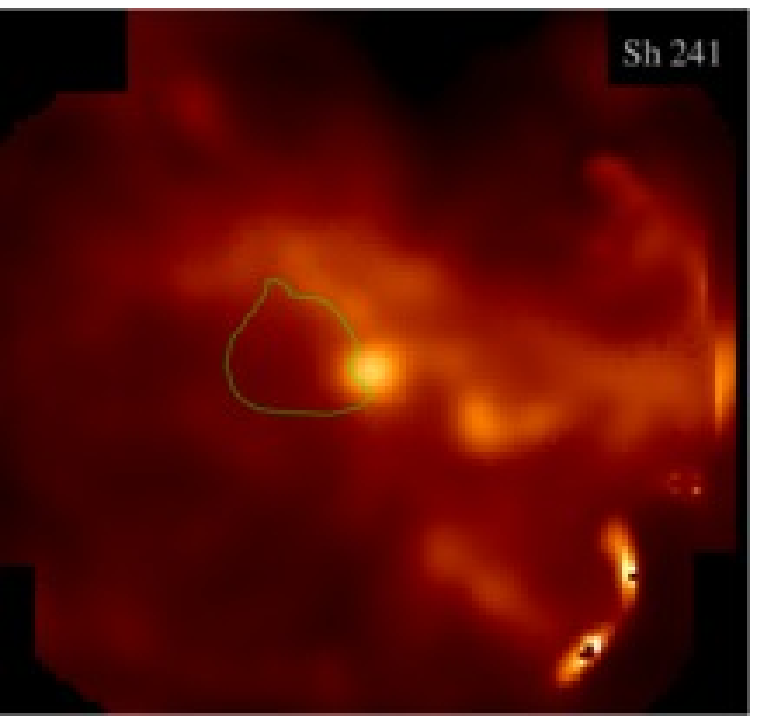}
	\hskip 0.00000001cm
	\includegraphics[width=3.4 in]{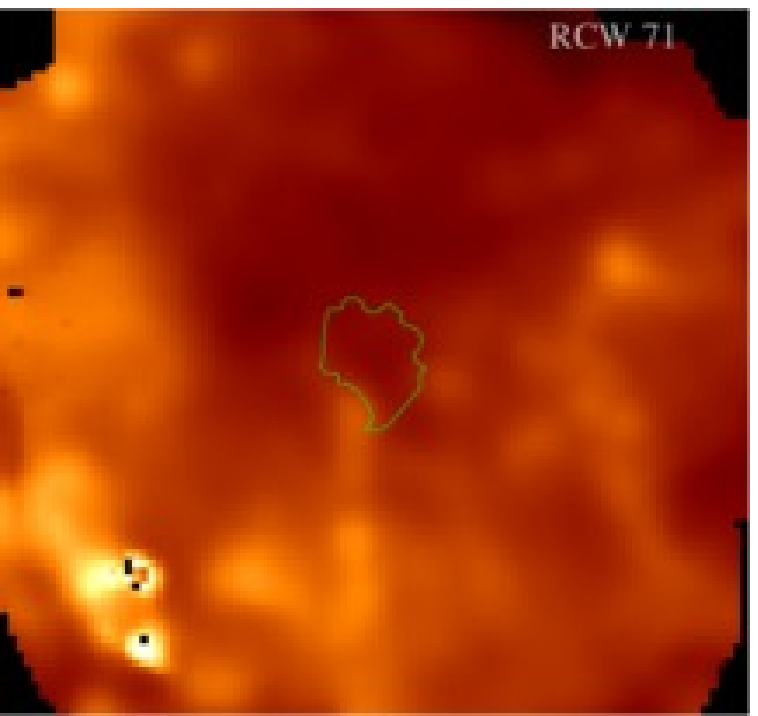}
	}
  \hskip 0.00000001cm
  \includegraphics[width=7 in]{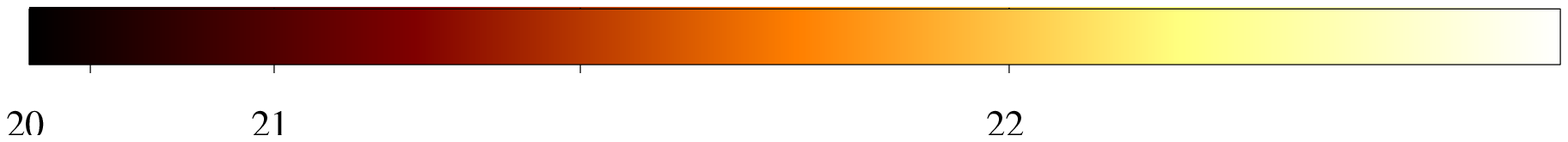}
\caption{Column density maps created using {\it Herschel} 350\,\micron\ data
and the temperatures shown in Figure~\ref{fig:tempmap}.  The scaling for all
panels is the same; the values range from $10^{20}\,{\rm cm^{-2}}$ to $10^{22.5}\,{\rm cm^{-2}}$.
The green curves show the approximate extent of the bubble interiors.  Enhancements are seen along the bubble PDRs.}

\label{fig:column}
\end{figure*}

\begin{figure*}[!ht]
  \ContinuedFloat
  \centering
  \subfloat{\includegraphics[width=3.4 in]{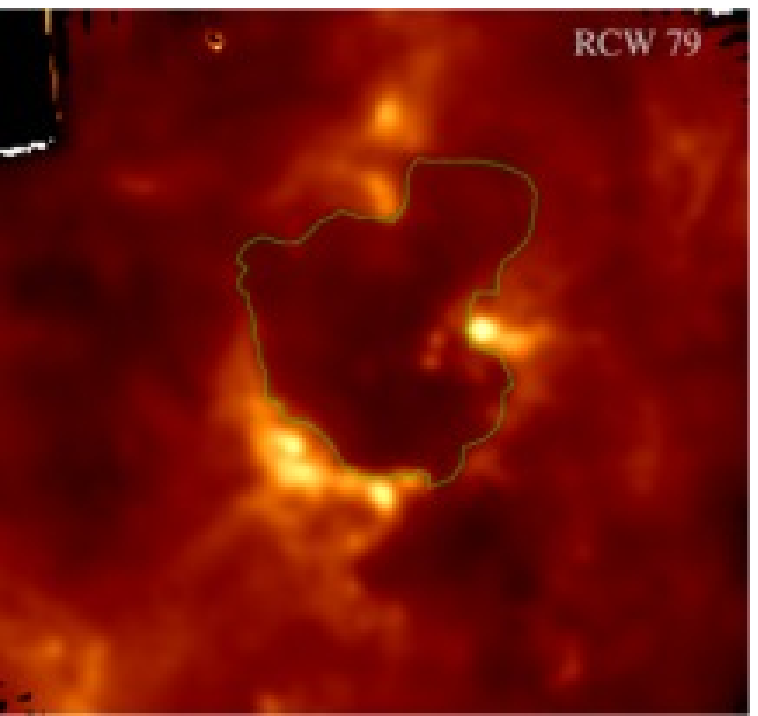}
    \hskip 0.00000001cm
    \includegraphics[width=3.4 in]{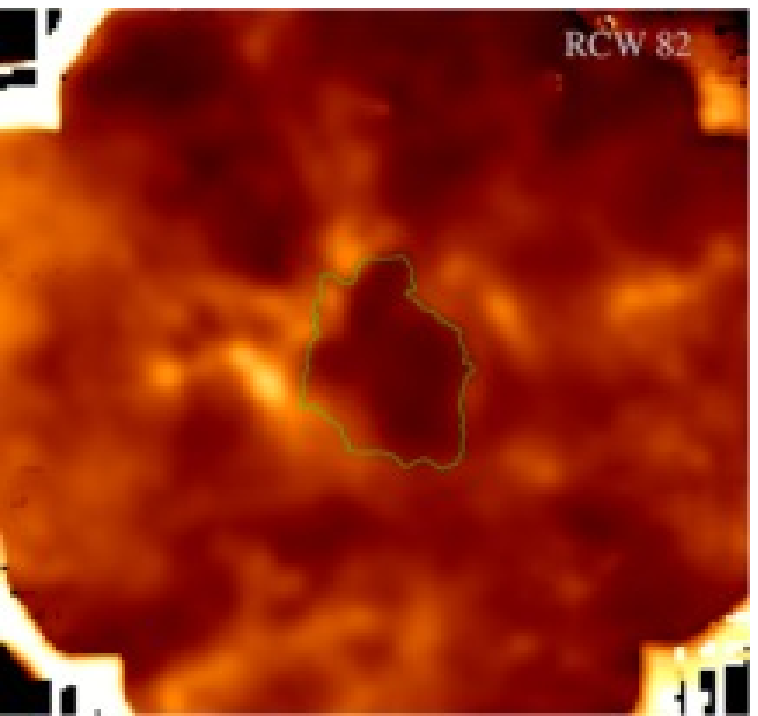}
  }
  \qquad
  \subfloat{\includegraphics[width=3.4 in]{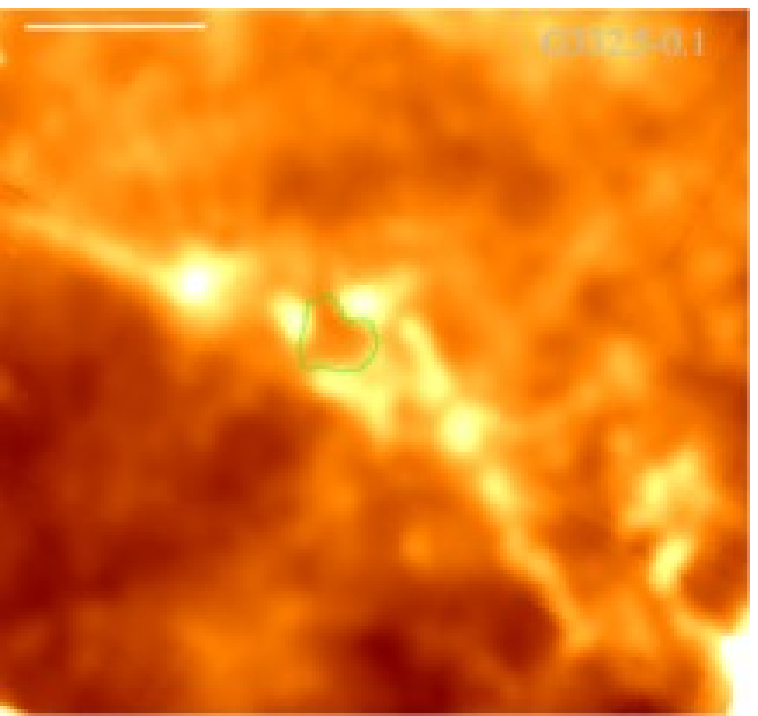}
    \hskip 0.00000001cm
    \includegraphics[width=3.4 in]{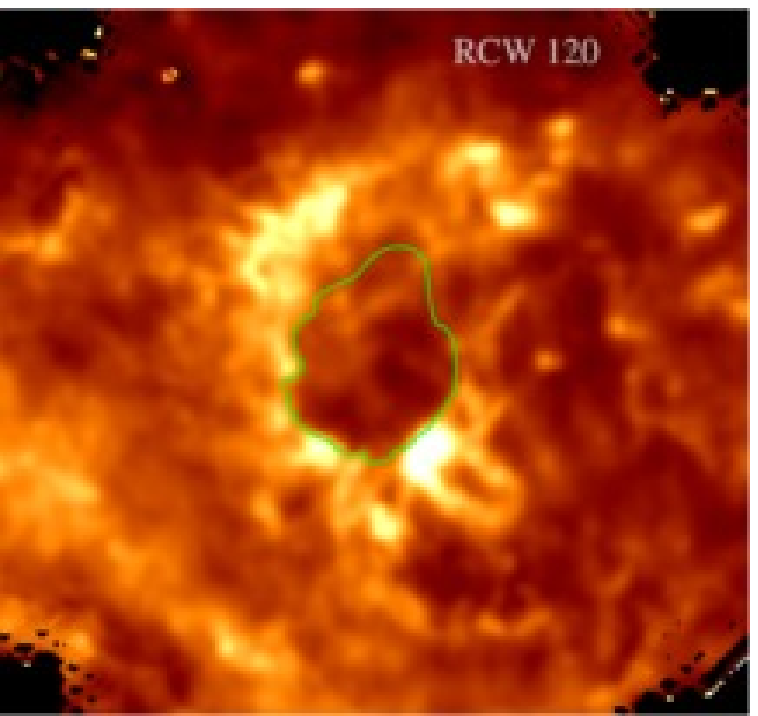}
  }
  \hskip 0.00000001cm
  \includegraphics[width=7 in]{column_colorbar_BB.eps}
  \begin{flushleft}
    \caption{ -- continued.}
  \end{flushleft}
\end{figure*}

We list the mean dust properties for each aperture classification in
Table~4, which shows the average dust temperature and
$\beta$-values for the $\beta=2$ trial and the $\beta$-free trial,
together with their 1-$\sigma$ standard deviations, the percentage of
apertures that have peak fluxes long-ward of 100\,\micron, the
percentage of apertures that have peak fluxes long-ward of
160\,\micron, and the number of apertures of a given classification.
The seventh and eigth columns dealing with the wavelength of the peak
flux provide a check against the results of the temperature fitting --
SEDs that peak long-ward of 160\,\micron\ have dust temperatures of
$\lesssim\!12$\,K while those that peak long-ward of 100\,\micron\ have
dust temperatures of $\lesssim\!21$\,K, assuming $\beta=2$.  The row
labeled ``All Filaments'' includes filaments of aperture
classification ``F'' and infrared dark filaments of aperture
classification ``F (ID)'' from Table~3.  The rows
labelled ``IRDC'' and ``Not IRDC'' contain only data from inner-Galaxy
\hii\ regions for which we have 8.0\,\micron\ data (RCW\,79, RCW\,82,
G332.5$-$0.1, and RCW\,120).

A difference in temperature has implications for subsequent star
formation.  The Jeans mass scales as
$M_J~\propto~T^{3/2}\,\rho^{-1/2},$ where $\rho$ is the mass volume
density, and therefore higher temperatures lead to higher values of
$M_J$.  The average temperature of the PDR classification, 25.6\,K, is
significantly higher than that of the filaments, 16.9\,K.  For the
above average temperatures, if the gas and dust share the same
temperature and the density in the PDR and the filaments is
comparable, the Jeans mass would be $\sim80$\% higher in the
\hii\ region PDRs compared to the filaments.  This would imply that on
average higher mass stars will form in the \hii\ region PDRs through
gravitational collapse compared to the filaments.  We stress, however, that the
density is a large unknown here and further observations are needed to
better constrain this parameter.

\begin{figure*}[!t]
\centering
\includegraphics[width=6.5 in]{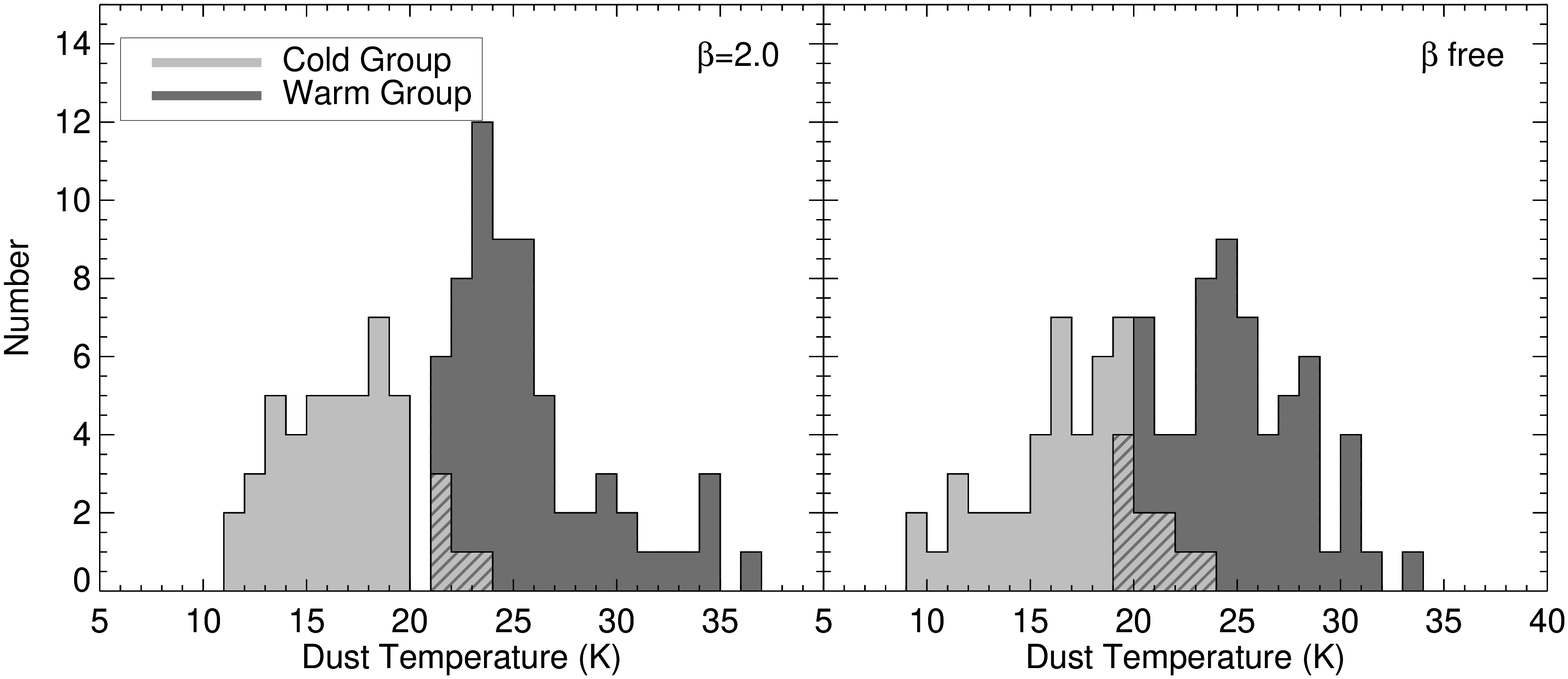}

\caption{Dust temperatures of warm and cold aperture groups.  Shown
  are the distributions for the apertures in the cold group
  (``Filament'' and ``IRDC'' classifications) in light gray and the
  apertures in the warm group (``Entire,'' ``Other \hiismall,'' and
  ``PDR'' classifications) in dark gray.  We do not include in
    this figure apertures with classification ``PS''.  Where the two
  distributions overlap, we show light gray with dark gray hatching.
  The left panel shows dust temperatures derived when $\beta$ was held
  fixed to a value of 2.0 while the right panel shows the dust
  temperatures derived when $\beta$ was allowed to vary.  It is
  evident that these two aperture groups have different dust
  temperature distributions.}

\label{fig:temp_dist}
\end{figure*}

Table~3 and Figure~\ref{fig:avg_seds} show that there are two broad
classifications of apertures that can be grouped by their similar
temperature values.  The first is a ``warm'' group characterized by
temperatures of $\sim 26$\,K that includes the ``Entire'', ``Other \hii'', and
``PDR'' aperture classifications.  There is no statistical difference
between all possible combinations of aperture classifications within
the ``warm'' group, as shown by a Kolmogorov-Smirnov (K-S) test which
assesses the probability that two distributions belong to the same
parent distribution.  The second ``cold'' group consists of apertures
classified as ``Filament'' or ``IRDC'' and is characterized by dust
temperatures of $\sim 17$\,K.  Some apertures in the ``PS''
classification have dust properties similar to those in the ``warm''
group, while some have dust properties more similar to those in the
``cold'' group; we do not include ``PS'' apertures in either group.  This likely reflects the range of evolutionary states
in this classification: the colder apertures of the ``PS'' class are
in an earlier evolutionary state than the warmer ones that may be
heated by a central young stellar object.
In Figure~\ref{fig:temp_dist}, we show graphically how the ``warm''
and ``cold'' aperture groups, which were defined according to
  their aperture classifications and not according to their dust
  temperatures, have different temperature distributions.  Roughly
half of the apertures in the ``warm'' group have an emission peak
long-ward of 100\,\micron\ while all the apertures in the ``cold''
group meet this criterion.  There are no apertures in the ``warm''
group whose wavelength of peak flux emission is long-ward of
160\,\micron, while 13\% of the apertures in the ``cold'' group meet
this criterion.



Filaments that are infrared dark at 8.0\,\micron\ are colder than
those that are not infrared dark at 8.0\,\micron.  When $\beta$ is
held fixed to 2.0, the ``IRDC'' filaments have a mean temperature of
$15.5\pm3.5$\,K while the ``Not IRDC'' filaments have a mean
temperature of $18.8\pm2.9$\,K; these values change to $14.7\pm3.6$\,K
and $17.8\pm2.2$\,K respectively when $\beta$ is allowed to vary.  A
K-S test shows that the temperatures derived for the two groups when
$\beta$ was held to 2.0 are significantly different.  This suggests
that IRDCs are colder versions of the filaments we identified.  The
detection of an IRDC relies on a strong IR background that the cloud
can absorb against and some of the filaments not seen in absorption
may simply be lacking sufficient background.  Cold dust temperatures
$T_d \le 12$\,K are also found in less massive filaments of nearby
complexes \citep{bontemps10, arzoumanian11} and thus such temperatures
are not restricted to the more massive Galactic IRDCs.

\begin{table*}\scriptsize
\centering
{\small \caption{Dust properties by aperture classification}}
\begin{tabular}{lccccccccc}
\hline\hline
& \multicolumn{2}{|c|}{$\beta = 2$} & \multicolumn{4}{|c|}{$\beta$ free} & & & \\
Class. & $<\!T_d\!>$ & $\sigma_{T_d}$ & $<\!T_d\!>$ & $\sigma_{T_d}$ & $<\!\beta\!>$ & $\sigma_\beta$ & $>\!100\micron$ & $>\!160\micron$ & N \\
& K & K & K & K & & & \% & \% & \\
\hline
${\rm Entire}$ & $25.3$ & $2.4$ & $26.5$ & $2.9$ & $ 1.9$ & $0.3$ & $\phn 50$ & $\phn 0$ & $           8$ \\
${\rm Other\ HII}$ & $26.1$ & $2.0$ & $26.2$ & $3.2$ & $ 2.0$ & $0.3$ & $\phn 37$ & $\phn 0$ & $           8$ \\
${\rm PS}$ & $21.3$ & $3.3$ & $22.2$ & $3.1$ & $ 1.8$ & $0.2$ & $\phn 83$ & $\phn 0$ & $          18$ \\
${\rm PDR}$ & $25.8$ & $4.0$ & $24.4$ & $3.3$ & $ 2.1$ & $0.4$ & $\phn 61$ & $\phn 0$ & $          49$ \\
${\rm All\ Filaments}$ & $16.8$ & $3.0$ & $16.8$ & $3.4$ & $ 2.1$ & $0.5$ & $100$ & $13$ & $          46$ \\
${\rm \ \ \ \ IRDC}$ & $15.5$ & $3.5$ & $14.7$ & $3.6$ & $ 2.2$ & $0.4$ & $100$ & $29$ & $          17$ \\
${\rm \ \ \ \ Not\ IRDC}$ & $18.8$ & $2.9$ & $17.8$ & $2.2$ & $ 2.2$ & $0.2$ & $100$ & $\phn 0$ & $           8$ \\

\hline
\label{tab:classes}
\end{tabular}
\end{table*}

\begin{figure}[!t]
\includegraphics[width=3.6 in]{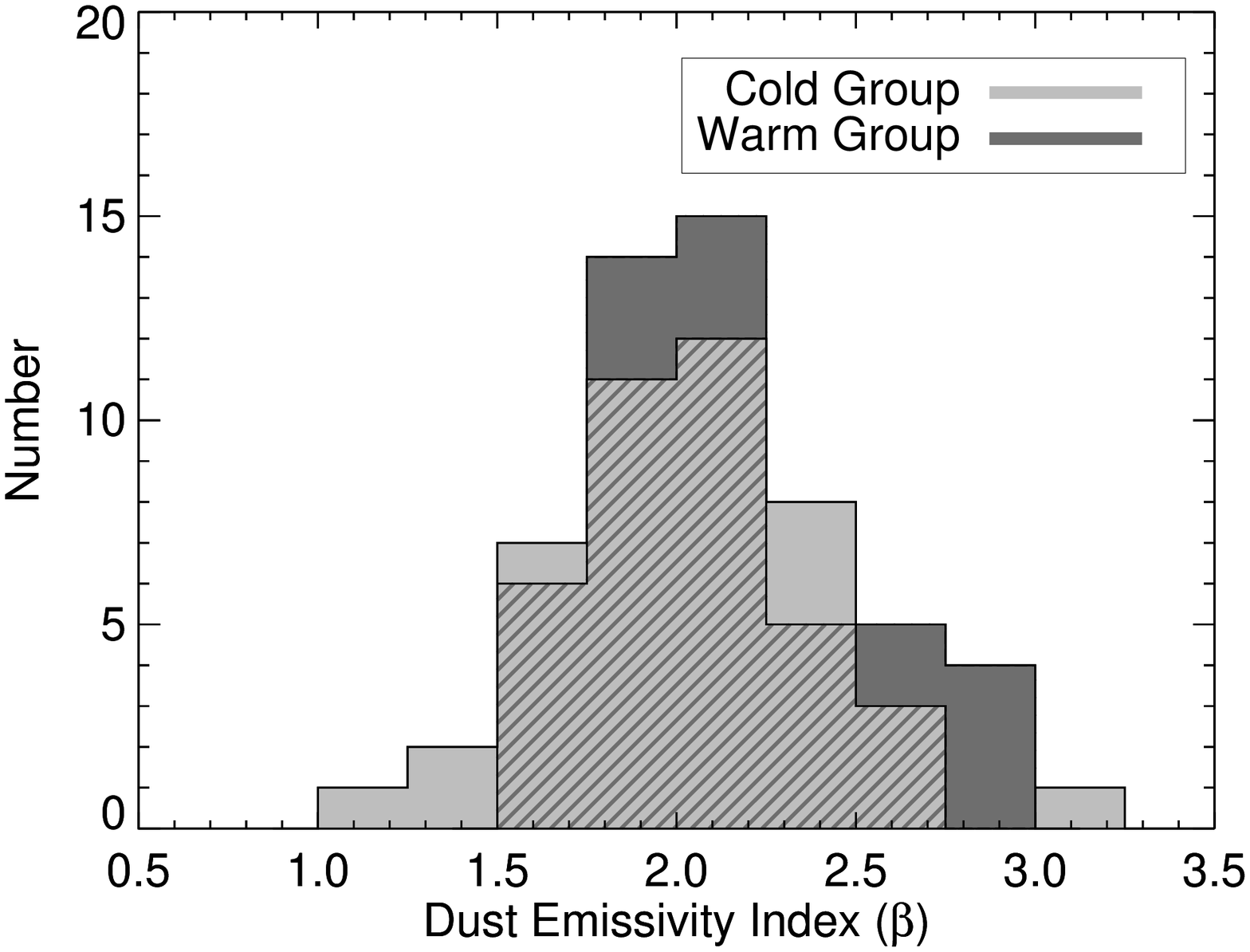}

\caption{Dust opacity spectral index $\beta$ of warm and cold aperture
  groups.  Shown are the distributions for the apertures in the cold
  group (``Filament'' and ``IRDC'' classifications) in light gray and
  the apertures in the warm group (``Entire,'' ``Other \hiismall,'' and ``PDR''
  classifications) in dark gray.  Where the two distributions overlap,
  we show light gray with dark gray hatching.  The two aperture groups
  have only slightly different dust emissivity index distributions.
  The typical error in $\beta$ for individual apertures is $\sim0.3$.}

\label{fig:beta_dist}
\end{figure}

As shown in Table~4, apertures in the ``warm'' and ``cold'' groups
share a similar distribution of $\beta$ values.  We show these
distributions graphically in Figure \ref{fig:beta_dist}.  Both groups
are centered about $\beta = 2.0$.  The average error in $\beta$ is
0.3, which likely causes some of the spread in the distribution.

\subsection{The \boldmath{$\beta-T_d$} Relationship\label{sec:beta_t}}
\begin{figure}
\centering
\includegraphics[width=3.6 in]{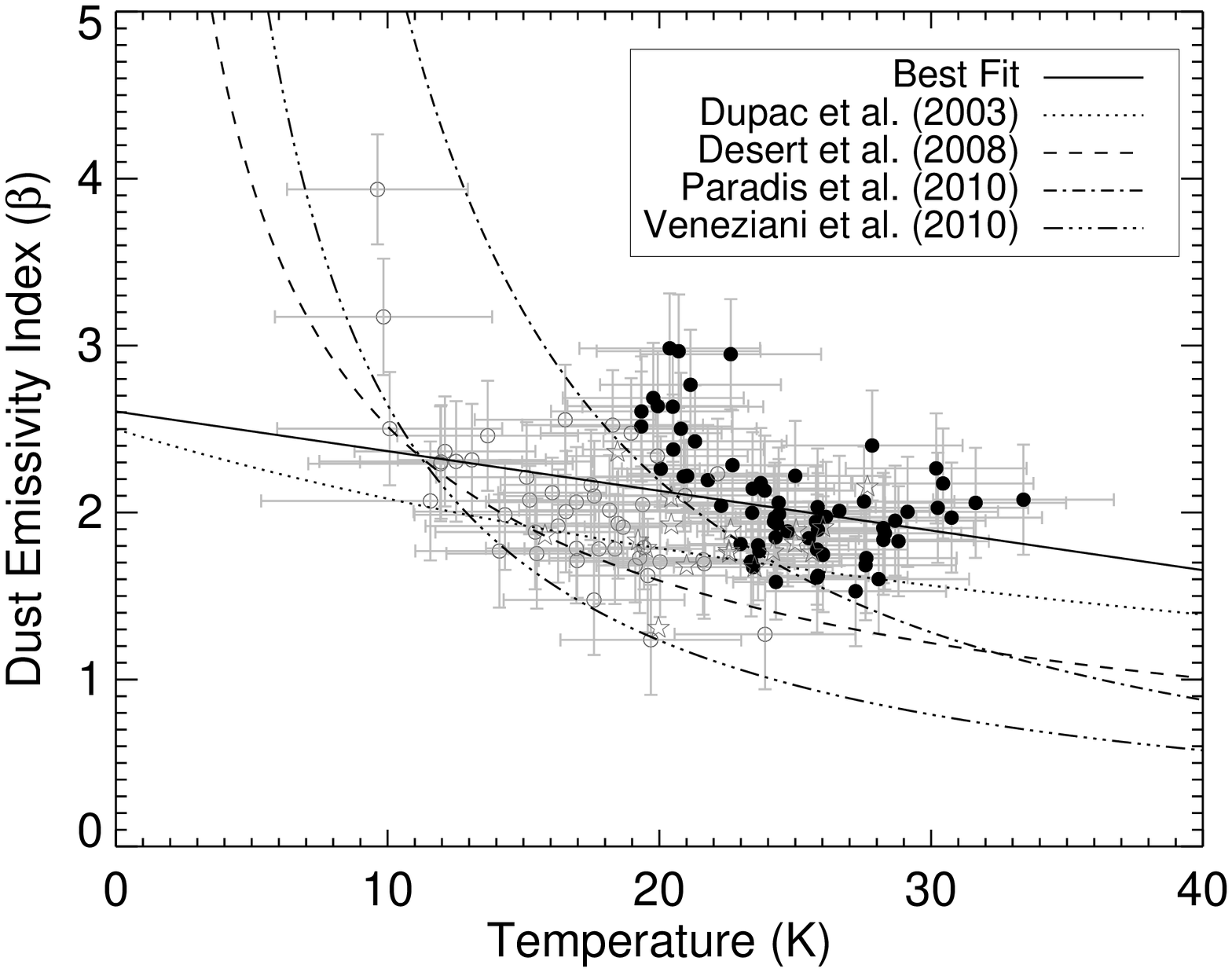}

\caption{The $\beta$-$T_d$ relationship derived from the aperture
  photometry.  Data points from the ``warm'' group are shown with
  filled circles while those from the ``cold'' group are shown with
  open circles.  Data points from ``PS'' apertures are shown with star
  symbols. We plot on top of the data the best fit line as well as the
  best fits lines found by \citet{dupac03}, \citet{desert08},
  \citet{paradis10}, and \citet{veneziani10}.}
\label{fig:t_vs_beta_all}
\end{figure}

As discussed in the Introduction, there is significant observational
evidence for an anti-correlation between $\beta$ and dust temperature.
It is therefore reasonable to expect the same correlation for our
data.  In Figure~\ref{fig:t_vs_beta_all} we plot the dust temperature
against $\beta$ for all apertures.  We do not find a strong
relationship between the two quantities. The data points in
Figure~\ref{fig:t_vs_beta_all} are coded such that apertures in the
``warm'' group are shown with filled circles and those in the ``cold''
group are shown with open circles.  The ``PS'' apertures are shown
with star symbols.  The trend lines found by \citet{dupac03},
\citet{desert08}, \citet{veneziani10}, and \citet{paradis10} are
over-plotted, as is the best-fit line to our data.  The trend lines
found by previous others do not agree with our trend line.  Most of
these previous studies have used different wavelengths to those used
here.

Using the same functional form as \citet{desert08}, our best-fit line
follows the relationship:
\begin{equation}
\beta = (4.3\pm0.5)\,T_d^{-0.2\pm0.04}\,.
\label{eq:fit}
\end{equation}
A linear fit follows the relationship:
\begin{equation}
\beta = 2.6\pm0.1 - (0.02\pm0.005) \, T_d\,
\label{eq:fit2}
\end{equation}
Both fits are rather poor and share the same reduced $\chi^2$ value of
2.2, with the same number of free parameters.  The linear fit in
Equation~\ref{eq:fit2} shows essentially no relation between the two
quantities.  Given that the linear relationship is the simpler of the
two forms, and both have the same statistical significance, we believe
it is more representative of the data shown here.  Amongst the other
studies considered, Equation~\ref{eq:fit2} shows the best agreement
with the relation of \citet{dupac03}.

It is interesting that the cold and warm groups alone appear more
consistent with a $\beta-T_d$ relationship than when they are
combined.  A fit of the form in Equation~\ref{eq:fit} to the cold
group has a reduced $\chi^2$ value of 1.4, while a fit to the warm
group using the same functional form has a reduced $\chi^2$ value of
2.2.  These fits, while better, are still rather poor.  We defer
further discussion of this point to a future publication.

%




\subsubsection{Simulations of the $\beta-T_d$ Relationship}
We perform simulations to better understand the effect of calibration
and photometric uncertainties on the derived $T_d$ and $\beta$-values.
According to their user manuals, the flux calibration for the PACS and
SPIRE instruments is stable in time and the calibration for the
photometric bands of a given detector is correlated.  We model four
different combinations of calibration errors: $+5\%$ for all PACS
bands and $+5\%$ for all SPIRE bands, $+5\%$ for PACS and $-5\%$ for
SPIRE, $-5\%$ for PACS and $+5\%$ for SPIRE, and $-5\%$ for PACS and
$-5\%$ for SPIRE.  We use $+5\%$ for all 870\,\micron\ calibration
errors.  We chose these values for the uncertainty estimates to
  assess how different combinations of uncertainties may affect the
  derivation of the $\beta-T$ relationship.  The values are not
  necessarily consistant with the instrumental and photometric
  uncertainties discussed in more detail below.  \citet{anderson12}
found using {\it Herschel} data that the photometric uncertainties are
in most cases dominated by the choice of background region.  We did
not rigorously estimate this source of uncertainty here but we assume
that the values are similar to those in \citet{anderson12}.  For the
photometric uncertainty, we take the average uncertainty they give for
their photometry of \hii\ regions: 7\%, 10\%, 11\%, 13\%, and 15\% for
the 70\,\micron, 160\,\micron, 250\,\micron, 350\,\micron, and
500\,\micron\ bands.  We assume the photometric uncertainty at
100\,\micron\ is 10\% and the error at 870\,\micron\ is 15\%.

We simulate the emission from dust that has a range of temperatures
from 10 to 35\,K in steps of 5\,K, in a manner similar to that of
\citet{shetty09a}.  For this simulation we use a single $\beta$ value
of 2.0.  For each $\beta-T_d$ combination, we calculate the fluxes of
a simulated gray-body SED.  We then randomly extract values from
Gaussian photometric error distributions with $1\sigma$-levels given
above.  We add the calibration uncertianties in quadrature to the
photometric errors.  The resultant synthetic SEDs have flux values
varied by representative calibration and photometric uncertainties.
We have 16 trials from the combinations of four wavelength options and
four calibration options.

\begin{figure}
\centering
\includegraphics[width=2.8 in]{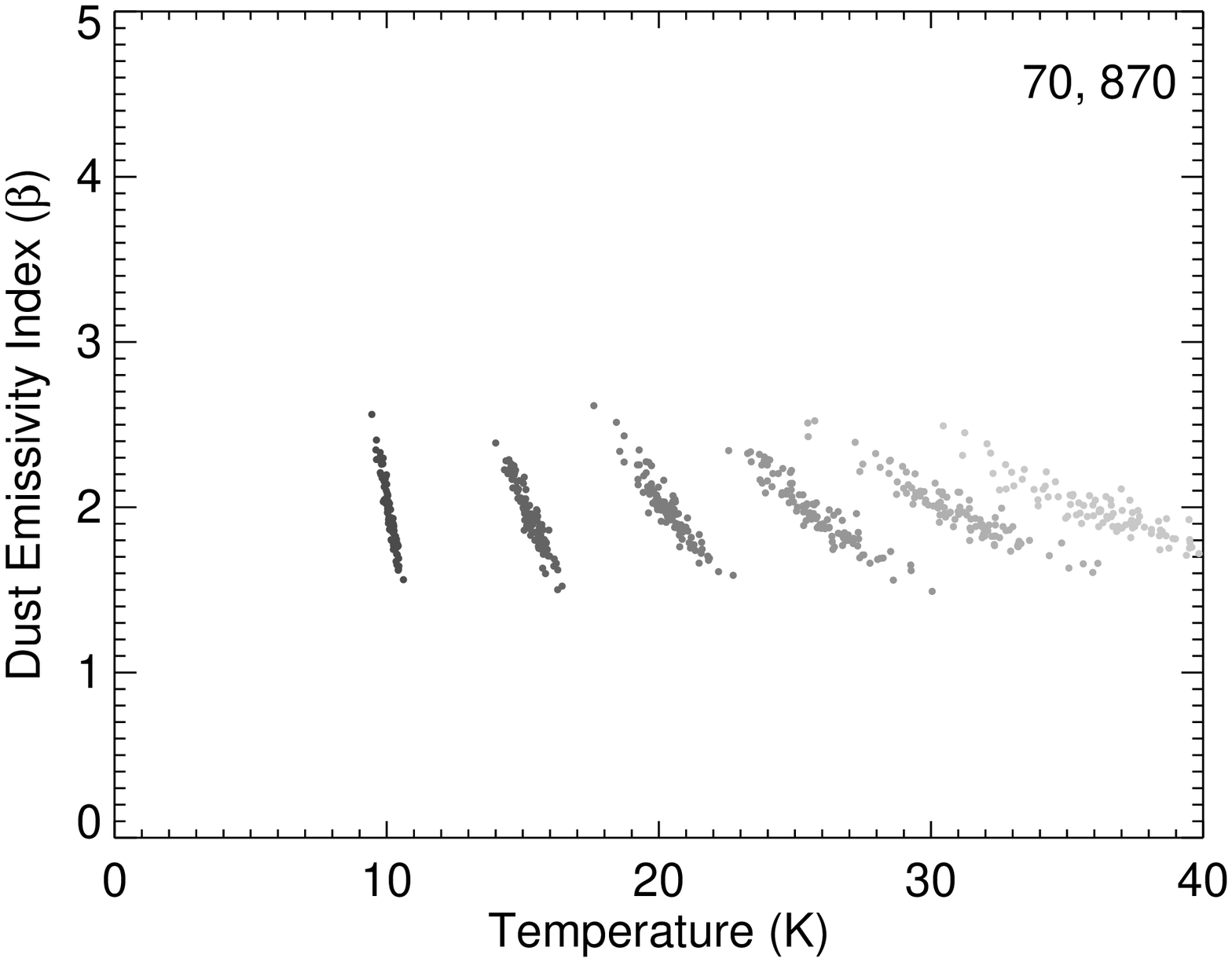}
\includegraphics[width=2.8 in]{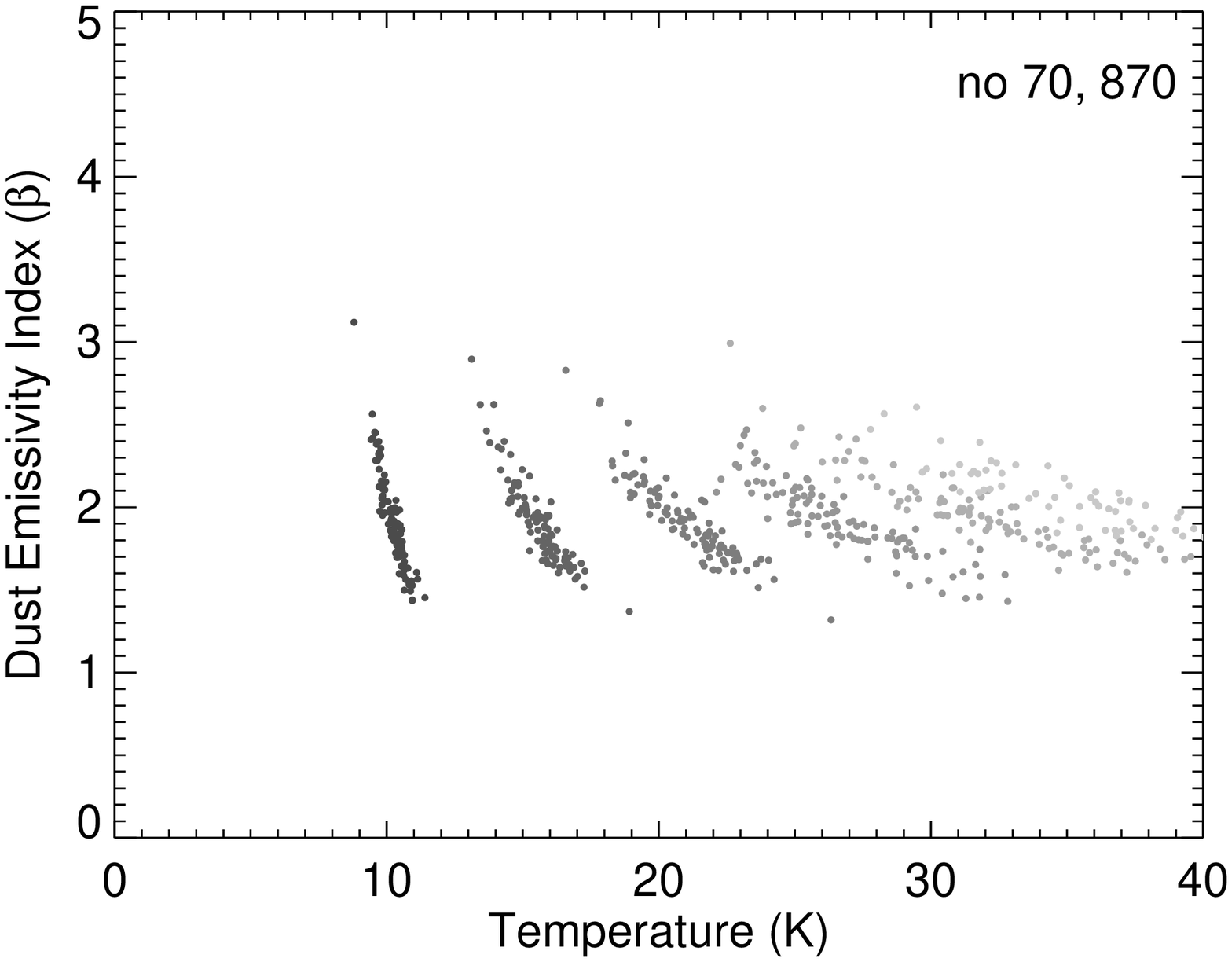}
\includegraphics[width=2.8 in]{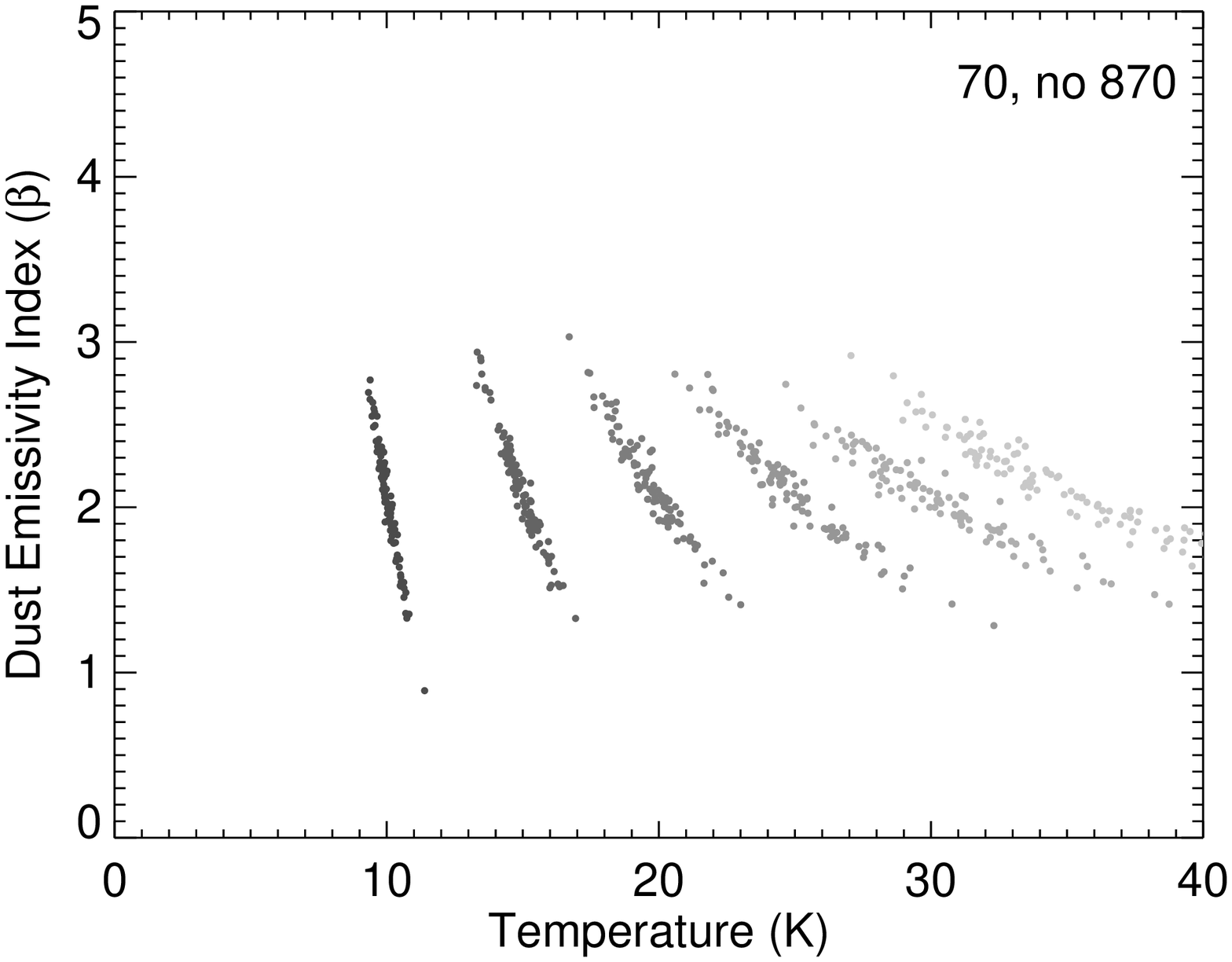}
\includegraphics[width=2.8 in]{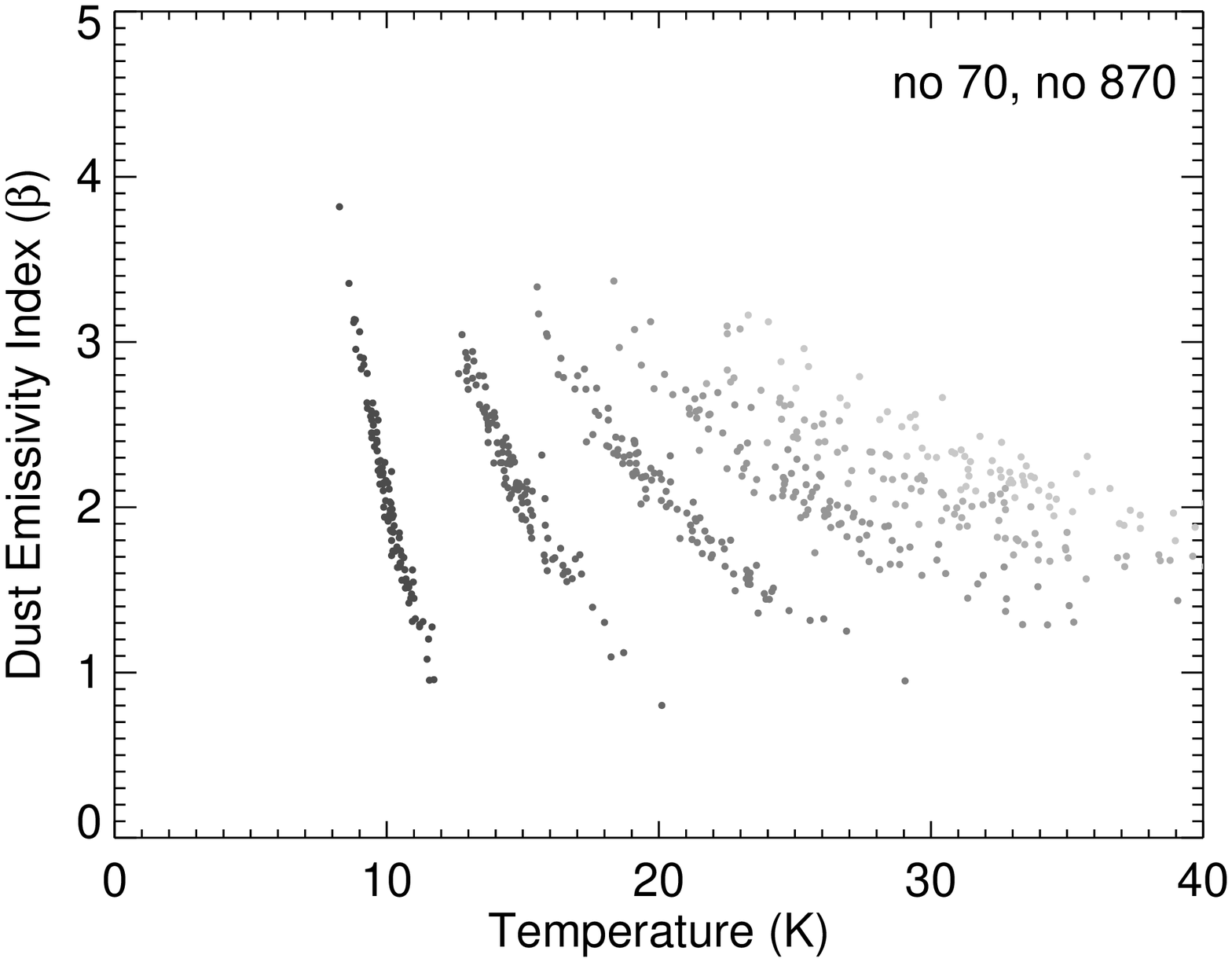}

\caption{Fits of $T_d$ and $\beta$ using simulated data.  Each data
  point is the result of an SED fit to simulated data accounting for
  calibration and photometric uncertainties.  The simulated SEDs were
  created with input temperature values from 10 to 35\,K in steps of 5\,K and
  $\beta=2.0$.  Points with a common input temperature share a common
  shade of gray.}

\label{fig:synthetic}
\end{figure}

We show the results of this modeling in Figure~\ref{fig:synthetic} for
the four trials with calibration errors of $+5\%$ for PACS and $-5\%$
for SPIRE.  Each data point in Figure~\ref{fig:synthetic} represents
the derived $T_d$ and $\beta$-values for a single simulated SED; there
are 600 data points in total, 100 for each input temperature.  Points
sharing a common input temperature also share a common shade of gray.
All four trials with difference calibration errors that share a
wavelength combination have results similar to that shown in
Figure~\ref{fig:synthetic}, although the scatter in the data points is
generally larger when the PACS and SPIRE calibration errors have
inverse signs.

A number of points are evident in Figure~\ref{fig:synthetic}.  First,
for each $\beta-T_d$ pair, a false relationship is produced.
Secondly, a large number of wavelengths helps greatly in reducing the
spread in the calculated values and the uncertainty in the derived
fit.  Excluding either the 70\,\micron\ or the 870\,\micron\ data
results in $\sim50\%$ larger fit uncertainties; excluding both
wavelengths results in $\sim100\%$ larger fit uncertainties in both
$\beta$ and $T_d$.
Finally, the most likely value found is that of the input value; there
is no systematic offset between the derived and input $\beta-T_d$
pair.

Could the $\beta-T_d$ relationship be falsely produced by our
analysis?  To better assess the effect of the {\it Herschel}
calibration and photometric uncertainties on our derived $T_d$ and
$\beta$-values, we repeat the simulation above using as input
temperatures the 129 temperature values from the $\beta$-fixed trial
derived in the aperture photometry (see Table~3), and $\beta=2.0$.
When constructing the simulated SEDs, we include data at the same
wavelengths as were included in our 129 aperture photometry fits: 42
data points with 70\,\micron\ and 870\,\micron\ data, 2 data points
with 70\,\micron\ but no 870\,\micron\ data, 16 data points with
870\,\micron\ data but no 70\,\micron\ data, and 69 data points with
neither 70\,\micron\ nor 870\,\micron\ data.  We use for the
calibration errors one value for all PACS bands and one value for all
SPIRE bands.  This calibration error is drawn randomly from a uniform
distribution, using as a maximum the percentage error given previously
in Section~\ref{sec:data}.  We treat the photometric errors in the
same way as described above and again add these two sources of error
in quadrature.  We simulate the 129 SEDs 1,\,000 times, using for each
of the 1,\,000 trials different calibration errors.  We fit both a
linear regression line and also a power law regression to each of the
1,\,000 sets of 129 points.

\begin{figure} \centering
\includegraphics[width=3 in]{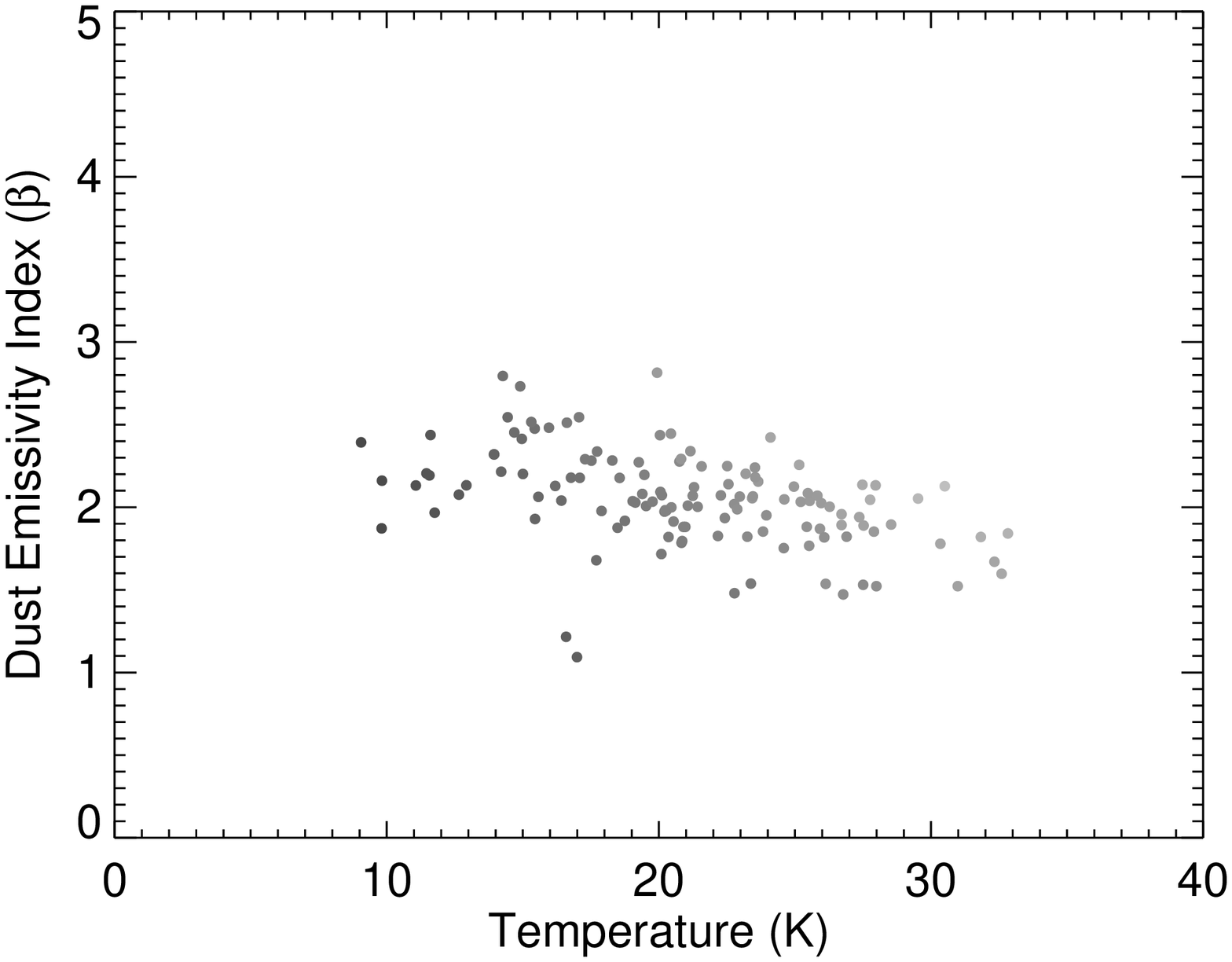}

\caption{Fits of $T_d$ and $\beta$ using simulated data, similar to
  Figure~\ref{fig:synthetic}.  We used the fitted temperatures from
  Table~3 to create the simulated data, and $\beta=2.0$.  Shown is one
  of 1,\,000 trials using different simulated calibration errors.
  Colors indicate the input temperature to the simulated SEDs: black
  is $\sim10$\,K and light gray is $\sim35$\,K.}

\label{fig:synthetic_real}
\end{figure}

In Figure~\ref{fig:synthetic_real} we show an example from one of the
1,\,000 trials: the trial whose linear regression fit parameters are
the closest to the median fit parameters from all 1,\,000 trials.  As
in Figure~\ref{fig:synthetic}, each data point in
Figure~\ref{fig:synthetic_real} is the fit to a simulated SED that
includes calibration and photometric errors.  If there were no sources
of error added, we would recover the 129 derived temperatures, all
with $\beta=2.0$.  The simulated temperature and $\beta$-values shown
in Figure~\ref{fig:synthetic_real} are visually similar to our derived
values shown in Figure~\ref{fig:t_vs_beta_all}.  The median linear fit
is:
\begin{equation}
\beta = 2.6\pm0.2-(0.02\pm0.01) \, T_d\,,
\label{eq:simfit}
\end{equation}
and the median power-law fit is
\begin{equation}
\beta = (4.0\pm1.3)\,T_d^{-0.2\pm0.1}\,.
\label{eq:simfit2}
\end{equation}
The simulated fits again show a weak inverse relationship between
$\beta$ and $T_d$, with the same values (within the errors) as what
was found in Equation~\ref{eq:fit} and Equation~\ref{eq:fit2}.

From Figures~\ref{fig:t_vs_beta_all} and \ref{fig:synthetic_real},
there appears to be less variation in $\beta$ in the simulated data
than in the real data.  The summed square of the difference in $\beta$
between the linear fit and values of $\beta$ in
Figure~\ref{fig:t_vs_beta_all} is 15.6.  The median summed squared
difference in the 1,\,000 fits to the synthetic data described above
is 13.0 with a large standard deviation of 8.2.  We therefore conclude
that most of the variation in $\beta$ can be explained by calibration
and photometric uncertainty but that there may also be variations in
$\beta$ not explained by these sources of uncertainty.


\subsubsection{Comparison with Previous Results Using These Data}
The lack of a strong relationship between $T_d$ and $\beta$ is in
contrast to the suggestion of \citet{anderson10a} and \citet{rodon10}.
Using an earlier processing of the same data employed here for
RCW\,120 and Sh-104 (except for the {\it Herschel} Hi-Gal
70\,\micron\ data for RCW\,120), these authors' results were
consistent with the existence of a relationship between $T_d$ and
$\beta-T$, although these works did not consider the error simulations
that we perform here.  The methods between their works and ours are
nearly identical and therefore it seems likely that the updated
processing of the {\it Herschel} data has caused some of the
differences.  The median absolute flux differences between our
processing and the processing used in \citet{anderson10a} for RCW\,120
(after rebinning to the same pixel grid and subtracting a background
such that 99.99\% of the pixels have positive values) are 2.3\%,
4.7\%, 0.7\%, 0.7\%, and 0.8\% for the 100\,\micron, 160\,\micron,
250\,\micron, 350\,\micron, and 500\,\micron\ data.  The median
absolute flux differences for Sh-104 are similar: 0.8\%, 1.2\%, 0.4\%,
0.3\%, and 0.5\% for the 100\,\micron, 160\,\micron, 250\,\micron,
350\,\micron, and 500\,\micron\ data.\footnote{We also checked the
  flux differences between the current {\it Herschel} processing data
  and a processing employing HIPE version 4.0 and {\it Scanamorphos}
  version 2.0 and find very similar results for both regions.
  Furthermore, the differences between the data processed with HIPE
  version 4.0 and the data used here are $\le 0.3\%$ for all
  photometric bands for both regions.  This indicates that the current
  processing is stable and we do not expect future versions of HIPE or
  {\it Scanamorphos} to significantly alter the flux values.}

The reprocessing has affected some derived values of $T_{\rm dust}$
and $\beta$, but these differences are generally within the previously
stated uncertainties.  Because we use Hi-Gal 70\,\micron\ data here in
place of the MIPSGAL \citep{carey09} 70\,\micron\ data used in
\citet{anderson10a} the differences in derived $T_d$ and
$\beta$-values reflect the combined effect of the new processing
(including changes to the calibration) and the different
70\,\micron\ data sets.  For the trial when $\beta$ was allowed to
vary, the median absolute temperature difference is 2.4\,K.  About
half of the apertures in common have temperature values within the
uncertainties given in \citet{anderson10a}.  The median absolute
difference in $\beta$ is 0.3, and over half of these differences are
within the uncertainties given in \citet{anderson10a}.  The data used
here for Sh-104 are the same as that of \citet{rodon10}, albiet
reprocessed.  We find that, for the trial when $\beta$ was free to
vary, the median absolute difference in temperature between what we
find here and that shown in \citet{rodon10} is 1.2\,K; in all cases
this median absolute temperature difference is less than the errors
given in \citet{rodon10}.  The median absolute difference in
$\beta$-value is 0.2 and for all apertures except for two the
differences are less than the errors given in \citet{rodon10}.

So why does our interpretation of the $\beta-T_d$ relationship differ
from that of \citet{anderson10a} and \citet{rodon10}?  Both of their
results were based on 15 apertures from one field while our result
here is based on 129 apertures from eight fields; we believe our
results are more robust.  Additionally, their $\beta-T_d$ fits were
dependant on results toward the ``interior'' of the bubbles, which we
have excluded from the present analysis.  Our thinking has evolved on
the issue of the bubble interiors and, given the range of dust
temperatures likely present there (see Section~\ref{sec:warm}), it
seems prudent to exclude the interiors from any analysis involving
single-temperature fits.  We note that the temperatures for the
``interior'' apertures in \citet{rodon10} have significantly higher
uncertainties than those of other apertures in the field.  We conclude
that the differences between the present work and the previous
analyses stem partly from differences in data processing, but also
from which apertures were considered and which 70\,\micron\ data were
used.  This demonstrates the caution required for future work on the
$\beta-T_d$ relationship.


\subsection{The Three-Dimensional Nature of Bubbles\label{sec:3d}}
There has recently been some debate on whether the types of \hii\
regions studied here are two-dimensional objects, ``rings,'' or
three-dimensional objects, ``bubbles.''  The $J=3\rightarrow2$ \co\
observations in \citet{beaumont10} show little emission toward the
bubble interiors.  The lack of detected CO emission in the bubble interiors
indicates that ``bubbles'' are flat objects.  \citet{anderson11},
however, suggest that bipolar bubbles would be much more common if
all bubbles were two-dimensional, an effect that is not observed.
Furthermore, these authors point out that the recombination line
widths of bubbles that are not bi-polar should be larger since the
ionized flows toward and away from the observer are largely unimpeeded;
such larger line widths are not observed.

The {\it Herschel} data shown here also support the idea that bubble
\hii\ regions are three-dimensional structures.  Toward the center of
all bubbles in our sample there is emission at all {\it Herschel}
wavelengths.  While at wavelengths $<\!100\,\micron$, such emission
comes mainly from a warmer dust component (see below), the longer
wavelength emission indicates the presense of cold dust.  If such
bubbles were in fact rings, we would not expect to find cold dust
toward their central regions.  \citet{anderson10a} found support for
the hypothesis of \citet{beaumont10} - namely little emission toward
center of RCW\,120.  This observation, however, relied on an earlier
processing of the {\it Herschel} data that removed emission 
around bright PDRs and thus from the bubble interiors.  The
current processing using {\it Scanamorphos} better reconstructs the
flux around and within the bubble interiors.

{\it What percentage of the total emission comes from the bubble
  interiors?}  Using aperture photometry, we answer this question by
comparing the flux from the bubble interiors against that of the
``Entire'' class defined earlier.  The four angularly largest regions
in our sample, Sh\,104, W5-E, RCW\,79, and RCW\,120 permit the most
accurate measurements, and we restrict the current analysis to these
regions.  We use the ``Entire'' apertures from earlier and define new
apertures at each wavelength for the interior regions that trace the
inner PDR boundary.  Because the choice of background aperture may
affect our measurements, we define four background apertures for each
wavelength of the four \hii\ regions.  We perform the aperture
photometry as before to determine the percentage of the total emission
detected towards the bubble ``interiors.''


Roughly $20\%$ of the total emission from the bubbles in our sample
comes from the interior.  This indicates that toward the interior of
the bubbles, we are detecting cold dust emission from the ``front''
and ``back'' sides of the bubble.  The average percentage of emission
from the interior for the four regions studied here at 100\,\micron,
160\,\micron, 250\,\micron, 350\,\micron and 500\,\micron\ is
$26\pm0.7\%$, $20\pm6\%$, $17\pm7\%$, $14\pm8\%$, and $12\pm7\%$,
respectively.  Much of the deviation in the above numbers comes from
Sh\,104, which has 10\% more flux from the interior at each wavelength
compared to the averages.  The exact interpretation of these numbers
requires modeling that is beyond the scope of this paper.  It is,
nevertheless, suggestive that {some bubble \hii\ regions are
three-dimensional objects.}

A higher percentage of the emission at lower wavelengths comes from
the bubble interiors.  Restricting our analysis to RCW\,79 and
RCW\,120, for which we have {\it Herschel} Hi-Gal observations at
70\,\micron\ and MIPSGAL observations at 24\,\micron, we find that the
contribution to the total flux from the interior is 60\% at
24\,\micron\ for RCW\,120 and 55\% for RCW\,79, and 33\% at
70\,\micron\ for RCW\,120 and 69\% for RCW\,79.  These
24\,\micron\ values are slightly higher than that of
\citet{deharveng10}, who found that about half the emission at
24\,\micron\ comes from the bubble interiors.

\subsection{The Warm Dust Component\label{sec:warm}}
\begin{figure*} \centering
\includegraphics[width=1.5 in]{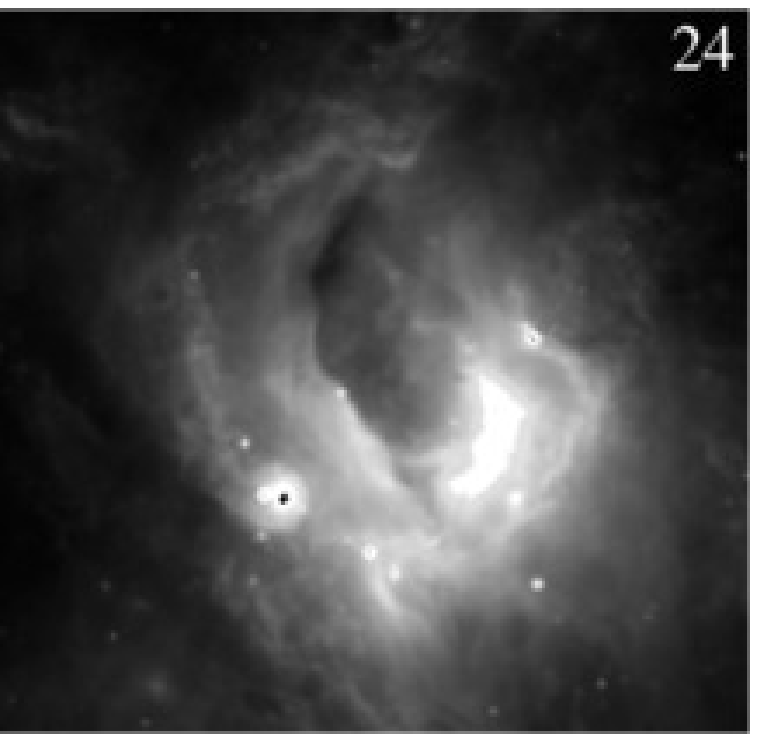}
\includegraphics[width=1.5 in]{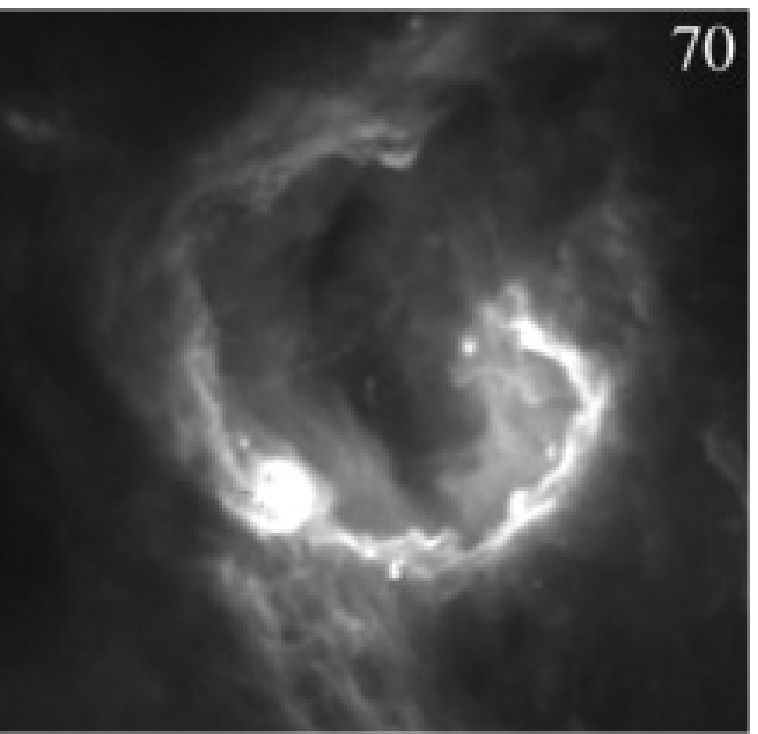}
\includegraphics[width=1.5 in]{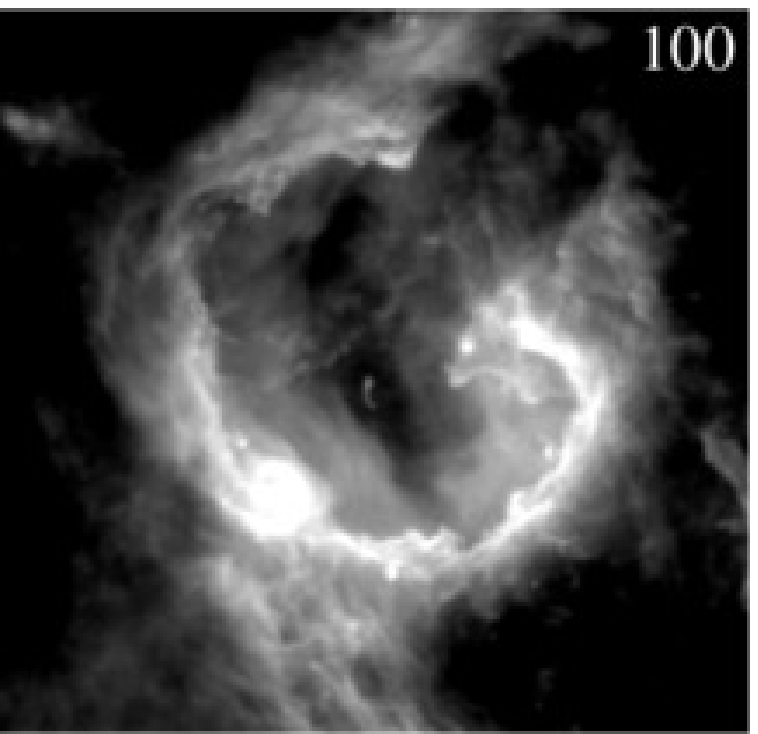}
\includegraphics[width=1.5 in]{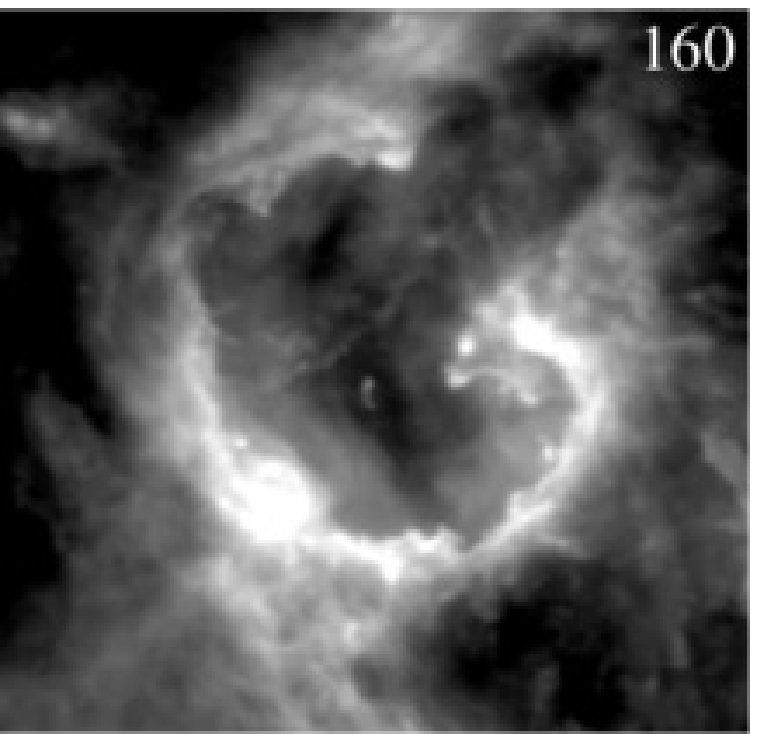}

\caption{The warm dust component in RCW\,79.  The warm dust is prominent at
24\,\micron\ and decreases in intensity relative to the emission from the
PDR at longer wavelengths.  The size of all fields is $20\arcmin$ square.}

\label{fig:warm}
\end{figure*}

Nearly all regions in our sample have a ``warm'' dust component not
accounted for in the current treatment (see
Figure~\ref{fig:avg_seds}).  This component has significantly less
flux compared to the cold component.  For a given flux, the mass
decreases with increasing temperature.  The lower flux and higher
temperature together imply that the warm component contains very
little mass compared to the cold component.  The warm component emits
strongly at 24\,\micron\ for bubble \hii\ regions
\citep[see][]{watson08, deharveng10}, but {\it Herschel} observations
also show emission at longer wavelengths, especially 70\,\micron\ and
100\,\micron, with the same spatial distribution seen at 24\,\micron
(Figure~\ref{fig:warm}).  This indicates that some fraction of the
emission in the {\it Herschel} PACS bands is due to the warm component.
This warm component has generally been attributed to stochastically heated
very small grains \citep[VSGs; ][]{draine07}.

Using the dust model {\it DustEM} \citep{compiegne11}, \citet{compiegne10} show the SED
{\it Herschel} Hi-Gal field at $\ell=59\degr$, a ``diffuse'' region
and a ``dense'' region.  In the ``diffuse'' field they found that the
VSG component contributes on average $35\%$ of the flux at
70\,\micron, 13\% of the flux at 100\,\micron, and 5\% of the flux at
160\,\micron.  In the ``dense'' field, these percentages decreased to
12\%, 4\%, and 2\%, respectively.  Neither region contains active star
formation.  Since nearly all of our measurements are toward dense
structures, it seems likely that the latter numbers are more
representative of our data.

{\it What effect does including data at 70\,\micron\ have on our
  derived temperatures?}  Of the data used here to derive dust
temperatures, the 70\,\micron\ data are most heavily affected by 
  a possible warmer dust contribution, caused in part by the emission
  from VSGs.  To answer this question, we perform two tests.  In the
first test, we recompute the temperature map for RCW\,120 including
the 70\,\micron\ data, and compare it to the map computed without the
70\,\micron\ data.  We find that including or excluding the
70\,\micron\ data point has minimal impact on the derived
temperatures: the average change per temperature map pixel is
$\sim4\%$, or 1.6\,K on average.  In our second test, we re-fit the
fluxes extracted with aperture photometry for RCW\,120.  The
temperatures change by $\sim6\%$ on average for RCW\,120 apertures,
compared to the temperatures derived when the 70\,\micron\ data are
included when $\beta$ is held to a value of 2.0.  When $\beta$ is
allowed to vary, the derived temperatures change by $\sim4\%$ and the
derived $\beta$ values change by $\sim8\%$.  In both cases, excluding
the 70\,\micron\ data has the effect of decreasing the derived
temperature by $<1$\,K on average, although the derived $\beta$-values
are on average neither higher nor lower.  Excluding data at
70\,\micron, however, increases the fit uncertainties
(cf. Figure~\ref{fig:synthetic}).  {\it We conclude that including
  data at 70\,\micron\ data has minimal impact on the derived
  temperatures and $beta$-values.}

We stress that the impact of excluding the 70\,\micron\ is lessened
because we also have data at 100\,\micron.  We re-fit the fluxes
extrated using aperture photometry for RCW\,120 excluding data at both
70\,\micron\ and 100\,\micron.  When $\beta$ is held fixed to a value
of 2.0, we find temperatures $\sim40\%$ different than when the
70\,\micron\ and 100\,\micron\ data were included.  For apertures with
previously-derived temperatures greater than $30$\,K, the temperature
difference is up to 100\%.  Many regions of the Hi-Gal survey will not
have access to {\it Herschel} data at 100\,\micron.  We caution that
the derived temperatures for regions $>30$\,K will not be
well-constrained if 70\,\micron\ data are excluded and
100\,\micron\ data are not available.

{\it What effect may the warmer dust component have on our
  derived temperatures?}  To answer this question, we re-fit the data
for RCW\,120 after decreasing the flux to account for the warmer
  dust.  To account for the largest possible contribution, we
adjusted the fluxes downward by the average values for the diffuse
field in \citet{compiegne10} (35\%, 13\%, and 5\% for 70\,\micron,
100\,\micron, and 160\,\micron) and re-fit the SED.  We find that the
derived temperatures change by $\sim5\%$ both when $\beta$ is fixed
and when it is allowed to vary, and that the derived $\beta$-values
change by $\sim5\%$.  As mentioned above, these probably
  over-estimate any contribution from a warmer dust component for the
dense structures here studied, and therefore the 5\% temperature
change should be regarded as an upper limit.  {\it We conclude that
  the warmer dust component, caused in part by VSGs, has minimal
    effect on the derived temperatures and $\beta$-values.}



\subsection{The Masses of Bubble H{\small II} Regions} 
\begin{figure} \centering
\includegraphics[width=3 in]{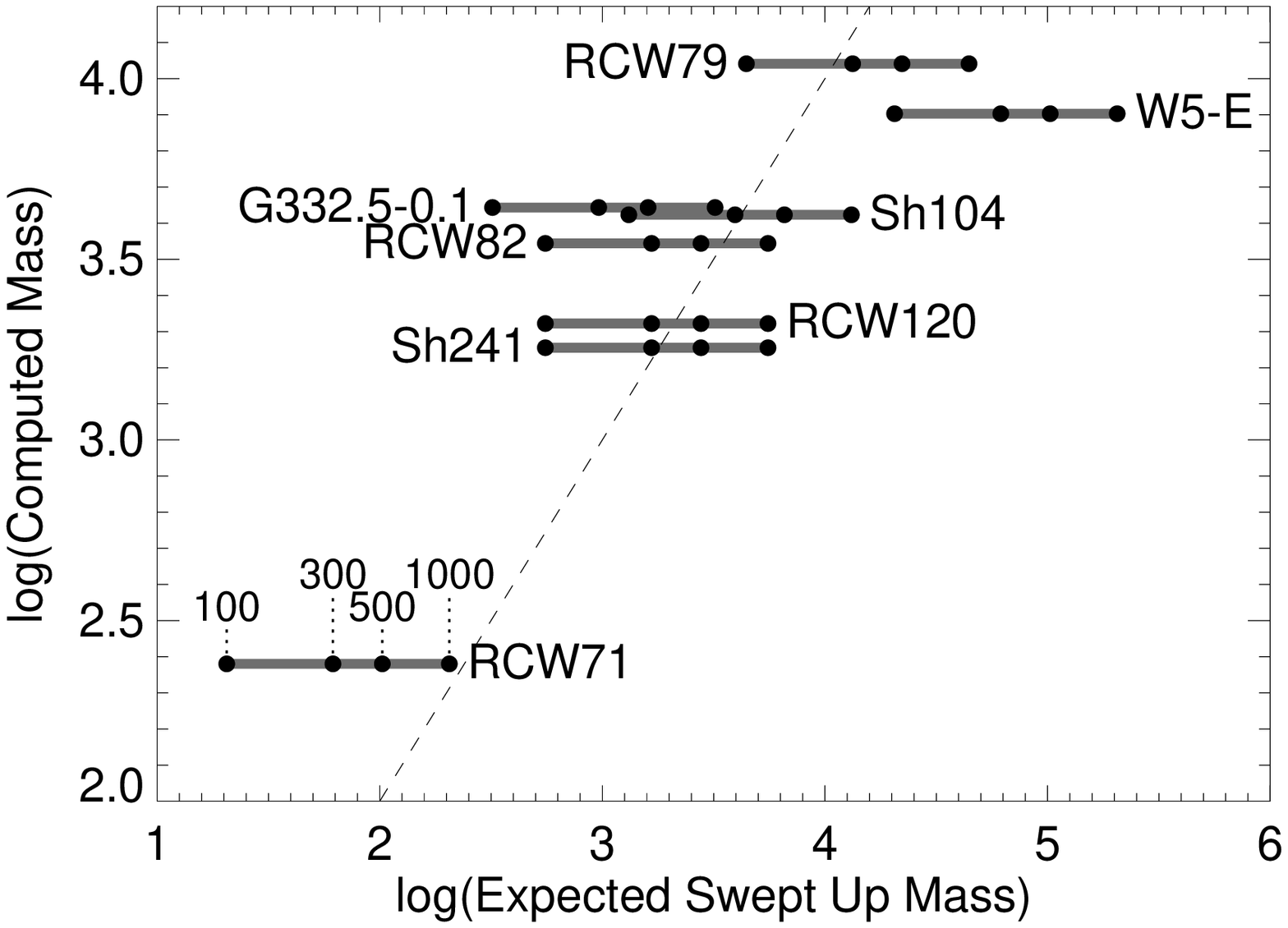}

\caption{The expected swept up mass versus the calculated mass for the
  eight bubble regions.  The expected mass is computed for initial
  ambient densities from 100 to 1,\,000\,cm$^{-3}$, assuming all
  material overtaken by the ionization front is now contained in the
  PDR of the region.  The dashed line shows a one-to-one
  relationship.}

\label{fig:mass}
\end{figure}

The collect and collapse process requires a massive shell of material
for the formation of second-generation stars.  With the calculated
masses, we may test whether the derived mass is similar to that
expected based on the size of the region.  We show in
Figure~\ref{fig:mass} the relationship between the total mass of the
\hii\ region from the ``Entire'' apertures and the expected mass based
on the physical size of the region, for a range of ambient densities
from 100 to 1,\,000\,cm$^{-3}$.  We calculated the expected mass by
assuming that all material overtaken by the ionization front during
the expansion of the \hii\ region is still associated with the region,
by neglecting the mass from the ionized plasma, and by assuming that
the \hii\ regions are spherical.  The expected mass is therefore the
dust density times the volume of the sphere times a gas-to-dust ratio
of 100 (the bubble diameters are given in Table~1).  The assumption
that the bubbles are spherically symmetric introduces uncertainty to
the calculations.  It is nevertheless a good first-order assumption
and is supported by our finding that the bubbles are likely
three-dimensional (see Section~\ref{sec:3d}).

Figure~\ref{fig:mass} shows that for most regions, the computed mass
is similar to the expected mass for reasonable values of the ambient
density. The largest exception is for W5-E, which has significantly
less mass than expected.  The expansion of W5-E may not have been
restricted, as is implied by its open geometry.  This region may have
formed in a filament oriented east-west, and its expansion to the
north and south may have progressed largely unimpeded
\citep{deharveng12}.  The opposite may be the case for RCW\,71 and
G332, which have more mass than expected and appear to have nearly
complete PDRs.

\section{Conclusions}
We have analyzed the dust properties of eight Galactic \hii\ regions
with a bubble morphology using {\it Herschel} data from the HOBYS and
``Evolution of Interstellar Dust'' key programs.  We used aperture
photometry to determine the dust temperature and dust emissivity index
$\beta$ for regions of interest within the fields of these eight
regions.  We also computed maps of the dust temperature and the total
column density.  Most locations show contributions from two dust
components.  Here we only determine temperatures for the colder of
these two components, which contains nearly all the dust mass.

Our conclusions are:
\begin{itemize}
\item All \hii\ regions in the sample are well-characterized with mean
  dust temperatures of 26\,K.

\item The PDRs of the \hii\ regions have mean temperatures of 26\,K.
  The temperature is an important parameter for investigating
  subsequent star formation since higher temperatures lead to higher
  values of the Jeans mass, all else being equal.  At the wavelengths
  sampled by {\it Herschel}, bubble \hii\ regions are dominated by the
  emission from their PDRs -- approximately 80\% of the total flux is
  contained within the PDR.  The remaining 20\% of the flux is from
  the interior of the bubbles, which suggests that the bubbles are
  three-dimensional structures rather than rings.

\item The apertures can be divided into two statistically distinct
  groups: a ``warm'' group averaging 26\,K and a cold group averaging
  17\,K.  The ``warm'' group contains apertures from the PDRs of
  \hii\ regions, as well as the total emission from the \hii\ regions
  themselves while the ``cold'' group contains apertures from cold
  filaments.

\item The coldest regions of the fields are spatially coincident with
  infrared dark clouds (IRDCs); these have mean dust temperatures of
  15\,K.  One third of the IRDCs here detected by {\it Herschel} have
  160\,\micron\ fluxes greater than their 100\,\micron\ fluxes.  This
  criterion can be used to locate clouds with similar temperatures
  that, because of the lack of mid-infrared background, do not appear
  infrared dark.  Filaments that are not detect in absorption at
  8.0\,\micron\ have mean dust temperatures of 19\,K.

\item We estimate that not accounting for the warmer dust component
  and including the data at 70\,\micron\ affects the derived
  temperatures by at most 5\% for the objects studied here.  We stress
  that the inclusion of data at 70\,\micron\ has less of an impact
  because we also have data at 100\,\micron\ -- the exclusion of data
  at both 70\,\micron\ and 100\,\micron\ results in large differences
  in temperature of $\sim40\%$.  This will be the situation for the
  majority of the {\it Herschel} Hi-Gal observations.  In this case,
  the inclusion of data at 70\,\micron\ is preferable to its
  exclusion.  Color corrections employed here change the derived
  temperatures by $\sim3\%$.

\item The total mass associated with these regions spans a wide range,
  from $\sim300$ to $\sim10,\,000$\,\msun.

\item We find only weak support for a relationship between $\beta$ and
  $T_d$, in contrast to what has been found by previous authors.  The
  slight correlation between $\beta$ and $T_d$ that we found is
  statistically similar to that caused by calibration and photometric
  uncertainties.  This result differs from what was suggested by
  earlier analyses of the data shown here, reduced using a different
  processing method.  That the $\beta-T_d$ relationship is so
  sensitive to the data processing method shows the difficulty in
  trying to understand it using {\it Herschel} data alone.

\end{itemize}

\clearpage
 \renewcommand{\thefootnote}{\alph{footnote}}
{\tiny
\longtab{3}{
\begin{longtable}{llccccccccccl}
\caption{Dust properties derived from aperture photometry}\\
\hline\hline \\ [-2ex]
& & & & & \multicolumn{2}{|c|}{$\beta = 2$} & \multicolumn{3}{|c|}{$\beta$ free} & & & \\
Region & Name\tablenotemark{a} & $l$ & $b$ & Size & $T_d$ & $M$ & $T_d$ & $\beta$ & $M$ & Class.\tablenotemark{b} \\
& & deg. & deg. & $\arcmin^2$ & K & \msun & K & & \msun & & & \\
\hline
Sh\,104&Entire& 74.765&$+0.624$&$75.0$&$  24.4\pm   0.8$&$4200$&$  27.6\pm   3.9$&$   1.7\pm   0.3$&$3300$&E \\
 &HII (UCHII)& 74.790&$+0.572$&$\phn 2.7$&$  28.1\pm   1.0$&$\phn 440$&$  28.7\pm   4.2$&$   2.0\pm   0.3$&$\phn 420$&H {\scriptsize II} \\
 &CS (CS)& 74.721&$+0.590$&$\phn 0.5$&$  24.1\pm   0.7$&$\phn \phn 89$&$  25.0\pm   3.2$&$   1.9\pm   0.3$&$\phn \phn 82$&PS \\
 &E. PDR& 74.766&$+0.565$&$\phn 1.2$&$  24.9\pm   0.8$&$\phn 140$&$  24.4\pm   3.0$&$   2.1\pm   0.3$&$\phn 150$&PDR \\
 &N. PDR 1 (PDR 2)& 74.812&$+0.615$&$\phn 0.9$&$  22.2\pm   0.6$&$\phn 120$&$  23.6\pm   2.8$&$   1.8\pm   0.3$&$\phn 110$&PDR \\
 &N. PDR 2 (PDR 3)& 74.818&$+0.657$&$\phn 2.0$&$  24.0\pm   0.7$&$\phn 110$&$  24.3\pm   3.0$&$   2.0\pm   0.3$&$\phn 110$&PDR \\
 &W. PDR 1 (PDR 4)& 74.787&$+0.679$&$\phn 2.6$&$  21.3\pm   0.6$&$\phn 300$&$  23.4\pm   2.7$&$   1.7\pm   0.3$&$\phn 240$&PDR \\
 &W. PDR 2 (PDR 6)& 74.728&$+0.671$&$\phn 2.4$&$  23.8\pm   0.7$&$\phn 160$&$  24.7\pm   3.1$&$   1.9\pm   0.3$&$\phn 150$&PDR \\
 &S. PDR 1 (PDR 7)& 74.706&$+0.624$&$\phn 1.4$&$  25.0\pm   0.8$&$\phn 160$&$  23.9\pm   2.9$&$   2.1\pm   0.3$&$\phn 170$&PDR \\
 &S. PDR 2 (PDR 8)& 74.736&$+0.565$&$\phn 1.2$&$  26.7\pm   0.9$&$\phn \phn 49$&$  26.6\pm   3.7$&$   2.0\pm   0.3$&$\phn \phn 50$&PDR \\
 &N. Filament 1& 74.865&$+0.705$&$\phn 0.8$&$  15.0\pm   0.5$&$\phn \phn 31$&$  13.7\pm   2.2$&$   2.5\pm   0.9$&$\phn \phn 41$&F \\
 &N. Filament 2& 74.822&$+0.704$&$\phn 2.4$&$  18.1\pm   0.4$&$\phn 180$&$  18.5\pm   1.7$&$   1.9\pm   0.3$&$\phn 170$&F \\
 &N. Filament 3& 74.843&$+0.618$&$\phn 1.3$&$  17.9\pm   0.4$&$\phn 110$&$  19.3\pm   1.9$&$   1.7\pm   0.3$&$\phn \phn 89$&F \\
 &N. Filament 4* (PDR 1)& 74.836&$+0.583$&$\phn 2.2$&$  19.9\pm   0.5$&$\phn 170$&$  21.6\pm   2.3$&$   1.7\pm   0.3$&$\phn 140$&F \\
 &N. Filament 5*& 74.767&$+0.783$&$\phn 0.8$&$  16.9\pm   0.6$&$\phn \phn 22$&$  17.8\pm   3.2$&$   1.8\pm   0.8$&$\phn \phn 19$&F \\
 &S. Filament 1*& 74.670&$+0.566$&$\phn 7.4$&$  21.6\pm   0.6$&$\phn 180$&$  20.9\pm   2.5$&$   2.1\pm   0.4$&$\phn 190$&F \\
 &S. Filament 2* (Outer)& 74.662&$+0.662$&$\phn 1.7$&$  14.8\pm   0.8$&$\phn \phn 39$&$  15.5\pm   3.4$&$   1.8\pm   1.2$&$\phn \phn 35$&F \\
W5-E&Entire&138.060&$+1.461$&$2515.3$&$  23.8\pm   0.7$&$8000$&$  24.2\pm   3.0$&$   1.9\pm   0.3$&$7700$&E \\
 &HII 1 (Sh 201)&138.481&$+1.637$&$\phn \phn 69.6$&$  26.9\pm   0.9$&$\phn 830$&$  28.3\pm   4.1$&$   1.9\pm   0.3$&$\phn 750$&H {\scriptsize II} \\
 &HII 2&137.696&$+1.533$&$\phn \phn 25.1$&$  24.9\pm   0.8$&$\phn 590$&$  25.8\pm   3.4$&$   1.9\pm   0.3$&$\phn 550$&H {\scriptsize II} \\
 &HII 3&137.735&$+1.458$&$\phn \phn 23.8$&$  23.8\pm   0.7$&$\phn 690$&$  26.0\pm   3.4$&$   1.8\pm   0.3$&$\phn 580$&H {\scriptsize II} \\
 &E. PS 1&138.323&$+1.509$&$\phn \phn \phn 2.4$&$  21.0\pm   0.6$&$\phn \phn 43$&$  22.6\pm   2.5$&$   1.8\pm   0.3$&$\phn \phn 37$&PS \\
 &E. PS 2 (BRC14)&138.298&$+1.560$&$\phn \phn \phn 4.4$&$  23.5\pm   0.7$&$\phn 440$&$  25.6\pm   3.3$&$   1.8\pm   0.3$&$\phn 370$&PS \\
 &N. PS 1 (BRC13)&138.142&$+1.684$&$\phn \phn \phn 1.6$&$  22.4\pm   0.6$&$\phn \phn 60$&$  24.2\pm   2.9$&$   1.8\pm   0.3$&$\phn \phn 51$&PS \\
 &W. PS 1&137.793&$+1.461$&$\phn \phn \phn 1.7$&$  23.4\pm   0.7$&$\phn \phn 49$&$  25.0\pm   3.1$&$   1.8\pm   0.3$&$\phn \phn 43$&PS \\
 &W. PS 2 (BRC12)&137.505&$+1.397$&$\phn \phn \phn 1.3$&$  21.9\pm   0.6$&$\phn \phn 18$&$  22.6\pm   2.6$&$   1.9\pm   0.3$&$\phn \phn 17$&PS \\
 &W. PS 3 (BRC12)&137.537&$+1.276$&$\phn \phn \phn 1.5$&$  21.2\pm   0.6$&$\phn \phn 40$&$  23.5\pm   2.7$&$   1.7\pm   0.3$&$\phn \phn 32$&PS \\
 &E. PDR 1 (BRC14)&138.308&$+1.568$&$\phn \phn 23.0$&$  22.6\pm   0.7$&$\phn 770$&$  25.8\pm   3.4$&$   1.6\pm   0.3$&$\phn 580$&PDR \\
 &E. PDR 2&138.319&$+1.671$&$\phn \phn \phn 3.7$&$  21.9\pm   0.6$&$\phn \phn 54$&$  23.7\pm   2.8$&$   1.8\pm   0.3$&$\phn \phn 46$&PDR \\
 &N. PDR 1 (BRC13)&138.187&$+1.727$&$\phn \phn 24.7$&$  21.6\pm   0.6$&$\phn 400$&$  23.0\pm   2.6$&$   1.8\pm   0.3$&$\phn 350$&PDR \\
 &N. PDR 2&138.031&$+1.762$&$\phn \phn \phn 3.9$&$  22.5\pm   0.6$&$\phn \phn 27$&$  21.0\pm   2.2$&$   2.2\pm   0.3$&$\phn \phn 32$&PDR \\
 &N. PDR 3&137.921&$+1.664$&$\phn \phn 19.2$&$  21.2\pm   0.6$&$\phn 170$&$  24.3\pm   2.9$&$   1.6\pm   0.3$&$\phn 130$&PDR \\
 &N. PDR 4&137.897&$+1.920$&$\phn \phn 43.2$&$  21.7\pm   0.6$&$\phn 160$&$  20.1\pm   2.0$&$   2.3\pm   0.3$&$\phn 200$&PDR \\
 &W. PDR 1&137.930&$+1.119$&$\phn 196.1$&$  23.1\pm   0.7$&$1100$&$  24.3\pm   3.0$&$   1.9\pm   0.3$&$1000$&PDR \\
 &W. PDR 2 (BRC12)&137.553&$+1.344$&$\phn \phn \phn 6.9$&$  23.8\pm   0.7$&$\phn \phn 67$&$  24.3\pm   3.0$&$   1.9\pm   0.3$&$\phn \phn 64$&PDR \\
 &N. Filament 1&138.256&$+1.850$&$\phn \phn \phn 1.1$&$  18.4\pm   0.4$&$\phn \phn \phn 7.6$&$  19.4\pm   2.0$&$   1.8\pm   0.4$&$\phn \phn \phn 6.6$&F \\
 &N. Filament 2&137.738&$+1.919$&$\phn \phn \phn 5.4$&$  15.8\pm   0.3$&$\phn \phn 54$&$  17.0\pm   1.4$&$   1.7\pm   0.4$&$\phn \phn 45$&F \\
 &E. Filament 1&138.596&$+1.707$&$\phn \phn 12.4$&$  17.5\pm   0.4$&$\phn \phn 67$&$  19.1\pm   1.8$&$   1.7\pm   0.3$&$\phn \phn 54$&F \\
 &E. Filament 2&138.594&$+1.584$&$\phn \phn \phn 4.7$&$  18.3\pm   0.4$&$\phn \phn 79$&$  18.7\pm   1.7$&$   1.9\pm   0.3$&$\phn \phn 75$&F \\
Sh\,241&Entire&180.886&$+4.114$&$16.5$&$  22.9\pm   0.7$&$1800$&$  27.2\pm   3.8$&$   1.5\pm   0.3$&$1300$&E \\
 &HII&180.828&$+3.993$&$\phn 2.6$&$  24.8\pm   0.8$&$\phn 350$&$  27.6\pm   3.9$&$   1.7\pm   0.3$&$\phn 290$&H {\scriptsize II} \\
 &N. PS*&180.820&$+4.189$&$\phn 0.4$&$  19.2\pm   0.5$&$\phn \phn 32$&$  21.0\pm   2.2$&$   1.7\pm   0.3$&$\phn \phn 26$&PS \\
 &W. PS*&180.859&$+4.032$&$\phn 0.5$&$  16.6\pm   0.4$&$\phn \phn 93$&$  20.0\pm   1.9$&$   1.3\pm   0.3$&$\phn \phn 58$&PS \\
 &S. PS*&180.918&$+4.017$&$\phn 0.9$&$  18.4\pm   0.4$&$\phn 110$&$  19.5\pm   1.9$&$   1.8\pm   0.3$&$\phn \phn 98$&PS \\
 &N. PDR 1&180.856&$+4.122$&$\phn 0.8$&$  23.8\pm   0.7$&$\phn 150$&$  25.8\pm   3.4$&$   1.8\pm   0.3$&$\phn 130$&PDR \\
 &N. PDR 2&180.858&$+4.106$&$\phn 0.7$&$  24.1\pm   0.8$&$\phn 150$&$  28.1\pm   4.0$&$   1.6\pm   0.3$&$\phn 110$&PDR \\
 &W. PDR&180.873&$+4.092$&$\phn 1.3$&$  22.5\pm   0.7$&$\phn 560$&$  25.8\pm   3.3$&$   1.6\pm   0.3$&$\phn 420$&PDR \\
 &W. Filament 1*&180.855&$+4.069$&$\phn 0.4$&$  19.2\pm   0.5$&$\phn \phn 39$&$  23.9\pm   3.3$&$   1.3\pm   0.4$&$\phn \phn 24$&F \\
 &W. Filament 2*&180.840&$+4.044$&$\phn 5.4$&$  18.5\pm   0.4$&$\phn 540$&$  20.0\pm   2.0$&$   1.7\pm   0.3$&$\phn 450$&F \\
 &W. Filament 3*&180.866&$+4.053$&$\phn 0.3$&$  15.4\pm   0.3$&$\phn 110$&$  17.6\pm   1.6$&$   1.5\pm   0.4$&$\phn \phn 79$&F \\
 &W. Filament 4*&180.874&$+4.047$&$\phn 0.7$&$  16.1\pm   0.4$&$\phn 140$&$  19.7\pm   1.9$&$   1.2\pm   0.3$&$\phn \phn 82$&F \\
 &S. Filament*&180.915&$+4.040$&$\phn 0.9$&$  17.3\pm   0.4$&$\phn 100$&$  18.4\pm   1.9$&$   1.8\pm   0.4$&$\phn \phn 87$&F \\
RCW\,71&Entire&302.799&$+1.287$&$21.6$&$  30.3\pm   1.2$&$240$&$  30.8\pm   5.0$&$   2.0\pm   0.3$&$230$&E \\
 &S. PS 1*&302.842&$+1.225$&$\phn 0.8$&$  15.3\pm   0.3$&$\phn 37$&$  15.8\pm   1.4$&$   1.9\pm   0.4$&$\phn 34$&PS \\
 &S. PS 2*&302.809&$+1.233$&$\phn 0.9$&$  18.2\pm   0.4$&$\phn 20$&$  19.2\pm   2.0$&$   1.8\pm   0.4$&$\phn 18$&PS \\
 &N. PDR&302.802&$+1.309$&$\phn 1.6$&$  34.5\pm   1.6$&$\phn 13$&$  30.2\pm   5.0$&$   2.3\pm   0.4$&$\phn 16$&PDR \\
 &W. PDR&302.780&$+1.284$&$\phn 0.7$&$  30.7\pm   1.2$&$\phn 15$&$  30.2\pm   4.9$&$   2.0\pm   0.3$&$\phn 15$&PDR \\
 &S. PDR 1&302.784&$+1.259$&$\phn 1.1$&$  28.3\pm   1.0$&$\phn 20$&$  27.5\pm   4.0$&$   2.1\pm   0.3$&$\phn 21$&PDR \\
 &S. PDR 2&302.807&$+1.261$&$\phn 0.8$&$  26.4\pm   0.9$&$\phn 24$&$  28.2\pm   4.1$&$   1.8\pm   0.3$&$\phn 21$&PDR \\
 &E. PDR 1&302.822&$+1.291$&$\phn 0.6$&$  34.9\pm   1.7$&$\phn \phn 8.7$&$  33.4\pm   6.2$&$   2.1\pm   0.4$&$\phn \phn 9.3$&PDR \\
 &E. PDR 2&302.810&$+1.277$&$\phn 0.6$&$  33.2\pm   1.5$&$\phn 19$&$  30.4\pm   4.9$&$   2.2\pm   0.3$&$\phn 22$&PDR \\
 &E. PDR 3&302.823&$+1.273$&$\phn 0.5$&$  32.6\pm   1.4$&$\phn \phn 6.9$&$  31.6\pm   5.6$&$   2.1\pm   0.4$&$\phn \phn 7.3$&PDR \\
 &W. Filament 1*&302.718&$+1.319$&$\phn 2.3$&$  16.1\pm   0.4$&$\phn 40$&$  16.9\pm   1.6$&$   1.8\pm   0.4$&$\phn 35$&F \\
 &W. Filament 2*&302.732&$+1.307$&$\phn 0.5$&$  18.3\pm   0.5$&$\phn \phn 5.8$&$  18.2\pm   2.5$&$   2.0\pm   0.6$&$\phn \phn 6.0$&F \\
 &W. Filament 3*&302.720&$+1.271$&$\phn 0.6$&$  14.3\pm   0.3$&$\phn 12$&$  14.3\pm   1.7$&$   2.0\pm   0.6$&$\phn 12$&F \\
 &W. Filament 4*&302.734&$+1.276$&$\phn 0.5$&$  17.3\pm   0.4$&$\phn \phn 8.9$&$  16.9\pm   2.0$&$   2.1\pm   0.5$&$\phn \phn 9.3$&F \\
 &N. Filament*&302.859&$+1.388$&$\phn 0.8$&$  12.8\pm   0.6$&$\phn \phn 9.8$&$   9.6\pm   1.9$&$   3.9\pm   1.6$&$\phn 29$&F \\
RCW\,79&Entire&308.709&$+0.628$&$182.4$&$  26.9\pm   0.9$&$11000$&$  28.8\pm   3.8$&$   1.8\pm   0.3$&$9700$&E \\
 &HII&308.750&$+0.548$&$\phn \phn 6.8$&$  27.5\pm   0.9$&$\phn 1700$&$  25.0\pm   2.0$&$   2.2\pm   0.2$&$2100$&H {\scriptsize II} \\
 &W. PS 1 (Cond. 4)&308.645&$+0.648$&$\phn \phn 1.4$&$  25.1\pm   0.8$&$\phn \phn 720$&$  26.0\pm   2.4$&$   1.9\pm   0.2$&$\phn 660$&PS \\
 &W. PS 2 (Cond. 6)&308.673&$+0.637$&$\phn \phn 1.1$&$  29.7\pm   1.1$&$\phn \phn \phn 73$&$  27.7\pm   3.1$&$   2.2\pm   0.2$&$\phn \phn 84$&PS \\
 &N. PDR 1 (Cond. 5)&308.729&$+0.725$&$\phn \phn 2.6$&$  23.4\pm   0.6$&$\phn \phn 440$&$  21.8\pm   1.7$&$   2.2\pm   0.2$&$\phn 520$&PDR \\
 &N. PDR 2&308.793&$+0.648$&$\phn 14.0$&$  29.2\pm   1.0$&$\phn \phn 660$&$  21.1\pm   1.6$&$   2.8\pm   0.2$&$1300$&PDR \\
 &W. PDR 1&308.630&$+0.558$&$\phn \phn 3.6$&$  34.3\pm   1.5$&$\phn \phn 170$&$  27.8\pm   2.9$&$   2.4\pm   0.2$&$\phn 250$&PDR \\
 &W. PDR 2&308.613&$+0.607$&$\phn \phn 4.2$&$  36.1\pm   1.6$&$\phn \phn 250$&$  22.6\pm   1.8$&$   2.9\pm   0.2$&$\phn 610$&PDR \\
 &W. PDR 3 (Cond. 7)&308.673&$+0.620$&$\phn \phn 0.9$&$  27.1\pm   1.0$&$\phn \phn \phn 69$&$  28.2\pm   3.2$&$   1.9\pm   0.2$&$\phn \phn 64$&PDR \\
 &S. PDR (Cond. 3)&308.690&$+0.533$&$\phn \phn 6.6$&$  25.2\pm   0.8$&$\phn 1400$&$  25.8\pm   2.3$&$   1.9\pm   0.2$&$1300$&PDR \\
 &2nd PDR E.&308.708&$+0.470$&$\phn 23.5$&$  24.3\pm   0.7$&$\phn 2000$&$  24.4\pm   3.0$&$   2.0\pm   0.3$&$2000$&PDR \\
 &2nd PDR W.&308.522&$+0.585$&$\phn 14.7$&$  23.1\pm   0.7$&$\phn \phn 610$&$  20.5\pm   2.1$&$   2.4\pm   0.4$&$\phn 800$&PDR \\
 &N. Filament (Cond. 8)&308.725&$+0.774$&$\phn \phn 6.9$&$  21.7\pm   0.5$&$\phn \phn 950$&$  18.3\pm   1.1$&$   2.5\pm   0.2$&$1400$&F \\
 &E. Filament*&308.892&$+0.698$&$\phn \phn 7.9$&$  24.0\pm   0.7$&$\phn \phn 130$&$  22.1\pm   2.4$&$   2.2\pm   0.3$&$\phn 160$&F \\
RCW\,82&Entire&310.980&$+0.408$&$62.2$&$  24.6\pm   0.7$&$3400$&$  21.3\pm   2.2$&$   2.4\pm   0.3$&$4700$&E \\
 &E. PS&311.034&$+0.382$&$\phn 1.8$&$  21.1\pm   0.5$&$\phn 430$&$  20.4\pm   1.4$&$   2.1\pm   0.2$&$\phn 470$&PS \\
 &N. PDR 1&311.003&$+0.462$&$\phn 2.2$&$  25.3\pm   0.8$&$\phn 200$&$  22.7\pm   1.9$&$   2.3\pm   0.2$&$\phn 250$&PDR \\
 &W. PDR 1&310.948&$+0.424$&$\phn 2.8$&$  29.7\pm   1.0$&$\phn 130$&$  20.4\pm   1.9$&$   3.0\pm   0.3$&$\phn 280$&PDR \\
 &W. PDR 2&310.928&$+0.445$&$\phn 2.1$&$  25.8\pm   0.8$&$\phn 190$&$  20.5\pm   1.7$&$   2.6\pm   0.3$&$\phn 310$&PDR \\
 &S. PDR&310.987&$+0.359$&$\phn 4.8$&$  25.5\pm   0.7$&$\phn 340$&$  19.8\pm   1.4$&$   2.7\pm   0.2$&$\phn 590$&PDR \\
 &E. PDR 1&311.026&$+0.413$&$\phn 1.5$&$  25.3\pm   0.7$&$\phn 160$&$  19.9\pm   1.4$&$   2.6\pm   0.2$&$\phn 270$&PDR \\
 &E. PDR 2&311.013&$+0.430$&$\phn 1.1$&$  23.0\pm   0.6$&$\phn 110$&$  19.3\pm   1.4$&$   2.5\pm   0.2$&$\phn 160$&PDR \\
 &N. Filament*&311.060&$+0.516$&$\phn 2.6$&$  21.9\pm   0.6$&$\phn \phn 79$&$  19.0\pm   1.8$&$   2.5\pm   0.4$&$\phn 110$&F (ID) \\
 &W. Filament*&310.850&$+0.486$&$\phn 4.0$&$  18.4\pm   0.4$&$\phn 260$&$  17.5\pm   1.4$&$   2.2\pm   0.3$&$\phn 300$&F \\
 &S. Filament 1*&310.927&$+0.277$&$\phn 2.7$&$  18.0\pm   0.6$&$\phn \phn 57$&$  17.6\pm   2.3$&$   2.1\pm   0.5$&$\phn \phn 60$&F \\
 &S. Filament 2*&310.896&$+0.379$&$\phn 2.5$&$  19.3\pm   0.5$&$\phn \phn 97$&$  16.5\pm   1.6$&$   2.6\pm   0.4$&$\phn 140$&F (ID) \\
 &E. Filament 1*&311.086&$+0.429$&$\phn 5.5$&$  19.7\pm   0.5$&$\phn 300$&$  19.4\pm   1.5$&$   2.0\pm   0.3$&$\phn 310$&F \\
 &E. Filament 2*&311.052&$+0.394$&$\phn 1.1$&$  13.0\pm   0.3$&$\phn 290$&$  12.1\pm   0.9$&$   2.4\pm   0.4$&$\phn 380$&F (ID) \\
G332.5$-$0.1&Entire&332.539&$-0.141$&$13.6$&$  25.9\pm   0.8$&$4500$&$  26.1\pm   3.3$&$   2.0\pm   0.3$&$4500$&E \\
 &HII 1&332.544&$-0.126$&$\phn 0.9$&$  29.2\pm   1.0$&$\phn 590$&$  29.1\pm   3.1$&$   2.0\pm   0.2$&$\phn 590$&UC \\
 &HII 2&332.445&$-0.178$&$\phn 4.4$&$  24.0\pm   0.7$&$\phn 330$&$  19.3\pm   1.4$&$   2.6\pm   0.2$&$\phn 540$&UC \\
 &W. PS&332.603&$-0.167$&$\phn 1.6$&$  20.0\pm   0.5$&$1200$&$  20.4\pm   1.4$&$   1.9\pm   0.2$&$1100$&PS \\
 &E. PS&332.467&$-0.130$&$\phn 1.8$&$  20.7\pm   0.5$&$\phn 850$&$  18.5\pm   1.2$&$   2.4\pm   0.2$&$1100$&PS \\
 &E. PDR&332.560&$-0.149$&$\phn 1.1$&$  22.6\pm   0.6$&$\phn 650$&$  20.9\pm   1.5$&$   2.2\pm   0.2$&$\phn 770$&PDR \\
 &S. PDR&332.526&$-0.155$&$\phn 1.4$&$  25.5\pm   0.8$&$\phn 650$&$  23.7\pm   2.0$&$   2.2\pm   0.2$&$\phn 750$&PDR \\
 &E. Filament*&332.648&$-0.188$&$\phn 3.2$&$  15.9\pm   0.3$&$\phn 700$&$  15.1\pm   1.1$&$   2.2\pm   0.3$&$\phn 810$&F (ID) \\
 &S. Filament*&332.404&$-0.150$&$\phn 5.0$&$  19.6\pm   0.5$&$\phn 820$&$  21.7\pm   1.9$&$   1.7\pm   0.3$&$\phn 660$&F (ID) \\
 &W. Filament&332.501&$-0.119$&$\phn 1.5$&$  22.5\pm   0.6$&$\phn 710$&$  19.9\pm   1.3$&$   2.3\pm   0.2$&$\phn 930$&F (ID) \\
RCW\,120&Entire&348.251&$+0.481$&$100.3$&$  23.9\pm   0.6$&$2100$&$  26.0\pm   2.4$&$   1.7\pm   0.3$&$1800$&E \\
 &Cond. 1 PS&348.184&$+0.482$&$\phn \phn 0.7$&$  20.9\pm   0.5$&$\phn 300$&$  22.6\pm   1.6$&$   1.8\pm   0.2$&$\phn 260$&PS \\
 &Cond. 1 (Cond. 1)&348.180&$+0.480$&$\phn \phn 4.7$&$  21.0\pm   0.5$&$\phn 760$&$  23.5\pm   1.6$&$   1.7\pm   0.2$&$\phn 600$&PDR \\
 &S. PDR 1 (Cond. 2)&348.226&$+0.413$&$\phn \phn 2.5$&$  24.1\pm   0.6$&$\phn 160$&$  25.5\pm   2.0$&$   1.8\pm   0.2$&$\phn 140$&PDR \\
 &E. PDR 1 (Cond. 3)&348.272&$+0.403$&$\phn \phn 2.6$&$  23.4\pm   0.6$&$\phn \phn 88$&$  23.4\pm   1.9$&$   2.0\pm   0.2$&$\phn \phn 88$&PDR \\
 &E. PDR 2 (Cond. 4)&348.302&$+0.429$&$\phn \phn 3.0$&$  22.6\pm   0.6$&$\phn 110$&$  22.3\pm   1.7$&$   2.0\pm   0.2$&$\phn 110$&PDR \\
 &W. PDR 1 (Cond. 7)&348.199&$+0.531$&$\phn \phn 0.7$&$  24.8\pm   0.7$&$\phn \phn 20$&$  23.4\pm   1.9$&$   2.1\pm   0.2$&$\phn \phn 23$&PDR \\
 &S. PDR 1 (Cond. 10)&348.203&$+0.426$&$\phn \phn 1.0$&$  26.2\pm   0.8$&$\phn \phn 38$&$  25.8\pm   2.2$&$   2.0\pm   0.2$&$\phn \phn 39$&PDR \\
 &W. PDR 2 (PDR West)&348.266&$+0.556$&$\phn \phn 0.8$&$  31.6\pm   1.2$&$\phn \phn \phn 6.0$&$  20.7\pm   1.6$&$   3.0\pm   0.2$&$\phn \phn 14$&PDR \\
 &E. PDR 3 (PDR East)&348.329&$+0.464$&$\phn \phn 1.2$&$  25.3\pm   0.7$&$\phn \phn 31$&$  20.8\pm   1.4$&$   2.5\pm   0.2$&$\phn \phn 47$&PDR \\
 &W. IRDC 1*&348.179&$+0.600$&$\phn \phn 1.5$&$  12.7\pm   0.3$&$\phn \phn 30$&$  12.0\pm   0.9$&$   2.3\pm   0.4$&$\phn \phn 37$&F (ID) \\
 &W. IRDC 2*&348.165&$+0.662$&$\phn \phn 2.3$&$  14.0\pm   0.3$&$\phn \phn 30$&$  13.1\pm   0.8$&$   2.3\pm   0.3$&$\phn \phn 37$&F (ID) \\
 &W. IRDC 3*&348.196&$+0.768$&$\phn \phn 1.8$&$  13.5\pm   0.2$&$\phn \phn 64$&$  14.1\pm   1.1$&$   1.8\pm   0.4$&$\phn \phn 55$&F (ID) \\
 &W. IRDC 4*&348.234&$+0.591$&$\phn \phn 1.8$&$  17.6\pm   0.4$&$\phn \phn 28$&$  19.6\pm   2.0$&$   1.6\pm   0.4$&$\phn \phn 22$&F (ID) \\
 &N. IRDC 1* (Cond. 6)&348.292&$+0.644$&$\phn \phn 2.0$&$  15.0\pm   0.3$&$\phn 160$&$  15.4\pm   0.8$&$   1.9\pm   0.2$&$\phn 150$&F (ID) \\
 &N. IRDC 2* (Cond. 9)&348.345&$+0.652$&$\phn \phn 1.5$&$  13.4\pm   0.2$&$\phn 110$&$  12.5\pm   0.6$&$   2.3\pm   0.2$&$\phn 140$&F (ID) \\
 &N. IRDC 3*&348.297&$+0.682$&$\phn \phn 1.4$&$  13.1\pm   0.6$&$\phn \phn 26$&$  11.9\pm   1.5$&$   2.3\pm   0.4$&$\phn \phn 36$&F (ID) \\
 &N. IRDC 4*&348.305&$+0.703$&$\phn \phn 0.8$&$  12.2\pm   0.2$&$\phn \phn 38$&$   9.9\pm   0.4$&$   3.2\pm   0.3$&$\phn \phn 88$&F (ID) \\
 &N. IRDC 5*&348.325&$+0.716$&$\phn \phn 1.4$&$  11.5\pm   0.4$&$\phn \phn 73$&$  10.1\pm   1.1$&$   2.5\pm   0.4$&$\phn 120$&F (ID) \\
 &N. IRDC 6*&348.476&$+0.583$&$\phn \phn 1.3$&$  11.8\pm   0.5$&$\phn \phn 61$&$  11.6\pm   1.4$&$   2.1\pm   0.4$&$\phn \phn 65$&F (ID) \\
 &E. IRDC* (Cond. 5)&348.375&$+0.487$&$\phn 10.6$&$  15.9\pm   0.3$&$\phn 520$&$  16.3\pm   0.9$&$   1.9\pm   0.2$&$\phn 490$&F (ID) \\
 &S. Filament 1 (S. Filament 1)*&348.121&$+0.473$&$\phn \phn 5.4$&$  16.6\pm   0.3$&$\phn 160$&$  16.6\pm   1.0$&$   2.0\pm   0.2$&$\phn 160$&F \\
 &S. Filament 2 (S. Filament 2)*&348.134&$+0.512$&$\phn \phn 4.4$&$  15.5\pm   0.3$&$\phn 140$&$  15.2\pm   0.9$&$   2.1\pm   0.3$&$\phn 140$&F \\
 &S. Filament 3 (Cond. 8)*&348.149&$+0.421$&$\phn \phn 3.3$&$  16.6\pm   0.3$&$\phn 160$&$  16.1\pm   0.9$&$   2.1\pm   0.2$&$\phn 180$&F \\

\hline
\label{tab:apphot}
\end{longtable}
\footnotetext[1]{The parenthetical names are from previous works:
  \citet{rodon10} for Sh\,104, \citet{sharpless59} and
  \citet{sugitani91} for W5-E, \citet{zavagno06} for RCW\,79, and
  \citet{zavagno07} and \citet{anderson10a} for RCW\,120.  The names
  of apertures where we used a local background value are marked with
  an asterisk.}
\footnotetext[2]{The classifications are: ``E'' for entire, {``H{\scriptsize II}''}
  for other H{\scriptsize II} regions in the field, ``PS'' for point source, ``PDR'' for apertures
  along the photodissociation regions, ``F'' for filaments, and ``F
  (ID)'' for IRDC filaments seen in absorption at 8.0\,\micron.}
}
}
\renewcommand{\thefootnote}{\arabic{footnote}}
\clearpage

\appendix
\section{Comments on Individual Regions\label{sec:individual}}
Below we comment briefly on the eight bubble regions in our sample.
The masses and temperatures quoted are those of the $\beta=2.0$ trial.

\subsection{Sh\,104}
We find a total associated mass for Sh\,104 of $\sim4,\,000$\,\msun,
which is in the middle of the range for the \hii\ regions investigated
here.  This calculated mass is two-thirds that found in
\citet{deharveng03} using molecular gas tracers.  As mentioned
previously, the interior of Sh\,104 has more emission at all
wavelengths than the other regions in the sample.  Roughly 30\% of the
total FIR emission from the region comes from the interior region.  We
hypothesize that this is a result of Sh\,104 being nearly complete -
some of the radiation that would otherwise escape is trapped within
the bubble.

In Figure~\ref{fig:herschel}, there is a clear difference in
temperature between Sh\,104 itself and the local ambient medium.  We
find that its PDR is well-characterized by temperatures of 25\,K.
Local filaments are for the most part $\sim20$\,K, although
``N. Filament~1'' and ``N. Filament~5''\ appear to be significantly
colder at $15$\,K and $17$\,K respectively.  The temperature map of
Sh\,104 (Figure~\ref{fig:tempmap}) is relatively smooth but there is a
trend for hotter temperatures toward the southwest, and colder
temperatures toward the north-west.  The warmest region in the field
is the separate \hii\ region at \lb = (73.790, +0.572).  It has a mean dust
temperature of $28$\,K and a total associated mass of
$\sim400$\,\msun.  The other source found along the PDR, ``CS'' has a
mass of $\sim100$\,\msun.

The column density map in Figure~\ref{fig:column} shows the highest
values, up to $10^{22}\,{\rm cm^{-2}}$, along the PDR.  Local
filaments have mean column densities of $>10^{21}\,{\rm cm^{-2}}$.
There are a number of prominent filaments that begin on the border of
Sh\,104 and lead radially away; these are prominent in the column
density map.

\subsection{W5-E}
The most surprising aspect of W5-E found here is that its dust
temperatures vary little across the field.  Figure~\ref{fig:herschel}
shows a narrow range of {\it Herschel} colors, which leads to
relatively small variations in dust temperature.  The dust temperature
of the entire region is $24$\,K.
The coldest regions of the field are to the east (``E.~Filament~1''
and ``E.~Filament~2'') and north (``N.~Filament~1'' and
``N.~Filament~2'') -- they are between 16 and 18\,K.  There are other
cold locations along the PDR, most notably behind the ionization front
in ``N.~PDR~1'' (BRC13) and ``E.~PDR~1~(BRC14).''  The total
associated mass of W5-E, $\sim8,\,000\,\msun$, places it as the second
most massive in the sample; it is the largest region in the sample in
terms of physical diameter.

There are three other \hii\ regions in the field of W5-E.  With the
exception of the separate \hii\ region on the border of in RCW\,79, these are the most
massive secondary \hii\ regions in our sample.  The bipolar
\hii\ region Sh\,201 at \lb\ = (138.481, +1.637) is the hottest
location in the field at $27$\,K and stands out clearly in
Figure~\ref{fig:tempmap}.  Its total associated mass is
$\sim800$\,\msun.  The other \hii\ regions on the eastern border
are $\sim25$\,K; they have total associated masses of
$\sim600$\,\msun.

W5-E is unique in our sample in that it displays a number of features
obviously affected by the impinging stellar radiation.  These features
are discussed in detail in \citet{deharveng12}.  The
bright-rimmed clouds BRC12 and BRC13, which harbor embedded sources,
have mean temperatures of 24\,K and 22\,K, respectively, while the
dust associated with the embedded sources has a temperature of
$\sim22$\,K.  The embedded sources are conspicious in the temperature
map (Figure~\ref{fig:tempmap}) as higher temperature regions within
cold temperature clumps.  The pillars detected with {\it Spitzer} seen
to the southwest are also cool and have an average temperature of
$\sim20$\,K.  Many of these pillars have embedded stars forming at their
``tips'' detected by {\it Herschel}.  This is consistent with a
scenario where the majority of the pillar is shielded from impinging
ratiation and is thus able to maintain the cold temperatures necessary
for subsequent star formation.






\subsection{Sh\,241}
The temperature map of Sh\,241 in Figure~\ref{fig:tempmap} shows
relatively little variation across the field.  The large
high-temperature region to the north is probably not real as it is a
region of very low intensity emission.  The bubble \hii\ region itself is
$23$\,K.  The prominant separate compact \hii\ region is slightly warmer,
$25$\,K while filaments average $\sim17$\,K.  The total mass for
Sh\,241 is $\sim2,\,000$\,\msun\ while that of the compact \hii\ region is
$\sim300$\,\msun.  Sh\,241 is ringed by a region of cold emission to
the north of temperature $\sim18$\,K.  This region has a higher than
average column density, $>10^{21.5}\,{\rm cm^{-2}}$.  The colder
emission appears to be part of a large filament running East-West that
also contains another \hii\ region and what appears to be cold
protostars.

\subsection{RCW\,71}
As a whole, RCW\,71 is the warmest region in our sample at 30\,K.
RCW\,71 is also the least massive region in the sample, with
a total associated mass of just $\sim200$\,\msun.  This low mass estimate
may be due to an inaccurate distance.  As mentioned in
\S\ref{sec:hii_sample}, the kinematic and spectroscopic distances for
this source do not agree.  We note, however, that the PDR of RCW\,71
has only marginally higher column density values than the background
(Figure~\ref{fig:column}), and therefore the low associated mass
estimate may be real.

The temperature map of RCW\,71 in Figure~\ref{fig:tempmap} shows high
temperatures in the PDR up to $\sim35$\,K (in apertures ``E.~PDR~1''
and ``E.~PDR~2'').  Material has accumulated into cold massive
filaments most prominently to the east, but also to the west.  The
coldest of these, ``N.~Filament'' has a temperature of $13$\,K --
it is one of the coldest regions in the sample.  The field of RCW\,71
also has one of the coldest point sources, ``S.~PS~1.''  It has a
temperature of $15$\,K.

\subsection{RCW\,79}
RCW\,79 has a total associated mass of $\sim10,\,000\,\msun$, which
makes it the most massive in our sample.  This mass does not include
the secondary PDRs to the south, which are themselves
$\sim3,\,000$\,\msun\ combined.  Our mass estimates are considerably
higher than that of \citet{zavagno06}, who found 2,\,000\,\msun\ for
the entire region using 1.2\,mm observations.  Their observations,
however, may not have been sensitive to the more diffuse emission
detected by {\it Herschel}.  The column density map in
Figure~\ref{fig:column} has values in the PDR of up to
$\sim10^{22.5}\,{\rm cm^{-2}}$ and looks very similar to the 1.2\,mm
maps in \citet{zavagno06}.  Along the south-eastern border there is
another \hii\ region that has a total associated mass of
$\sim2,\,000\,\msun$.  This is the most massive secondary \hii\ region
in our sample -- it is significantly more massive than RCW\,71 and
approximately the same mass as RCW\,120.

RCW\,79 is interesting in that there is significant patches of cold
material ringing the PDR to the south and to the west.  These colder
areas are apparent in Figure~\ref{fig:tempmap}.  The coldest objects
in the field, the ``N. Filament'' and the ``E. Filament'' are
$22$\,K and $24$\,K, respectively.  There is ongoing star
formation in the field.  Within ``E.~Filament'' there are two compact
objects and there are compact sources detected in the ``S.~PDR''
aperture.

\subsection{RCW\,82}
Like RCW\,71, the PDR of RCW\,82 is filamentary, but its temperature
of $25$\,K is cooler than that of RCW\,71.  The total associated
mass of RCW\,82 is $\sim3,\,000$\,\msun.  This is significantly less
than what was found by \citet{pomares09} from CO observations.
The temperature in the field of RCW\,82 seen in
Figure~\ref{fig:tempmap} shows little variation.  To the west, RCW\,82
has what appears to be a second PDR (``W.~PDR~2'').  The IRDCs seen in
the field range from $\sim20$\,K for ``N.~Filament'' to $<15$\,K for ``E.~Filemant~2''.  As for Sh\,104, there are a number of cool
filaments leading radially away from RCW\,82; these have column
density values of $\sim10^{21.5}\,{\rm cm^{-2}}$.  The eastern PDR of
RCW\,82 has a number of cold locations.  These are evident in the
colors seen in Figure~\ref{fig:herschel_regions}, and also in the
temperature map in Figure~\ref{fig:tempmap}.

\subsection{G332.5$-$0.1}
G332.5$-$0.1 is angularly small and located along a prominent IRDC.
The total associated mass of G332.5$-$0.1 is
$\sim4,\,000$\,\msun\ which is on the higher end of the range of
\hii\ regions masses studied here.  The IRDC itself has a temperature
of up to $\sim20$\,K, which is warmer than other IRDCs here studied.
Indeed, as seen in Figure~\ref{fig:tempmap}, the IRDC is not
well-separated in temperature from the surrounding region.  The column
density distribution (Figure~\ref{fig:column}) shows the IRDC
prominently - it has column density values of $\sim10^{22}\,{\rm
  cm^{-2}}$.

Within the IRDC there are two prominent condensations: ``W.~PS'' and
``E.~PS''.  These condensations have masses of $\sim1,\,000$\,\msun.  The PDR has massive condensations to the
north, south, and east.  The northern condensation contains an
\hii\ region, with a total associated mass of of $\sim600$\,\msun.
This location is the hottest in the field, $29$\,K.

\subsection{RCW\,120}
The wide range of colors seen Figure~\ref{fig:herschel}
  indicates that RCW\,120 also has a large range of dust
temperatures, compared to the other \hii\ regions in the sample.
RCW\,120 has the largest concentration of IRDCs in its surroundings of
any object in our sample.  These IRDCs are the coldest objects found
in our analysis, averaging $13$\,K.  The largest of the IRDCs,
``E. IRDC,'' has a mass of $\sim500$\,\msun and has numerous
protostars detected within it.  This IRDC has a temperature derived
from aperture photometry of $16$\,K.  This temperature is almost
certainly affected by the numerous embedded protostars which have not
been removed when calculating temperatures.  The temperature map
(Figure~\ref{fig:tempmap}) shows temperatures of $\le15$\,K for
quiescent parts of the cloud where there are no detected protostars.
There are two large protostars detected in the ``E.~IRDC'' aperture
that are visible in the temperature map -- they have temperatures of
$\sim18$\,K.

A point source is detected within Condensation~1 at 70\,\micron\ and
100\,\micron\ \citep{zavagno10a} for which we find a total associated
mass of $\sim300$\,\msun.  The entire condensation has a mass of
$\sim800$\,\msun.  It is easily the most massive condensation in the
field.  In fact, it's mass is one third that of the total associated
mass for the RCW\,120 region, which is only $\sim2,\,000$\,\msun.  Using
the same 870\,\micron\ APEX-LABOCA data used here, \citet{deharveng09}
calculated a mass for the source within Condensation~1 of
140-250\,\msun\ and a mass for the entire condensation of
460-800\,\msun; the range of values comes from estimates of the dust
temperature from 30\,K to 20\,K.

The column density map of RCW\,120 (Figure~\ref{fig:column})
highlights the numerous filamentary structures.  These are prominent
in the source, but are also detected leading away from the ``E.~IRDC''
region.  The PDR and the local IRDCs have roughly the same value of
$N_{\rm H}, \sim10^{22}\,{\rm cm^{-2}}$.

\citet{anderson10a} found that for RCW\,120, the ionization front
appears ``patchy'' and that there are numerous locations where hot
dust is detected beyond the ionization front.  They hypothesized that
these locations represent holes in the PDR where radiation may escape
and heat the ambient medium.  Such radiation may cause an increase in
pressure which would aid in collapsing existing pre-stellar clumps.
We confirm their results here; the same three warmer locations can
again be identified beyond the ionization front of RCW\,120.  It is
interesting, however, that RCW\,120 remains the clearest example of
this phenomenon.  This may be due to its proximity to the Sun.

\vskip 20pt
We wish to thank the referee for a close reading of this
  manuscript, and especially for comments that led to significant
  improvements of the simulations in Section~\ref{sec:beta_t}.
SPIRE has been developed by a consortium of institutes led by Cardiff
Univ. (UK) and including Univ. Lethbridge (Canada); NAOC (China); CEA,
LAM (France); IFSI, Univ. Padua (Italy); IAC (Spain); Stockholm
Observatory (Sweden); Imperial College London, RAL, UCL-MSSL, UKATC,
Univ. Sussex (UK); Caltech, JPL, NHSC, Univ. Colorado (USA). This
development has been supported by national funding agencies: CSA
(Canada); NAOC (China); CEA, CNES, CNRS (France); ASI (Italy); MCINN
(Spain); SNSB (Sweden); STFC (UK); and NASA (USA).\\ PACS has been
developed by a consortium of institutes led by MPE (Germany) and
including UVIE (Austria); KU Leuven, CSL, IMEC (Belgium); CEA, LAM
(France); MPIA (Germany); INAF-IFSI/OAA/OAP/OAT, LENS, SISSA (Italy);
IAC (Spain). This development has been supported by the funding
agencies BMVIT (Austria), ESA-PRODEX (Belgium), CEA/CNES (France), DLR
(Germany), ASI/INAF (Italy), and CICYT/MCYT (Spain).\\ This research
has made use of NASA’s Astrophysics Data System Bibliographic Services
and the SIMBAD database operated at CDS, Strasbourg,
France.\\ L.D.A. acknowledges support by the ANR Agence Nationale for
the research project ``PROBeS'', number ANR-08-BLAN-0241.

\bibliographystyle{aa} 
\bibliography{/home/loren/papers/ref.bib} 

\end{document}